\documentclass[aps,prx,superscriptaddress,twocolumn,longbibliography]{revtex4-2}
\usepackage{units}
\usepackage{amsmath}
\usepackage{amsthm}
\usepackage{amssymb}
\usepackage{graphicx}
\usepackage{color}
\usepackage{xcolor}
\usepackage{bbold}

\usepackage[colorlinks=true, urlcolor=blue, citecolor=blue,linkcolor=blue,citebordercolor={1 0 0},linkbordercolor={0 0 1}]{hyperref}

\usepackage[linesnumbered, ruled,vlined]{algorithm2e}

\graphicspath{{./Images/},{./ImagesAppendix/}}

\renewcommand{\eqref}[1]{Eq.~(\ref{#1})} 

\theoremstyle{plain}

\theoremstyle{plain}

\ifx\proof\undefined
\newenvironment{proof}[1][\protect\proofname]{\par
	\normalfont\topsep6\p@\@plus6\p@\relax
	\trivlist
	\itemindent\parindent
	\item[\hskip\labelsep\scshape #1]\ignorespaces
}{%
	\endtrivlist\@endpefalse
}
\providecommand{\proofname}{Proof}
\fi
\theoremstyle{plain}

\theoremstyle{remark}

\newcommand{\bra}[1]{\langle #1|}
\newcommand{\ket}[1]{|#1 \rangle}
\makeatother
\newcommand{\braket}[2]{\langle #1 \vert #2 \rangle}
\newcommand{\abs}[1]{\left|#1\right|}
\newcommand{\idg}[1]{{\bfseries #1)}}

\newcommand\numberthis{\addtocounter{equation}{1}\tag{\theequation}}
\providecommand{\factname}{Fact}
\providecommand{\theoremname}{Theorem}
\providecommand{\claimname}{Claim}
\providecommand{\lemmaname}{Lemma}
\providecommand{\definitionname}{Definition}

\definecolor{KB}{rgb}{0.4,0.3,0.9}

\definecolor{THc}{rgb}{0.9,0.3,0.2}

\theoremstyle{definition}

\newcommand{\subfigimg}[3][,]{%
	\setbox1=\hbox{\includegraphics[#1]{#3}}
	\leavevmode\rlap{\usebox1}
	\rlap{\hspace*{2pt}\raisebox{\dimexpr\ht1-0.5\baselineskip}{{\bfseries \large\textsf{#2}}}}
	\phantom{\usebox1}
}

\begin{document}

\title{Optimal training of variational quantum algorithms without barren plateaus}

\author{Tobias Haug}
\email{thaug@ic.ac.uk}
\affiliation{QOLS, Blackett Laboratory, Imperial College London SW7 2AZ, UK}
\author{M. S. Kim}
\affiliation{QOLS, Blackett Laboratory, Imperial College London SW7 2AZ, UK}
\begin{abstract}
Variational quantum algorithms (VQAs) promise efficient use of near-term quantum computers. However, training VQAs often requires an extensive amount of time and suffers from the barren plateau problem where the magnitude of the gradients vanishes with increasing number of qubits.
Here, we show how to optimally train VQAs for learning quantum states. Parameterized quantum circuits can form Gaussian kernels, which we use to derive adaptive learning rates for gradient ascent. We introduce the generalized quantum natural gradient that features stability and optimized movement in parameter space. Both methods together outperform other optimization routines in training VQAs. Our methods also excel at numerically optimizing driving protocols for quantum control problems. 
The gradients of the VQA do not vanish when the fidelity between the initial state and the state to be learned is bounded from below. We identify a VQA for quantum simulation with such a constraint that thus can be trained free of barren plateaus.
Finally, we propose the application of Gaussian kernels for quantum machine learning.
\end{abstract}

\maketitle

\section{Introduction}
Quantum computers promise to tackle important problems that can be difficult for classical computers. 
To effectively use quantum computers available in the near future~\cite{preskill2018quantum,bharti2021noisy} variational quantum algorithms (VQAs) have been put forward.
These VQAs solve tasks by iteratively updating a parameterized quantum circuit (PQC) with a classical optimization routine in a feedback loop~\cite{peruzzo2014variational,kandala2017hardware,mcclean2016theory,cerezo2020variational}.
However, training of such VQAs can take an extensive amount of iterations, yielding a long time until the algorithm converges.
Further, VQAs are often impeded by the barren plateau problem where the variance of the gradients vanish exponentially with increasing number of qubits~\cite{mcclean2018barren}, specific cost functions~\cite{cerezo2021cost}, entanglement~\cite{marrero2020entanglement} and noise~\cite{wang2020noise}.
Further, the optimization routine of VQAs was shown to be NP-hard even for problems that are easy for classical computers~\cite{bittel2021training}.
Quantum algorithms that can avoid training PQCs to circumvent the barren plateau problem have been proposed~\cite{bharti2020quantum,bharti2020quantum2,bharti2020iterative,haug2020generalized,lau2021nisq,lim2021fastforwarding}.
To improve training of VQAs, one can use quantum geometric information to define the quantum natural gradient (QNG)~\cite{stokes2020quantum,koczor2019quantum,yamamoto2019natural}. However, the QNG can be unstable without regularization~\cite{van2020measurement,wierichs2020avoiding,gacon2021simultaneous} and suffers from barren plateaus as well~\cite{haug2021capacity}. Further, the learning rates for training with the QNG are so far chosen only in a heuristic manner.
A core challenge in quantum technology is learning how to control quantum systems efficiently. This task is routinely encountered in quantum control theory~\cite{d2007introduction} for finding protocols that prepare a quantum state using a controllable quantum system. For smaller quantum systems, numerical methods are employed to find good protocols by simulating the quantum system on classical computers~\cite{machnes2011comparing}.
VQAs learn quantum states by employing quantum computers and PQCs. Learning quantum states is an important subroutine in VQAs such as the projected variational quantum dynamics method~\cite{otten2019noise,barison2021efficient}, variational fast forwarding~\cite{gibbs2021long}, learning of scramblers~\cite{holmes2020barren}, quantum circuit born machines~\cite{benedetti2019generative}, quantum generative adversarial networks~\cite{huang2020experimental}, quantum autoencoders~\cite{romero2017quantum} and excited state calculations~\cite{higgott2019variational}.

Here, we show how to optimally train VQAs to prepare target quantum states.
We find that the fidelity between two states forms an approximate Gaussian kernel in respect to the distance in parameter space of the PQC, with the quantum Fisher information metric (QFIM) as the weight matrix.
With this result, we enhance the gradient ascent algorithm. We derive adaptive learning rates that are adjusted for every iteration of gradient ascent. Further, we introduce the generalized quantum natural gradient (GQNG) that interpolates between standard gradient and QNG. We find a type of GQNG that is stable without regularization while providing good movement in parameter space.
By combining adaptive learning rates and the QNG, we find that the training of VQAs and quantum control problems requires a substantially lower number of iterations compared to other optimization techniques.
We analytically derive the variance of the gradient of the VQA. 
When the fidelity between the initial state and the state to be learned has a lower bound, the variance of the gradient has also a lower bound that is independent of the number of qubits. Thus, the gradient will not vanish with increasing number of qubits. We show a type of VQA, the projected variational quantum dynamics method or restarted quantum dynamics method~\cite{otten2019noise,barison2021efficient}, that fulfills this condition and thus can be trained free of barren plateaus.
Our methods can improve various VQAs and numerical quantum control techniques, and promise implementation on near-term quantum computers.
As further application, our results are directly relevant for quantum machine learning. The Gaussian kernel can be realized by hardware-efficient PQCs to run quantum machine learning algorithms on the current quantum hardware.

\begin{figure*}[htbp]
	\centering
	\includegraphics[width=0.7\textwidth]{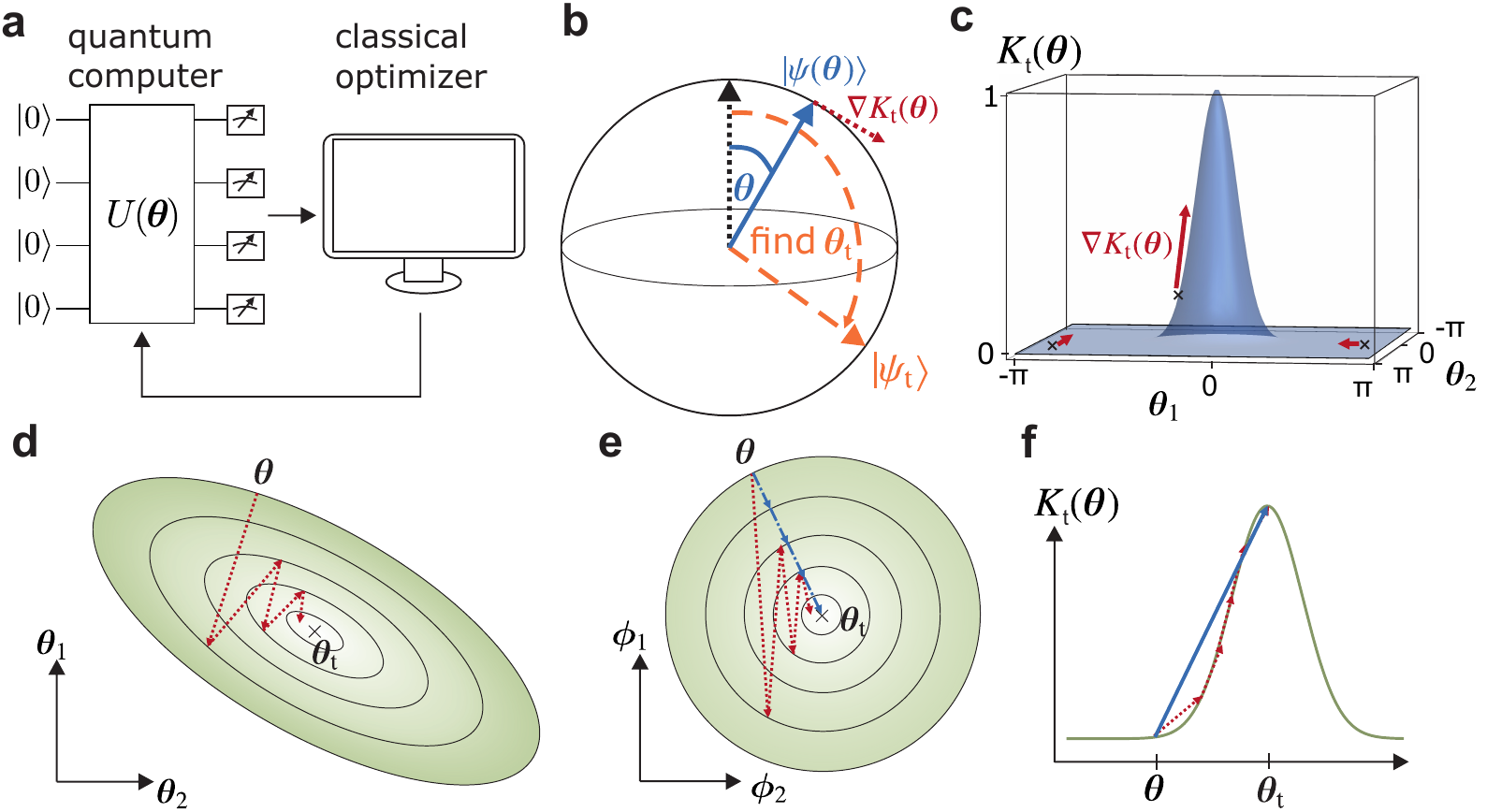}
	\caption{
	\idg{a} The variational quantum algorithm (VQA) consists of a parameterized quantum circuit (PQC) that generates the quantum state $\ket{\psi(\boldsymbol{\theta})}=U(\boldsymbol{\theta})\ket{0}$ with unitary $U(\boldsymbol{\theta})$ and parameters $\boldsymbol{\theta}$, as well as a classical optimization routine. Measurements on the quantum state are used to calculate the cost function, which is then optimized by the classical optimizer in a feed back loop by adjusting the parameters $\boldsymbol{\theta}$.
	\idg{b} VQA to represent the target state $\ket{\psi_\text{t}}$ using $\ket{\psi(\boldsymbol{\theta})}$. Goal is to find target parameters $\boldsymbol{\theta}_\text{t}=\text{argmax}_{\boldsymbol{\theta}}K_\text{t}(\boldsymbol{\theta})$ that approximate the target state by maximizing the fidelity $K_\text{t}(\boldsymbol{\theta})=\abs{\braket{\psi_\text{t}}{\psi(\boldsymbol{\theta})}}^2$. Training is performed using the gradient $G_0(\boldsymbol{\theta})=\nabla K_\text{t}(\boldsymbol{\theta})$, which points in the direction of steepest increase of fidelity. 
	\idg{c} The landscape of the fidelity $K_\text{t}(\boldsymbol{\theta})$ as function of $\boldsymbol{\theta}$ often has barren plateaus, where the fidelity and its gradients are exponentially small within most of the parameter space. However, as long as the initial quantum state of the VQA is guaranteed to have a lower bounded fidelity, then the magnitude of the gradient does not vanish even for many qubits and the barren plateaus can be avoided~(Eq.\ref{eq:var_grad_bound}).
	\idg{d} Gradient ascent optimizes the fidelity by updating $\boldsymbol{\theta}'=\boldsymbol{\theta} +\alpha G_0(\boldsymbol{\theta})$. As the fidelity landscape as function of $\boldsymbol{\theta}$ is in general not euclidean, standard gradient ascent does not take the fastest path. 
	\idg{e} By using quantum geometric information about the parameter space with the quantum Fisher information metric (QFIM) $\mathcal{F}(\boldsymbol{\theta})$, the parameter space can be transformed to get the quantum natural gradient (QNG) which moves in the best direction (see solid blue curve). To yield stable gradients in practice, the QNG requires regularization. The generalized quantum natural gradient (GQNG) (\eqref{eq:general_gradient}) interpolates between standard gradient and QNG, and can be stable without regularization.
	\idg{f} The learning rate $\alpha$ for the gradient update is normally a fixed heuristic learning rate (dashed red curves). The fidelity of PQCs can form Gaussian kernels, which is used to calculate adaptive learning rates (\eqref{eq:update_add}) for each gradient update (blue solid curve).
	}
	\label{fig:sketch}
\end{figure*}

\section{Model}
A unitary $U(\boldsymbol{\theta})$ parameterized by the $M$-dimensional parameter vector $\boldsymbol{\theta}\in\mathbb{R}^M$ generates the quantum state $\ket{\psi(\boldsymbol{\theta})}=U(\boldsymbol{\theta})\ket{0}$ consisting of $N$ qubits (see Fig.\ref{fig:sketch}a).
Our goal is to learn the target parameters $\boldsymbol{\theta}_\text{t}$ that approximate a given target state $\ket{\psi_\text{t}}$ (see Fig.\ref{fig:sketch}b). This is achieved by the optimization task $\boldsymbol{\theta}_\text{t}=\text{argmax}_{\boldsymbol{\theta}}K_\text{t}(\boldsymbol{\theta})$, where $K_\text{t}(\boldsymbol{\theta})=\abs{\braket{\psi_\text{t}}{\psi(\boldsymbol{\theta})}}^2$ is the fidelity.
A common approach to optimize the fidelity is standard gradient ascent. Here, one calculates the gradient of the fidelity $\nabla K_\text{t}(\boldsymbol{\theta})$, which is the direction of largest increase for the fidelity. Then, the parameter $\boldsymbol{\theta}$ is iteratively updated with the rule $\boldsymbol{\theta}'=\boldsymbol{\theta}+\alpha \nabla K_\text{t}(\boldsymbol{\theta})$, where $\alpha$ is the learning rate. 
After a number of iterations, the quality of the found approximate solution $\boldsymbol{\theta}_\text{t}'$ is measured with the infidelity
\begin{equation}\label{eq:infidelity}
\Delta K_\text{t}(\boldsymbol{\theta}_\text{t}')=1-K_\text{t}(\boldsymbol{\theta}_\text{t}')=1-\abs{\braket{\psi_\text{t}}{\psi(\boldsymbol{\theta}_\text{t}')}}^2\,.
\end{equation}
However, training VQAs is plagued by the following issues.
First, the gradient often becomes exponentially small with increasing number of qubits due to the barren plateau problem (see Fig.\ref{fig:sketch}c).
Second, the update rule assumes that the parameter space is euclidean, i.e. the fidelity varies at the same rate for every parameter direction. However, in general this is not the case as can be seen in Fig.\ref{fig:sketch}d. Then, the standard gradient is not the optimal direction for the parameter update. 
This can be aided by using the QFIM $\mathcal{F}(\boldsymbol{\theta})$.  For $\ket{\psi}=\ket{\psi(\boldsymbol{\theta})}$, it is given by~\cite{meyer2021fisher}
\begin{equation}\label{eq:quantumFisher}
\mathcal{F}_ {ij}(\boldsymbol{\theta})=4[\braket{\partial_i \psi}{\partial_j \psi}-\braket{\partial_i \psi}{\psi}\braket{\psi}{\partial_j \psi}]\,,
\end{equation} 
where $\partial_j\ket{\psi}$ is the gradient in respect to the $j$-th element of $\boldsymbol{\theta}$. For pure states, the QFIM tells us the change of fidelity for small parameter variations $\text{d}\boldsymbol{\mu}$~\cite{liu2019quantum,meyer2021fisher,haug2021capacity}
\begin{equation}\label{eq:fidelity_QFIM}
\mathcal{K}(\boldsymbol{\theta},\boldsymbol{\theta}+\text{d}\boldsymbol{\mu})=\abs{\braket{\psi(\boldsymbol{\theta})}{\psi(\boldsymbol{\theta}+\text{d}\boldsymbol{\mu})}}^2= 1-\frac{1}{4}\text{d} \boldsymbol{\mu}^{\text{T}}\mathcal{F}(\boldsymbol{\theta})\text{d} \boldsymbol{\mu}\,.
\end{equation}
The QFIM contains information about how the geometry of the parameter space is related to the geometry of quantum states.
The QNG $G_1(\boldsymbol{\theta})=\mathcal{F}^{-1}(\boldsymbol{\theta})\nabla K_\text{t}(\boldsymbol{\theta})$~\cite{stokes2020quantum,yamamoto2019natural,koczor2019quantum} utilizes this information to construct gradient updates which take the best path in the parameter space (see Fig.\ref{fig:sketch}e).
However, when applied in practice the QNG requires an additional hyperparameter for regularization to take care of the ill-conditioned inverse of the QFIM~\cite{gacon2021simultaneous,wierichs2020avoiding,van2020measurement}.
Third, the learning rate $\alpha$ is generally not known and is often determined only heuristically by setting it to a small constant (see Fig.\ref{fig:sketch}f). 
In the following, we show how to optimally perform gradient ascent by using the QNG and adaptive learning rates, as well as the conditions to avoid barren plateaus.

We propose that the fidelity as a function of parameter $\boldsymbol{\theta}$ can be approximated as a Gaussian kernel.
In particular, we have two quantum states of the PQC $\ket{\psi(\boldsymbol{\theta})}$ and $\ket{\psi(\boldsymbol{\theta}')}$ with parameters $\boldsymbol{\theta}$, $\boldsymbol{\theta}'$ and a distance in parameter space $\Delta \boldsymbol{\theta}=\boldsymbol{\theta}-\boldsymbol{\theta}'$. Then, the fidelity is given by
\begin{equation}\label{eq:kernel}
\mathcal{K}(\boldsymbol{\theta},\boldsymbol{\theta}')=\abs{\braket{\psi(\boldsymbol{\theta})}{\psi(\boldsymbol{\theta}')}}^2\approx\text{exp}[-\frac{1}{4}\Delta \boldsymbol{\theta}^{\text{T}}\mathcal{F}(\boldsymbol{\theta})\Delta \boldsymbol{\theta}]\,,
\end{equation}
where this approximation is valid within a distance $\abs{\Delta \boldsymbol{\theta}}<\epsilon_\text{G}$ with $\epsilon_\text{G}>0$. Here, the QFIM $\mathcal{F}(\boldsymbol{\theta})$ plays the role of the weight matrix of the kernel.
The first order approximation $\mathcal{K}(\boldsymbol{\theta},\boldsymbol{\theta}')\approx 1-\frac{1}{4}\Delta \boldsymbol{\theta}^{\text{T}}\mathcal{F}(\boldsymbol{\theta})\Delta \boldsymbol{\theta}$ returns \eqref{eq:fidelity_QFIM} as expected~\cite{liu2019quantum,meyer2021fisher}.
We show an exactly solvable PQC with Gaussian kernel in Appendix~\ref{app:gaussian}. In Fig.\ref{fig:kernel_vargrad}a, we find that various types of expressive PQCs match well with the Gaussian kernel, which improves with increasing number of qubits. 
We note that for large parameter norm $\Delta \boldsymbol{\theta}^{\text{T}}\mathcal{F}(\boldsymbol{\theta})\Delta \boldsymbol{\theta}$ the fidelity deviates from \eqref{eq:kernel} and eventually reaches the constant value $\langle \mathcal{K}_\text{rand}\rangle=\frac{1}{2^N}$ given by effectively random states~\cite{brown2010convergence,mcclean2018barren}. 
We can use this result to derive an upper bound for $\epsilon_\text{G}$. Assuming a simple QFIM $\mathcal{F}=\frac{I}{b}$ with $I$ being the identity and $b>0$ some constant, we find $\epsilon_\text{G}<2\sqrt{bN\log(2)}$.

We now replace the standard gradient $\nabla K_\text{t}(\boldsymbol{\theta})$ in gradient ascent with the GQNG
\begin{equation}\label{eq:general_gradient}
G_\beta(\boldsymbol{\theta})=\mathcal{F}^{-\beta}(\boldsymbol{\theta})\nabla K_\text{t}(\boldsymbol{\theta})\,,
\end{equation}
where $\beta\in[0,1]$.
As special cases, we have the standard gradient for $G_0(\boldsymbol{\theta})=\nabla K_\text{t}(\boldsymbol{\theta})$ and the QNG for $G_1(\boldsymbol{\theta})=\mathcal{F}^{-1}(\boldsymbol{\theta})\nabla K_\text{t}(\boldsymbol{\theta})$~\cite{stokes2020quantum,yamamoto2019natural}. 
The inverse of the QFIM $\mathcal{F}^{-1}(\boldsymbol{\theta})$ is highly sensitive to small changes in parameter $\boldsymbol{\theta}$ as often $\mathcal{F}(\boldsymbol{\theta})$ is ill conditioned due to small eigenvalues of the QFIM. This causes the QNG to become unstable and gradient ascent cannot converge.
A common approach to regularize the QNG is by adding the identity matrix to the QFIM before inversion $\mathcal{F}'=\mathcal{F}+\epsilon_\text{R}I$, where $I$ is the identity matrix and $\epsilon_\text{R}$ is a small hyperparameter. This hyperparameter has to be determined heuristically and may depend on the particular QFIM~\cite{gacon2021simultaneous,wierichs2020avoiding,van2020measurement}. 
We now show when regularization $\epsilon_\text{R}>0$ is necessary for the GQNG.
For arbitrary $\beta$ and $\epsilon_\text{R}=0$, the fidelity after one iteration of gradient ascent $\boldsymbol{\theta}'=\boldsymbol{\theta}+\alpha G_\beta(\boldsymbol{\theta})$ with arbitrary learning rate $\alpha$ is given by (derivation in Appendix~\ref{app:gqng})
\begin{align*}
K_\text{t}(\boldsymbol{\theta}')=&K_\text{t}(\boldsymbol{\theta})
\exp[-\frac{1}{4}(\alpha^2 \nabla K_\text{t}^\text{T}(\boldsymbol{\theta})\mathcal{F}^{1-2\beta}(\boldsymbol{\theta}) \nabla K_\text{t}(\boldsymbol{\theta})+\\
&2\alpha\Delta\boldsymbol{\theta}^\text{T}\mathcal{F}^{1-\beta}(\boldsymbol{\theta}) \nabla K_\text{t}(\boldsymbol{\theta}))]\,.\numberthis\label{eq:update_stable}
\end{align*}
For $\beta>\frac{1}{2}$, Eq.\ref{eq:update_stable} contains the QFIM with negative exponent $\mathcal{F}^{1-2\beta}$. In this case, an ill-conditioned QFIM can lead to an excessively large parameter update that moves us away from the target parameters instead of bringing us closer. Here, the QFIM has to be regularized with a parameter $\epsilon_\text{R}>0$ in order to ensure that it is not ill-conditioned. 
For $\beta\le\frac{1}{2}$ the QFIM appears only with non-negative exponents in Eq.\ref{eq:update_stable}. Even if the QFIM is ill-conditioned, the updated fidelity is not affected by it. Thus, the GQNG with $\beta=\frac{1}{2}$ and $G_\frac{1}{2}(\boldsymbol{\theta})=\mathcal{F}^{-\frac{1}{2}}\nabla K_\text{t}(\boldsymbol{\theta})$ is intrinsically stable without the need of further regularization and we can set $\epsilon_\text{R}=0$.

Next, we present adaptive learning rates for gradient ascent (derivation in Appendix~\ref{app:learning}). The goal is to choose learning rates adaptively at ever iteration of gradient ascent such that the fidelity increases as much as possible. The initial update rule is given by $\boldsymbol{\theta}_1=\boldsymbol{\theta}+\alpha_1 G_\beta(\boldsymbol{\theta})$ with the learning rate
\begin{equation}\label{eq:update_exact}
\alpha_1=\frac{2\sqrt{-\log(K_\text{t}(\boldsymbol{\theta}))}}{\sqrt{G_\beta(\boldsymbol{\theta})^\text{T} \mathcal{F}(\boldsymbol{\theta})G_\beta(\boldsymbol{\theta})}}\,.
\end{equation}
If the PQC is unable to represent the target state perfectly, i.e. $\text{max}_{\boldsymbol{\theta}}\abs{\braket{\psi(\boldsymbol{\theta})}{\psi_t}}^2=K_0<1$, where $K_0$ is the maximal possible fidelity, we can adjust the learning rate using $K_\text{t}(\boldsymbol{\theta}_1)$.
Then, the updated parameter $\boldsymbol{\theta}_\text{t}'$ is given by
$\boldsymbol{\theta}_\text{t}'=\boldsymbol{\theta}+\alpha_\text{t}G_\beta(\boldsymbol{\theta})$ with the adaptive learning rate
\begin{equation}\label{eq:update_add}
\alpha_\text{t}=\frac{1}{2}\left(\frac{4}{\alpha_1 G_\beta(\boldsymbol{\theta})^\text{T}\mathcal{F}(\boldsymbol{\theta}) G_\beta(\boldsymbol{\theta})}\log\left(\frac{K_\text{t}(\boldsymbol{\theta}_1)}{K_\text{t}(\boldsymbol{\theta})}\right)+\alpha_1\right)\,.
\end{equation}
This concludes one iteration of adaptive gradient ascent. By setting $\boldsymbol{\theta}=\boldsymbol{\theta}_\text{t}'$ and repeating above steps, the next iteration of adaptive gradient ascent is performed.

\begin{figure}[htbp]
	\centering
	\subfigimg[width=0.24\textwidth]{a}{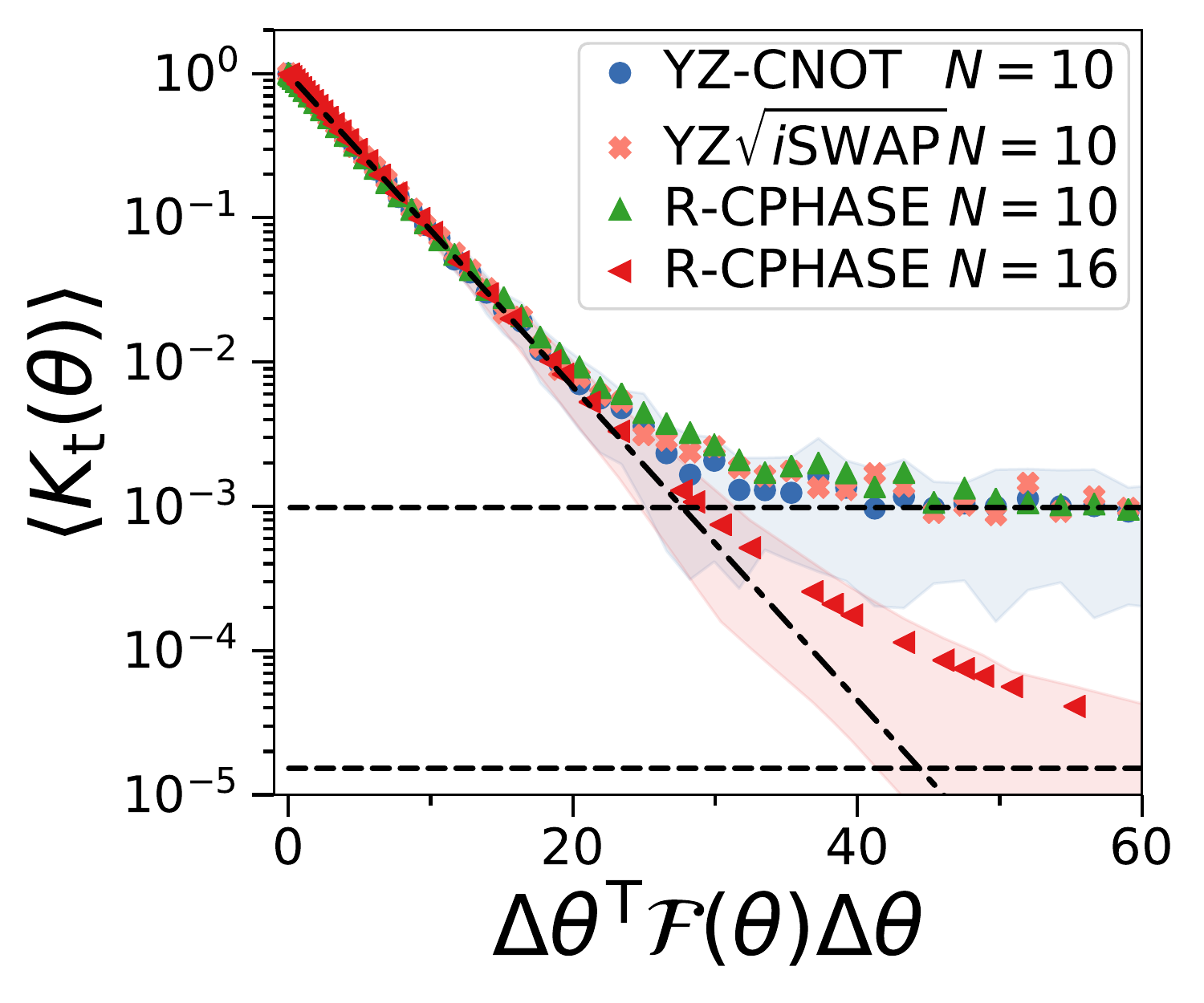}\hfill
	\subfigimg[width=0.24\textwidth]{b}{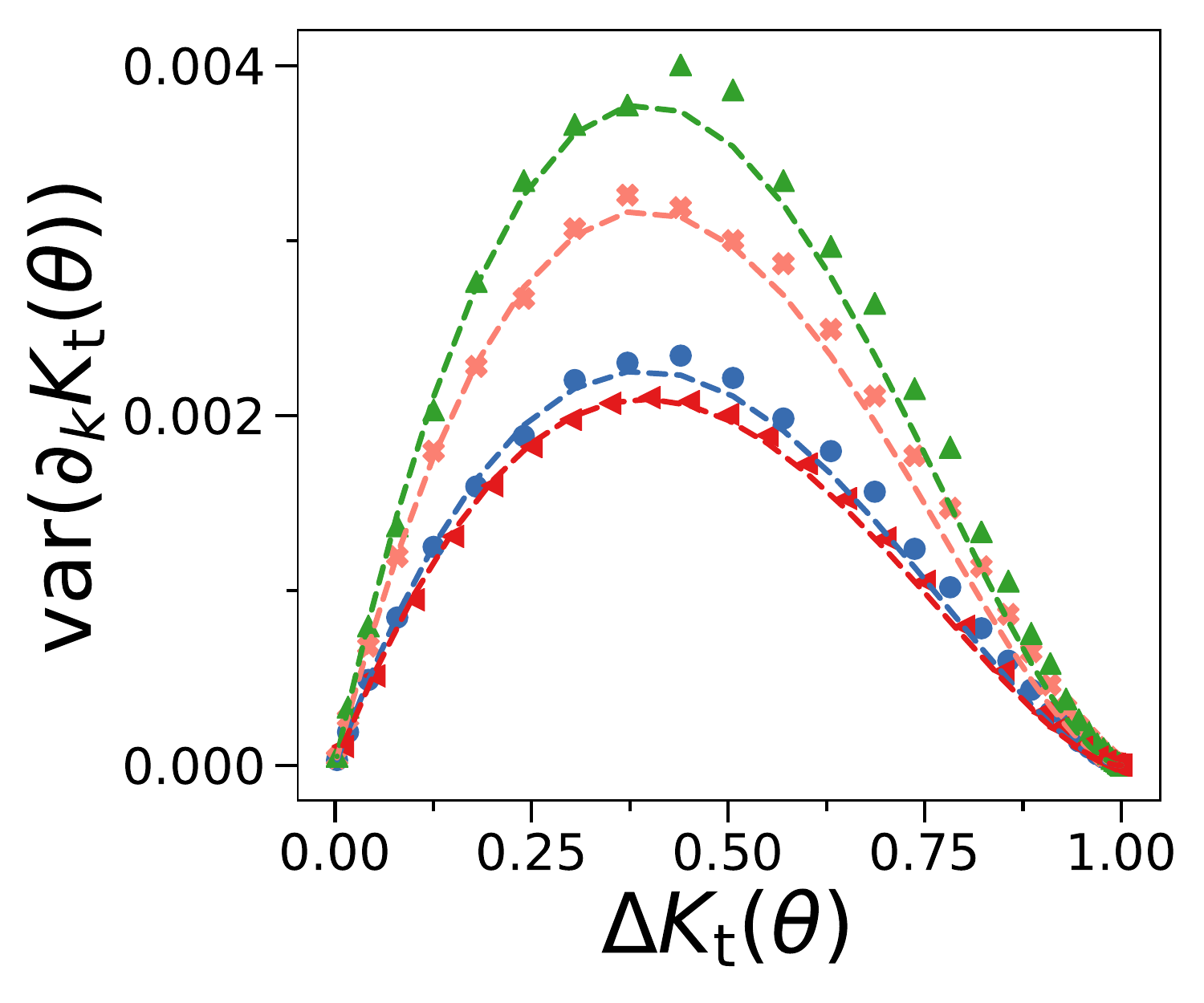}
	\caption{\idg{a} Average fidelity $\langle K_\text{t}(\boldsymbol{\theta})\rangle$ as function of parameter norm $\Delta \boldsymbol{\theta}^\text{T}\mathcal{F}(\boldsymbol{\theta})\Delta \boldsymbol{\theta}$, with distance $\Delta \boldsymbol{\theta}=\boldsymbol{\theta}-\boldsymbol{\theta}_\text{t}$ and target parameters $\boldsymbol{\theta}_\text{t}$. Shaded area is the 20-th and 80-th percentile of the fidelity. We find a good match with the Gaussian kernel (\eqref{eq:kernel}, dash-dotted line). For large norm, we see the fidelity converges to the fidelity given by random states $\langle \mathcal{K}_\text{rand}\rangle=\frac{1}{2^N}$ (dashed lines). We use three different PQCs with randomized parameters, which are defined in Appendix~\ref{app:pqc}. Number of layers $p=20$ for R-CPHASE $N=10$, $p=16$ for $N=16$, else $p=10$. Average over 50 random instances of $\boldsymbol{\theta}_\text{t}$.
	\idg{b} Variance of gradient $\text{var}(\partial_k K_\text{t}(\boldsymbol{\theta}))$ against infidelity $\Delta K_\text{t}(\boldsymbol{\theta})$ for different types of PQCs. Dashed lines are the analytic formula~\eqref{eq:var_grad} for the variance. }
	\label{fig:kernel_vargrad}
\end{figure}

For a given fidelity $K_\text{t}(\boldsymbol{\theta})$, we can derive the variance of the gradient analytically (see Appendix~\ref{app:variance})
\begin{align*}
&\text{var}(\partial_k K_\text{t}(\boldsymbol{\theta}))=\langle\langle (\partial_k K_\text{t}(\boldsymbol{\theta}))^2\rangle_{\Delta\boldsymbol{\theta}}\rangle_k - \langle\langle \partial_k K_\text{t}(\boldsymbol{\theta})\rangle_{\boldsymbol{\theta}}\rangle_k^2\\
&=\frac{1}{M}\frac{\text{Tr}(\mathcal{F}(\boldsymbol{\theta})^2)}{\text{Tr}(\mathcal{F}(\boldsymbol{\theta}))} K_\text{t}(\boldsymbol{\theta})^2\log\left[\frac{ K_0}{K_\text{t}(\boldsymbol{\theta})}\right]\,,\numberthis\label{eq:var_grad}
\end{align*}
where the average is first taken over distance  $\Delta\boldsymbol{\theta}=\boldsymbol{\theta}-\boldsymbol{\theta}_\text{t}$ and then over the gradient indices $k$.
The variance is maximized for fidelity $K_\text{t}(\boldsymbol{\theta})\approx0.6K_0$. For a fixed fidelity $K_\text{t}(\boldsymbol{\theta})$, the variance decreases linearly with number of parameters $M$ and is independent of qubit number $N$.  
With $\text{Tr}(\mathcal{F}(\boldsymbol{\theta})^2)\ge\frac{\text{Tr}(\mathcal{F}(\boldsymbol{\theta}))^2}{M}$, we give the lower bound of the variance
\begin{align*}
\text{var}(\partial_k K_\text{t}(\boldsymbol{\theta}))\ge\frac{\text{Tr}(\mathcal{F}(\boldsymbol{\theta}))}{M^2} K_\text{t}(\boldsymbol{\theta})^2\log\left[\frac{ K_0}{K_\text{t}(\boldsymbol{\theta})}\right]\,.\numberthis\label{eq:var_grad_lower}
\end{align*}
We find a good match between \eqref{eq:var_grad} and simulated results in Fig.\ref{fig:kernel_vargrad}b for different types of PQCs, which improves with more qubits. It starts deviating from the analytic result when the variance of the gradient becomes close to the one given by a state sampled from a deep PQC with $\text{var}(\nabla K_\text{t}(\boldsymbol{\theta}_\text{rand}))=\frac{1}{2^{2N+1}}$~\cite{mcclean2018barren} (see Appendix~\ref{app:data_gradient}).
In Fig.\ref{fig:var_step}a, we show the variance of the gradient for different types of PQCs and fidelities as a function of the number of qubits $N$, where we keep the number of layers $p$ fixed. We find a good match with \eqref{eq:var_grad}.

\begin{figure}[htbp]
	\centering	
	\subfigimg[width=0.24\textwidth]{a}{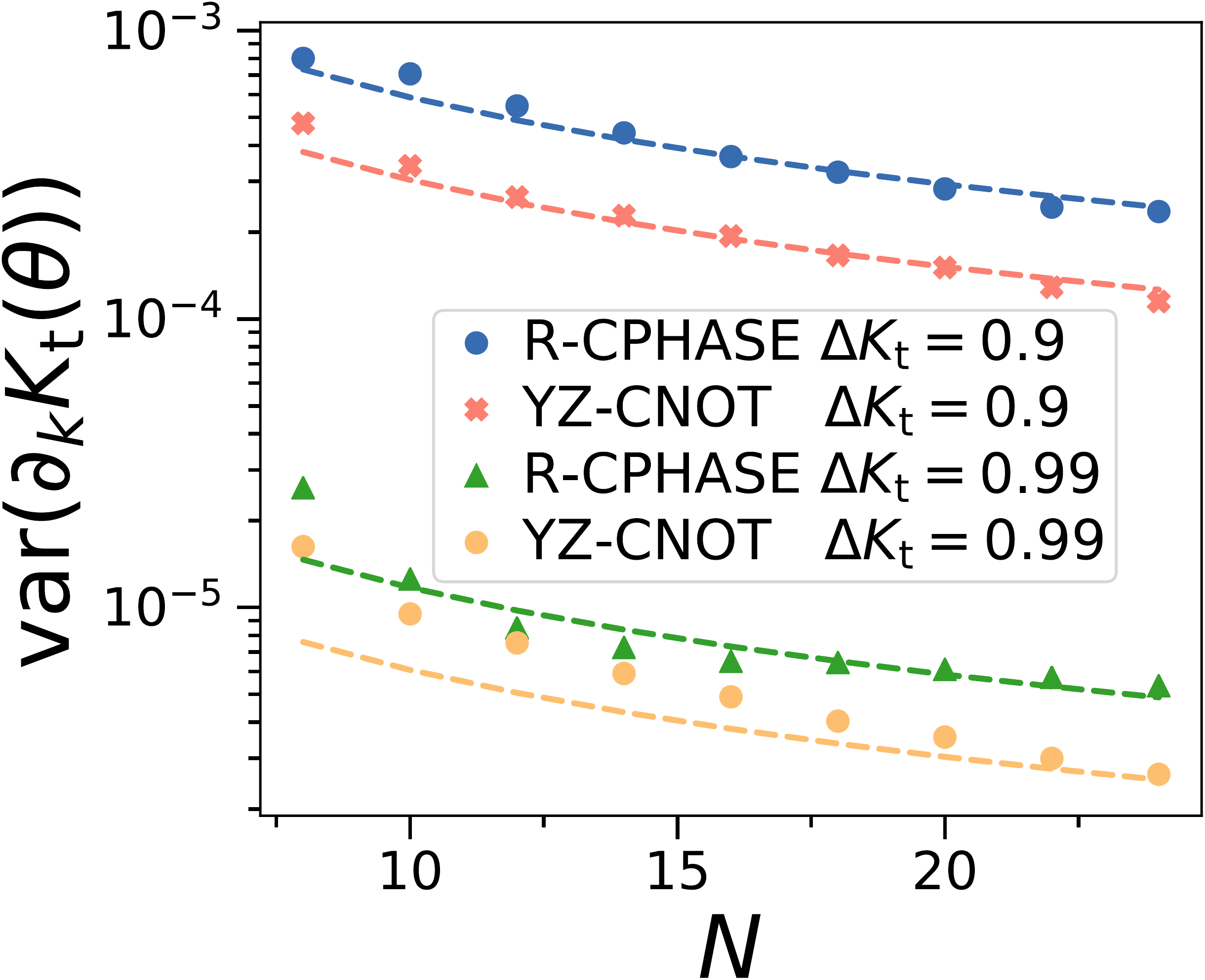}\hfill
	\subfigimg[width=0.24\textwidth]{b}{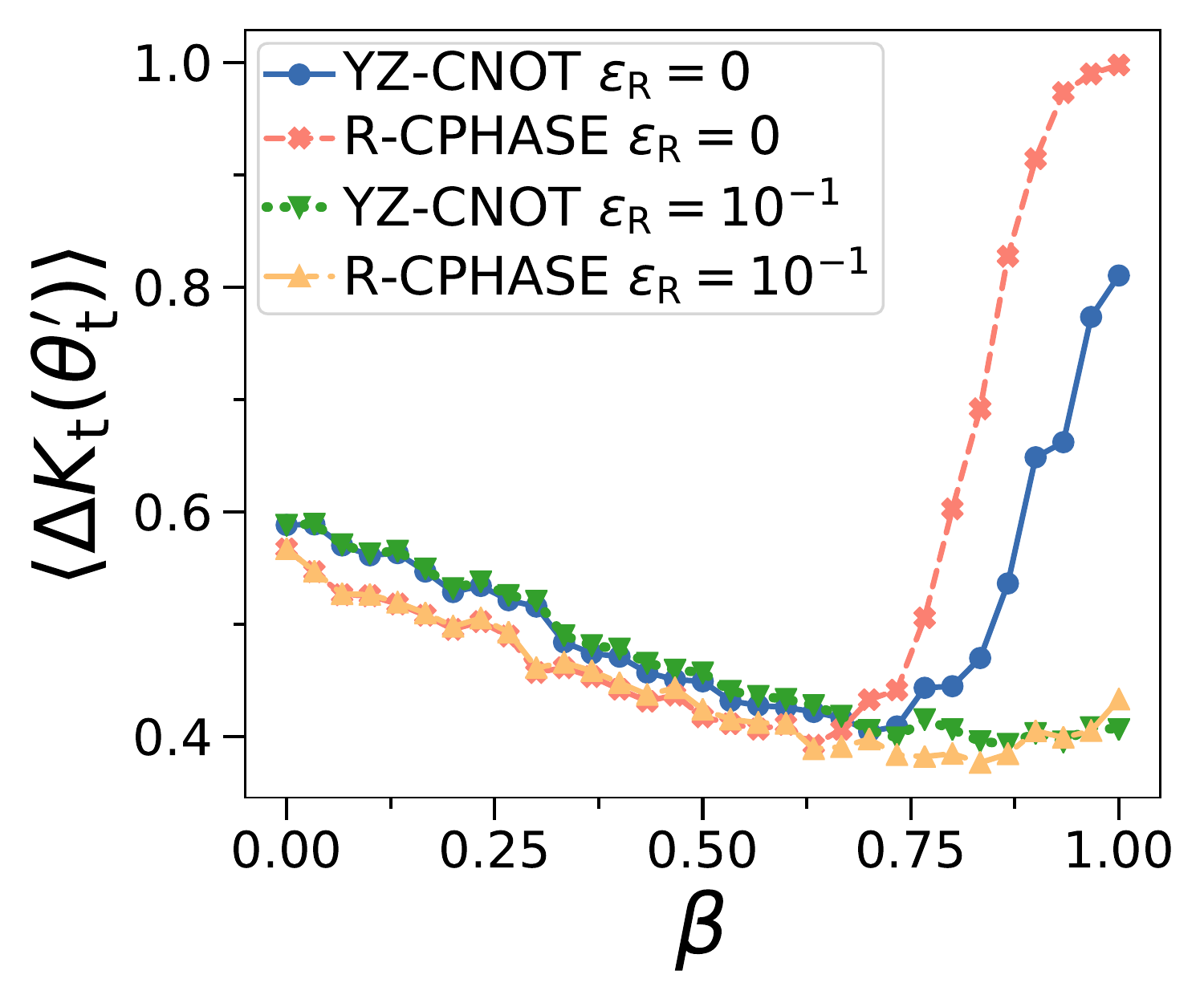}
	\caption{
	\idg{a} Variance of gradient $\text{var}(\partial_k K_\text{t}(\boldsymbol{\theta}))$ against number of qubits $N$ for different types of PQCs and infidelities $\Delta K_\text{t}(\boldsymbol{\theta})$. The number of layers is $p=20$ for R-CPHASE and $p=10$ for YZ-CNOT. Dashed lines are the analytic formula~\eqref{eq:var_grad} for the variance of the gradient. 
    \idg{b} Average infidelity $\langle \Delta F(\boldsymbol{\theta}_\text{t}')\rangle$ after one iteration of gradient ascent  with adaptive learning rate (\eqref{eq:update_add}) plotted against exponent $\beta$ of the GQNG (\eqref{eq:general_gradient}) for different types of PQCs and regularization parameter $\epsilon_\text{R}$. Initial infidelity is $\Delta K_\text{t}(\boldsymbol{\theta})=0.9$. 
	}
	\label{fig:var_step}
\end{figure}

\section{Results}
We now demonstrate the performance of adaptive gradient ascent using numerical simulations~\cite{yao,johansson2012qutip} with various types of expressive PQCs~\cite{haug2021capacity,sim2019expressibility}.
In Fig.\ref{fig:var_step}b, we plot the average infidelity after one iteration of gradient ascent $\langle\Delta K_\text{t}(\boldsymbol{\theta}_\text{t}')\rangle$ (\eqref{eq:infidelity}) against $\beta$ of the GQNG (\eqref{eq:general_gradient}). 
Initially, the infidelity decreases with increasing $\beta$ as more information from the QFIM is used. Without regularization $\epsilon_\text{R}=0$,  we observe a sharp increase of infidelity for $\beta>0.6$ due to the ill-conditioned inverse of the QFIM. For larger $\beta$, this is mitigated by a non-zero regularization parameter $\epsilon_\text{R}=10^{-1}$. Note that the regularization leads to a small increase in infidelity for $\beta\le\frac{1}{2}$.

\begin{figure}[htbp]
	\centering
	\subfigimg[width=0.24\textwidth]{a}{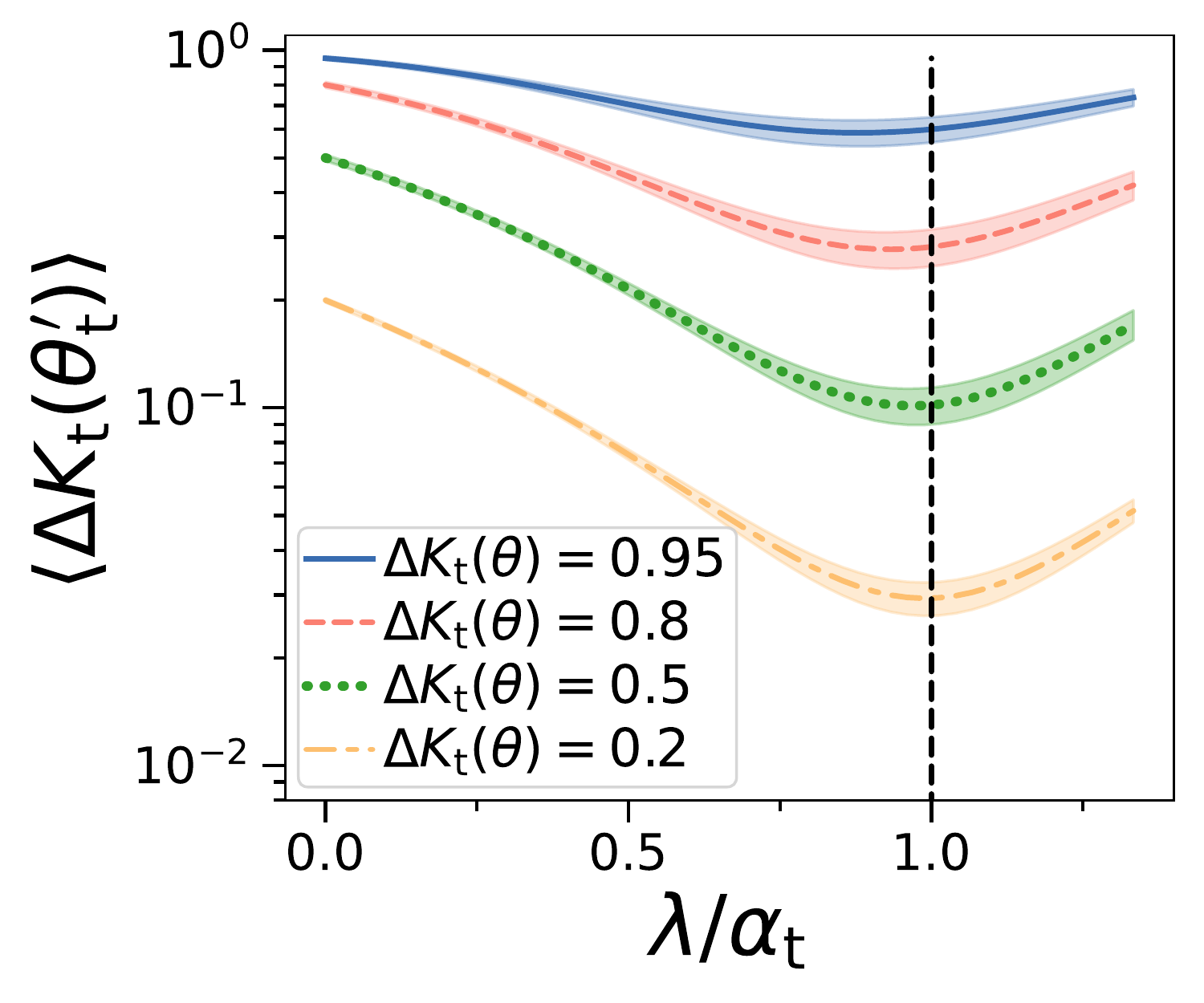}\hfill
	\subfigimg[width=0.24\textwidth]{b}{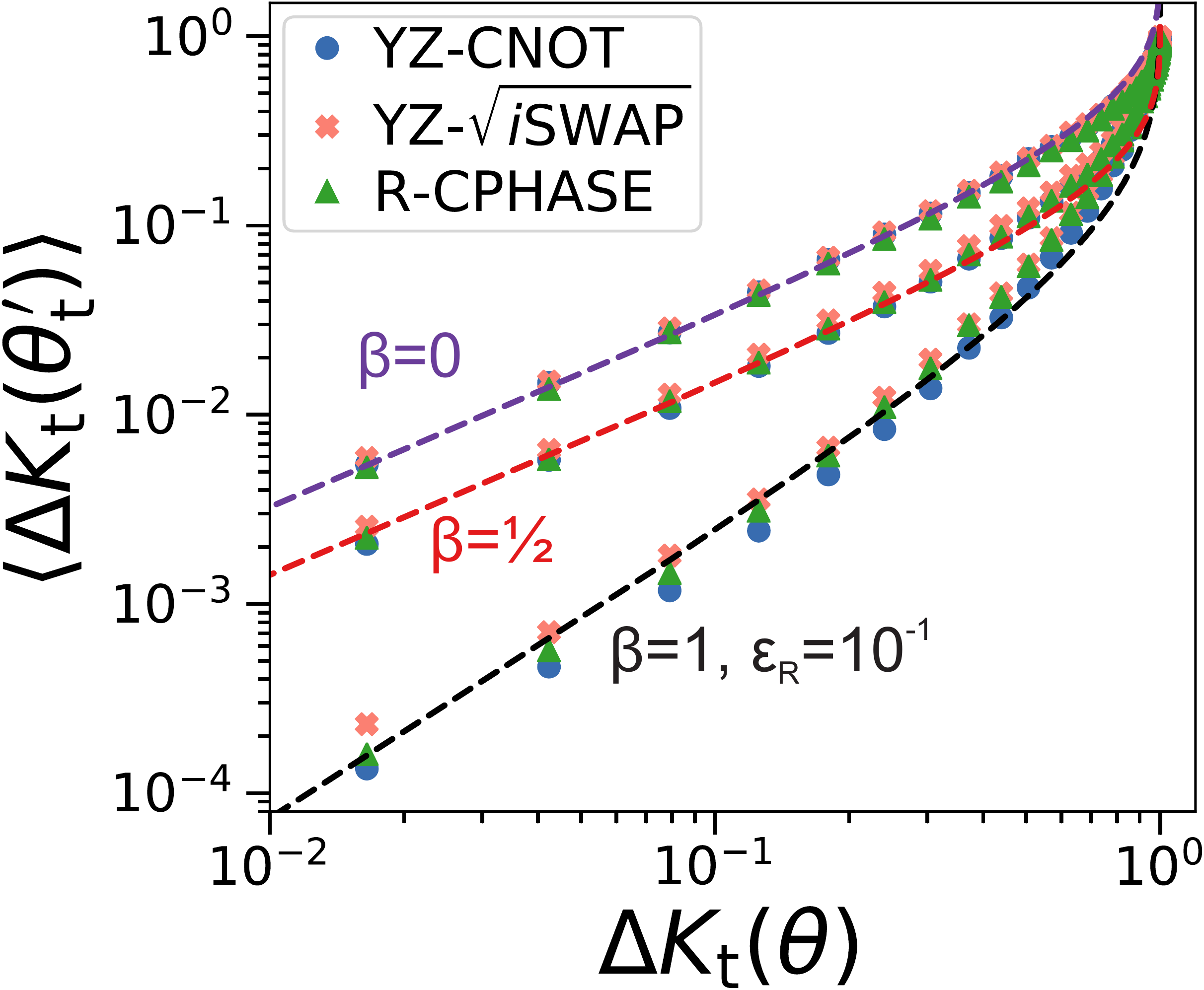}
	\caption{
	\idg{a} Average infidelity $\langle\Delta K_\text{t}(\boldsymbol{\theta}_\text{t}')\rangle$ (\eqref{eq:infidelity}) plotted against learning rate $\lambda$ using the gradient ascent update $\boldsymbol{\theta}_\text{t}'=\boldsymbol{\theta}+\lambda G_\frac{1}{2}(\boldsymbol{\theta})$ with GQNG. $\lambda$ is normalized in respect to adaptive learning rate $\alpha_\text{t}$ (\eqref{eq:update_add}), shown as vertical dashed line. Curves show various initial infidelities $\Delta K_\text{t}(\boldsymbol{\theta})$, with the shaded area being the standard deviation of $\Delta K_\text{t}(\boldsymbol{\theta}_\text{t}')$. Infidelity is averaged over 50 random instances of $\boldsymbol{\theta}_\text{t}$ for the YZ-CNOT PQC. 
	\idg{b} Average infidelity $\langle \Delta K_\text{t}(\boldsymbol{\theta}_\text{t}') \rangle$ after one iteration of adaptive gradient ascent against initial infidelity $ \Delta K_\text{t}(\boldsymbol{\theta})$. We show the regular gradient ($\beta=0$, upper curves), GQNG ($\beta=\frac{1}{2}$, center curves) and QNG ($\beta=1$, regularization $\epsilon_\text{R}=10^{-1}$, lower curves) for various types of PQCs. The red and black curves are fits with $\Delta K_\text{t}(\boldsymbol{\theta}_\text{t}') =c[\frac{1}{4}\Delta\boldsymbol{\theta}^\text{T}\mathcal{F}\Delta\boldsymbol{\theta}]^\nu=-c\log^\nu[1- \Delta K_\text{t}(\boldsymbol{\theta}) ]$ with $\nu=1$ for $\beta=0$, $\beta=\frac{1}{2}$ and $\nu=1.5$ for $\beta=1$. The scaling factor is $c(\beta=1)=0.072$, $c(\beta=\frac{1}{2})=0.14$ and $c(\beta=0)=0.32$.
	}
	\label{fig:scale}
\end{figure}

In Fig.\ref{fig:scale}a, we show the average infidelity after one iteration of gradient ascent with the GQNG as function of learning rate $\lambda$ for different initial infidelities $\Delta K_\text{t}(\boldsymbol{\theta})$. We find that $\alpha_\text{t}$ as calculated by \eqref{eq:update_add} gives nearly the best learning rate even for larger infidelities. We find the same result when using the regular gradient or the QNG, as well as for target states that cannot be perfectly represented by the PQC (see Appendix~\ref{app:data_learning}).

In Fig.\ref{fig:scale}b, we show $\langle \Delta K_\text{t}(\boldsymbol{\theta}_\text{t}') \rangle$ after one iteration of adaptive gradient ascent against initial infidelity $ \Delta K_\text{t}(\boldsymbol{\theta})$ with the standard gradient ($\beta=0$), the GQNG ($\beta=\frac{1}{2}$) and the QNG ($\beta=1$ with regularization $\epsilon_\text{R}=10^{-1}$). We find  the QNG outperforms GQNG and standard gradient at smaller initial infidelities.
We numerically find that the data is fitted with $ \Delta K_\text{t}(\boldsymbol{\theta}_\text{t}') =c[\frac{1}{4}\Delta\boldsymbol{\theta}^\text{T}\mathcal{F}(\boldsymbol{\theta})\Delta\boldsymbol{\theta}]^\nu=-c\log^\nu[1- \Delta K_\text{t}(\boldsymbol{\theta}) ]$, with $\nu=1$ for $\beta=0$, $\beta=\frac{1}{2}$ and $\nu=1.5$ for the QNG with $\beta=1$.
Thus, QNG shows a better scaling for training compared to GQNG and standard gradient.

\begin{figure}[htbp]
	\centering
	\subfigimg[width=0.24\textwidth]{b}{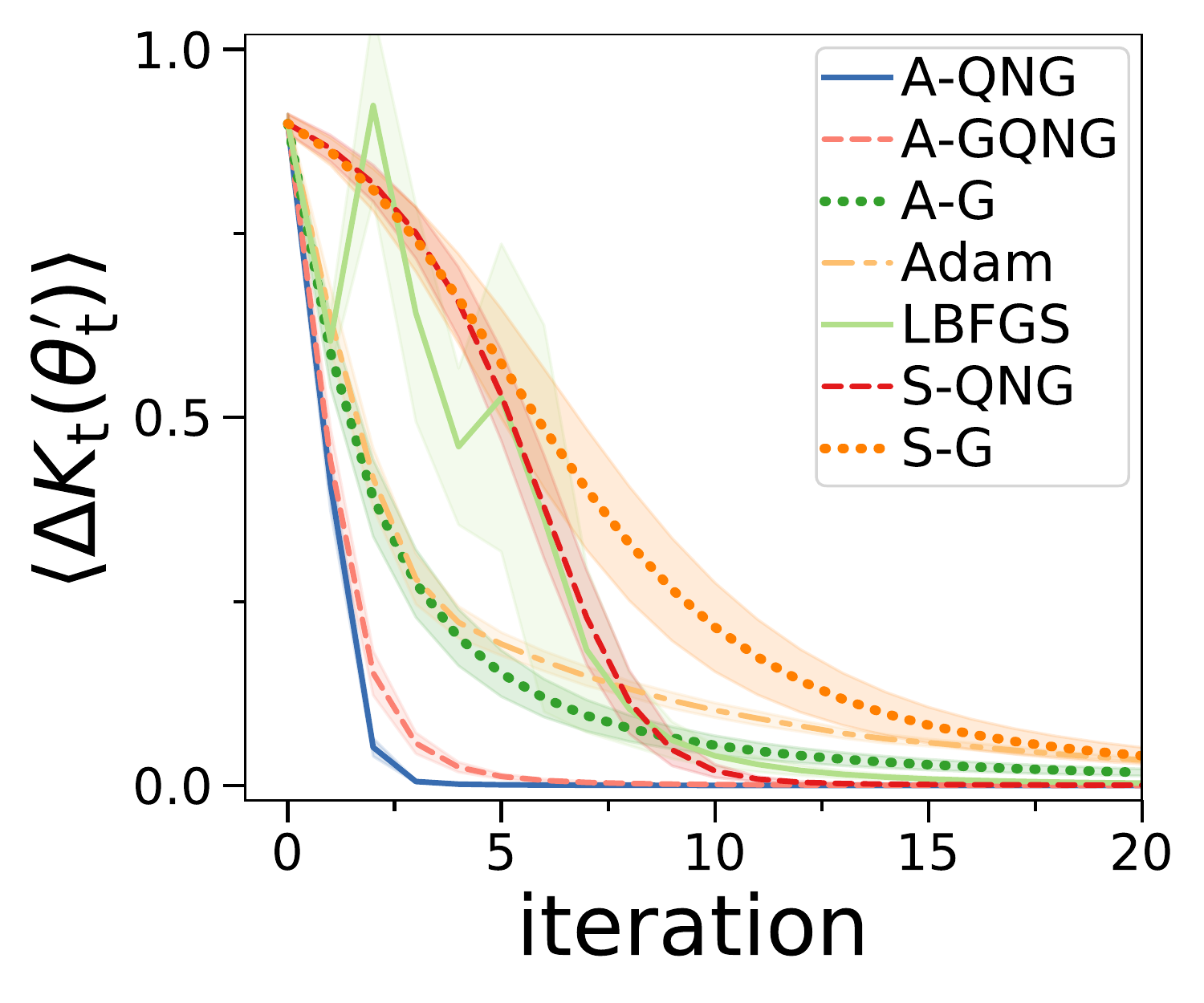}\hfill
	\subfigimg[width=0.24\textwidth]{b}{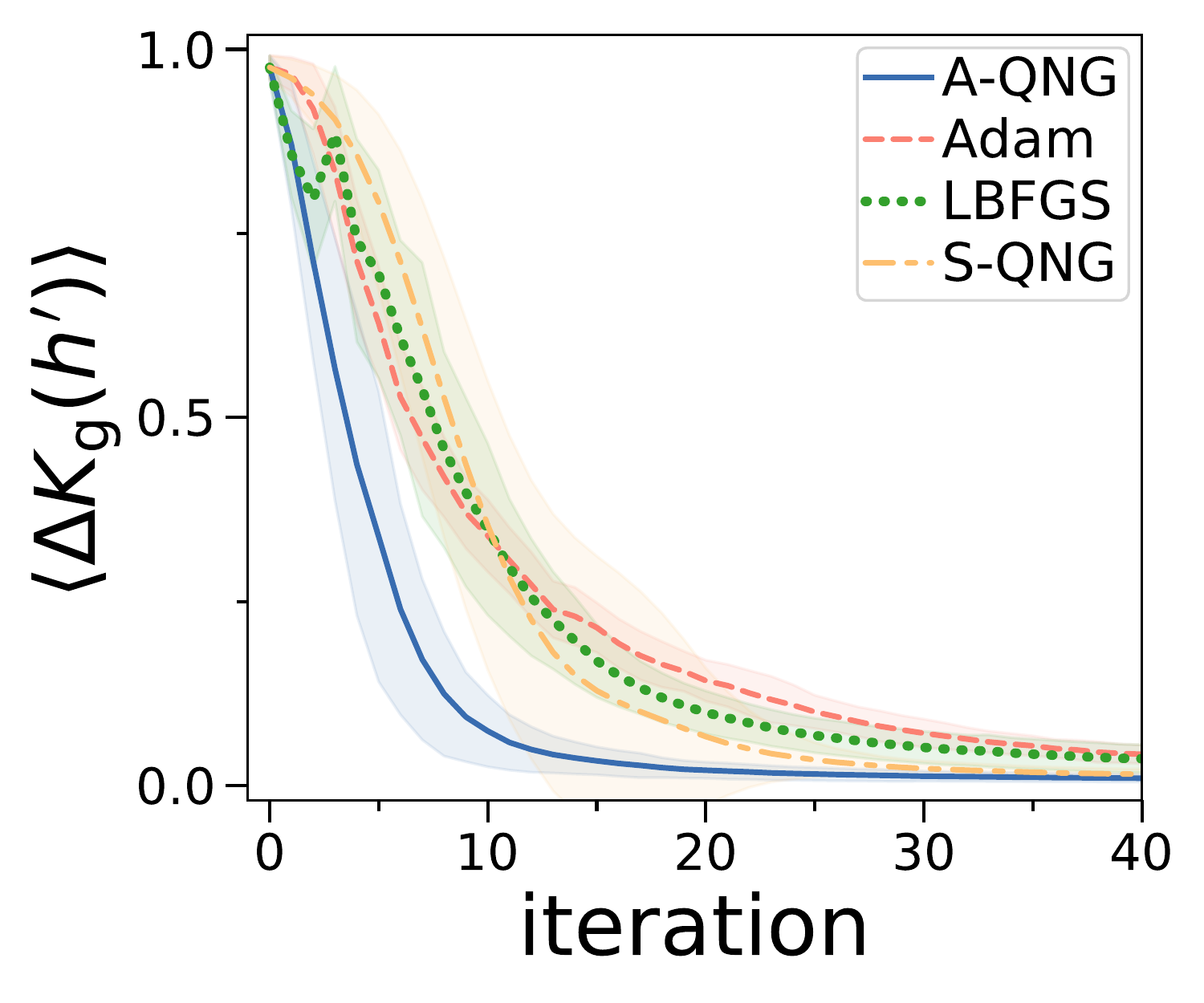}
	\caption{
	\idg{a} Training VQA. Average infidelity $\langle \Delta K_\text{t}(\boldsymbol{\theta}_\text{t}') \rangle$ against number of iterations for optimizing the VQA. Shaded area is the standard deviation over 50 instances of training. We compare different optimization methods against each other. We find that adaptive gradient ascent with QNG (A-QNG, $\beta=1$, regularization $\epsilon_\text{R}=10^{-1}$) performs best, followed by adaptive GQNG (A-GQNG, $\beta=\frac{1}{2}$, $\epsilon_\text{R}=0$). Standard optimization methods such as Adam ($\alpha=0.1$) and LBFGS perform a comparable to adaptive gradient ascent with the regular gradient (A-G). Non-adaptive QNG (S-QNG, $\alpha=1$, $\epsilon_\text{R}=10^{-1}$) is initially worse, but for more iterations outperforms the methods that do not use the QFIM. 
	Initial infidelity is $\Delta K_\text{t}(\boldsymbol{\theta})=0.9$, training is averaged over 50 random instances of $\boldsymbol{\theta}_\text{t}$, PQC is YZ-CNOT, $N=10$ and $p=10$.
	\idg{b} Optimizing control problem. Average infidelity $\langle \Delta K_\text{g}(h') \rangle$ against number of iterations for optimizing  driving parameters $h'$. We use driving Hamiltonian \eqref{eq:control} with $g=1$, $N=6$, $\Delta t=1$ and $T=d=16$. The goal is to find the driving protocol that evolves the zero state to the ground state of \eqref{eq:ising} with $g=1$ and $h=1$. We average the training data over 20 instances of initially random protocols with $h_n^p\in[-1,1]$.
	}
	\label{fig:training}
\end{figure}

In Fig.\ref{fig:training}a, we study the training of VQAs. We plot $\langle \Delta K_\text{t}(\boldsymbol{\theta}_\text{t}') \rangle$ against number of iterations of gradient ascent. We compare our adaptive training method with other established methods. Adaptive learning rates with QNG (A-QNG with $\beta=1$ and regularization $\epsilon_\text{R}=10^{-1}$) or GQNG (A-GQNG with $\beta=\frac{1}{2}$) outperforms other investigated methods and provides more than one order of magnitude smaller infidelites. It consistently performs well for different random instances, demonstrating a very low standard deviation in fidelity. 
We note that gradient ascent with adaptive learning rate and standard gradient (A-G with $\beta=0$) still performs well, comparable to Adam and LBFGS. 
See Appendix~\ref{app:vqa_data} for further training data with other PQCs, different initial infidelities and logarithmic plots.

Finally, we apply our methods to a quantum control problem. Our goal is to prepare the ground state $\ket{\psi_\text{g}}$ of the $N$-qubit one-dimensional Ising model with a transverse field $h$ and longitudinal field $g$
\begin{equation}\label{eq:ising}
H_0=\sum_{i=1}^N( \sigma^x_n\sigma^x_{n+1}+h\sigma^z_n+g\sigma^x_n)\,.
\end{equation}
We choose a time-dependent driving Hamiltonian $H(t)$ with arbitrary control over the transverse field $h_n(t)$ for each qubit $n$
\begin{equation}\label{eq:control}
H(t)=\sum_{i=1}^N( \sigma^x_n\sigma^x_{n+1}+h_n(t)\sigma^z_n+g\sigma^x_n)\,.
\end{equation}
We initially start from the all zero state $\ket{\psi(0)}=\ket{0}$ and evolve the quantum state in time with the time-dependent Schr\"odinger equation $\partial_t\ket{\psi(t)}=H(t)\ket{\psi(t)}$. Now, we want to find the driving parameters $h_n(t)$ that prepare the ground state with high fidelity $K_\text{g}(h')=\abs{\braket{\psi_\text{g}}{\psi(T)}}^2$ after a time $T$. To numerically solve this task, we split the driving amplitudes $h_n(t)$ into $d$ equal-sized timesteps of size $\Delta t$, where the driving parameters $h_n^p$ for a given qubit $n$ and timestep $p$ are held constant. Thus, we have a time-dependent driving protocol of the form $h_n(0\le t <\Delta t)=h_n^1$, $h_n(\Delta t\le t <2\Delta t)=h_n^2$, \dots, $h_n((d-1)\Delta t\le t <d\Delta t)=h_n^d$. 
Initially, we start with randomly chosen driving parameters $h_n^p$, where the set of driving parameters $h'$ consists of $dN$ parameters. 
We optimize the fidelity $K_\text{g}(h')$ by numerically calculating the gradient of the fidelity in respect to $h_n^p$ using the finite difference method. 
In Fig.\ref{fig:training}b, we compare our adaptive QNG method against other optimisation techniques. We find that adaptive QNG finds a high fidelity driving protocol with a lower number of iterations compared to other methods such as LBFGS. We show further results for optimizing quantum control problems in Appendix~\ref{app:control_data}.

\section{Discussion}
We showed how to optimally train a PQC to represent a target quantum state. We found that the fidelity of quantum states can form Gaussian kernels. With this relation we derive adaptive learning rates to adjust the magnitude of the gradient at every step. 
The QNG provides the best direction for gradient ascent.
We apply these methods to train different expressive PQCs that are able to represent a wide range of quantum states and are commonly used for VQAs~\cite{haug2021capacity,sim2019expressibility}.
Combining adaptive learning rates and the QNG, we yield a lower number of iterations with an order of magnitude improvement in accuracy compared to other methods. 
This approach can be directly applied to a large range of VQAs which have a subroutine that requires learning quantum states using the fidelity~\cite{otten2019noise,barison2021efficient,gibbs2021long,holmes2020barren,benedetti2019generative,jones2019variational,romero2017quantum}. 
Future work can extend our methods to other cost functions such as Hamiltonians or the Hilbert-Schmidt test~\cite{khatri2019quantum}. Adaptive learning rates and QNG combined could provide speed ups in training the variational quantum eigensolver for finding the ground states of Hamiltonians and molecules~\cite{peruzzo2014variational}.
The QFIM $\mathcal{F}$ needed for our method can be calculated on quantum computers using the shift-rule~\cite{mari2021estimating,meyer2021fisher}. Approximations for a more efficient calculation on quantum computer exist~\cite{stokes2020quantum,cerezo2021sub,gacon2021simultaneous,beckey2020variational,van2020measurement} as well as improved methods for classical computers~\cite{jones2020efficient}.

The QNG uses the inverse of the QFIM, which can be ill-conditioned. To stabilize the QNG, it can be regularized by adding the identity matrix with a small hyperparameter to the QFIM before inversion~\cite{van2020measurement,wierichs2020avoiding,gacon2021simultaneous}. We show when regularization is necessary with the GQNG (\eqref{eq:general_gradient}). For $\beta=\frac{1}{2}$, the GQNG $G_\frac{1}{2}(\boldsymbol{\theta})=\mathcal{F}^{-\frac{1}{2}}\nabla K_\text{t}(\boldsymbol{\theta})$ does not require regularization while using quantum geometric information as much as possible. For $\beta\le\frac{1}{2}$, the instability due to the inverse of the QFIM is canceled out by the quantum geometry of the PQC. Only for $\beta>\frac{1}{2}$ such as the QNG with $\beta=1$, regularization has to be applied to take care of the ill-conditioned QFIM. 
While the QNG performs better for small infidelities, it performs comparable to the GQNG for larger infidelities. As the GQNG is inherently stable for $\beta=\frac{1}{2}$, the GQNG may be a good choice for initial training steps or for noisy systems.

We also applied our methods to numerically find control protocols for quantum systems~\cite{glaser2015training,bastidas2020fully}. Here, a common task is to find driving protocols that generate a particular quantum state. Our adaptive QNG method can find solutions with far less number of iterations compared to other gradient based optimisation methods such as LBFGS, which are commonly applied in numerical solvers for quantum control such as GRAPE~\cite{khaneja2005optimal,machnes2011comparing,johansson2012qutip}. While our method requires the calculation of the QFIM, the additional computational effort may be compensated by the lower number of iterations and an efficient computation of the QFIM on classical computers~\cite{jones2020efficient}. Further, other quantum control methods like GRAPE are known to get often stuck in local minima, which the QNG is more adapt at avoiding~\cite{wierichs2020avoiding}.

A common feature of the cost function in many PQCs is that they form a narrow gorge, i.e. the cost function deviates from its mean only in an exponentially small parameter space~\cite{cerezo2021cost,arrasmith2021equivalence}. These narrow gorges directly imply the existence of barren plateaus, i.e. the variance of the gradient vanishes exponentially with number of qubits, making training difficult for quantum computers~\cite{mcclean2018barren,arrasmith2021equivalence}.
Training a PQC to represent a random state $\ket{\psi_\text{rand}}\in\mathcal{SU}(2^N)$ using the fidelity as cost function suffers from barren plateaus, which appear independently of the depth of the PQC~\cite{cerezo2021cost}. 
The Gaussian kernel gives us now a functional description of the narrow gorge and allows us to calculate the variance of the gradients. 
We can ensure trainability and avoid the barren plateau problem by demanding that the state to be learned $\ket{\psi_\text{t}}$ is not random, but has a lower bounded fidelity with the initial state $K_\text{t}(\boldsymbol{\theta})\ge\gamma$, where $\gamma$ is the lower bound.
We further assume that we are not too close to the optimal solution with $K_\text{t}(\boldsymbol{\theta})<0.6K_0$ where $K_0$ is the maximal possible fidelity.
Then,~\eqref{eq:var_grad_lower} tells us that the variance of the gradient is lower bounded by 
\begin{equation}\label{eq:var_grad_bound}
\text{var}(\nabla K_\text{t}(\boldsymbol{\theta}))\ge\frac{\text{Tr}(\mathcal{F})}{M^2}\log(\frac{K_0}{\gamma})\gamma^2\,.
\end{equation}
In particular, the lower bound of the variance is independent of the number of qubits $N$ of the quantum computer and therefore the gradient does not vanish even when we scale up $N$.
For training VQAs, a large enough $\gamma$ is essential to ensure sufficiently large gradients. Here, we calculate $\gamma$ explicitly for a type of VQA, the projected variational quantum dynamics method~\cite{otten2019noise,barison2021efficient}.
This algorithm simulates the time evolution of a quantum state $\ket{\psi(\boldsymbol{\theta})}$ for a Hamiltonian $H$ (see Appendix~\ref{app:pvqd} for training examples). It evolves the PQC using a single Trotter step with time $\Delta t$ and then variationally learns to represent the evolved state $\ket{\psi_\text{t}}=\exp(-iH\Delta t)\ket{\psi(\boldsymbol{\theta})}$ via the fidelity as cost function. This process is then repeated $N_\text{T}$ times to evolve for a total time $T=N_\text{T}\Delta t$. 
Here, we find that the fidelity is lower bounded by the time-energy uncertainty~\cite{anandan1990geometry}
\begin{equation}
\gamma=\abs{\bra{\psi(\boldsymbol{\theta})}\exp(-iH\Delta t)\ket{\psi(\boldsymbol{\theta})}}^2\ge1-\frac{1}{4}(\Delta E\Delta t)^2\,,
\end{equation}
where $\Delta E$ is the difference between the largest and smallest eigenenergies of $H$ whose eigenstates have non-zero overlap with $\ket{\psi(\boldsymbol{\theta})}$ (see Appendix~\ref{app:fidelity_time}). By choosing a small enough $\Delta t$, one can ensure sufficiently large gradients and that the VQA is free of barren plateaus. Thus, this VQA could be run on near-term quantum computers to simulate the dynamics of extensive many-body systems that are beyond the reach of classical simulation methods.
To reach this goal, future work has to engineer PQCs that can represent the time evolved states for a large number of qubits. Adaptively chosen PQCs~\cite{grimsley2019adaptive,zhang2021mutual} or PQCs tailored to the specific problem could enhance the representation power. Hybrid quantum states, which are linear combination of quantum states generated using the problem Hamiltonian, could systematically create an ansatz suited for the problem~\cite{bharti2020quantum2,bharti2020iterative,haug2020generalized}.
We note that the training of other VQAs without a bounded fidelity $\gamma$ is expected to be difficult, as most likely training will be stuck in the barren plateau. In this case, other types of cost function may provide a way out of the barren plateau~\cite{cerezo2021cost}.

As a tangential observation, we note that \eqref{eq:var_grad} enables us to calculate $\text{Tr}(\mathcal{F}^2)$ by measuring the variance of the gradients.

Finally, our results have direct implications for quantum machine learning, which utilizes quantum features to enhance machine learning. Here, data is embedded as parameters of the PQC and the fidelity of the quantum states is used as a nonlinear feature map~\cite{biamonte2017quantum,schuld2019quantum,schuld2021effect}. The Gaussian (or radial basis function) kernel we provide is such a nonlinear feature map,
which has important applications in various classical machine learning models such as support vector machines~\cite{goodfellow2016deep}.
In a quantum setting, these kernels have been so far only proposed with coherent states~\cite{chatterjee2016generalized,schuld2021quantum,otten2020quantum}. As advantage compared to coherent states, PQCs can adjust the QFIM and the weight matrix of the Gaussian kernel by changing the structure of the PQC~\cite{haug2021capacity}.
Given that these kernels can be efficiently computed classically, they are unlikely to provide a direct quantum advantage from data in machine learning~\cite{schuld2019quantum,schuld2021quantum,huang2020power}. However, with increasing distance in parameter space the approximation as Gaussian is less accurate and the kernel may acquire the features needed to provide quantum advantage~\cite{huang2020power}. The kernel could be immediately realized with current quantum hardware for tasks in quantum machine learning~\cite{havlivcek2019supervised,blank2020quantum,schuld2019quantum,schuld2021effect} and Gaussian processes~\cite{rasmussen2005gaussian,otten2020quantum} to provide novel applications of quantum computers.

Python code for the numerical calculations are available on Github~\cite{haug2021optimalquantumlearning}.

\medskip
\begin{acknowledgments}
We acknowledge discussions with Giuseppe Carleo, Kiran Khosla, Christopher Self and Alistair Smith. This work is supported by a Samsung GRC project and the UK Hub in Quantum Computing and Simulation, part of the UK National Quantum Technologies Programme with funding from UKRI EPSRC grant EP/T001062/1. 
\end{acknowledgments}
\bibliography{OptimalTraining}

\begin{thebibliography}{65}%
\makeatletter
\providecommand \@ifxundefined [1]{%
 \@ifx{#1\undefined}
}%
\providecommand \@ifnum [1]{%
 \ifnum #1\expandafter \@firstoftwo
 \else \expandafter \@secondoftwo
 \fi
}%
\providecommand \@ifx [1]{%
 \ifx #1\expandafter \@firstoftwo
 \else \expandafter \@secondoftwo
 \fi
}%
\providecommand \natexlab [1]{#1}%
\providecommand \enquote  [1]{``#1''}%
\providecommand \bibnamefont  [1]{#1}%
\providecommand \bibfnamefont [1]{#1}%
\providecommand \citenamefont [1]{#1}%
\providecommand \href@noop [0]{\@secondoftwo}%
\providecommand \href [0]{\begingroup \@sanitize@url \@href}%
\providecommand \@href[1]{\@@startlink{#1}\@@href}%
\providecommand \@@href[1]{\endgroup#1\@@endlink}%
\providecommand \@sanitize@url [0]{\catcode `\\12\catcode `\$12\catcode
  `\&12\catcode `\#12\catcode `\^12\catcode `\_12\catcode `\%12\relax}%
\providecommand \@@startlink[1]{}%
\providecommand \@@endlink[0]{}%
\providecommand \url  [0]{\begingroup\@sanitize@url \@url }%
\providecommand \@url [1]{\endgroup\@href {#1}{\urlprefix }}%
\providecommand \urlprefix  [0]{URL }%
\providecommand \Eprint [0]{\href }%
\providecommand \doibase [0]{https://doi.org/}%
\providecommand \selectlanguage [0]{\@gobble}%
\providecommand \bibinfo  [0]{\@secondoftwo}%
\providecommand \bibfield  [0]{\@secondoftwo}%
\providecommand \translation [1]{[#1]}%
\providecommand \BibitemOpen [0]{}%
\providecommand \bibitemStop [0]{}%
\providecommand \bibitemNoStop [0]{.\EOS\space}%
\providecommand \EOS [0]{\spacefactor3000\relax}%
\providecommand \BibitemShut  [1]{\csname bibitem#1\endcsname}%
\let\auto@bib@innerbib\@empty
\bibitem [{\citenamefont {Preskill}(2018)}]{preskill2018quantum}%
  \BibitemOpen
  \bibfield  {author} {\bibinfo {author} {\bibfnamefont {J.}~\bibnamefont
  {Preskill}},\ }\bibfield  {title} {\bibinfo {title} {Quantum computing in the
  nisq era and beyond},\ }\href@noop {} {\bibfield  {journal} {\bibinfo
  {journal} {Quantum}\ }\textbf {\bibinfo {volume} {2}},\ \bibinfo {pages} {79}
  (\bibinfo {year} {2018})}\BibitemShut {NoStop}%
\bibitem [{\citenamefont {Bharti}\ \emph {et~al.}(2021)\citenamefont {Bharti},
  \citenamefont {Cervera-Lierta}, \citenamefont {Kyaw}, \citenamefont {Haug},
  \citenamefont {Alperin-Lea}, \citenamefont {Anand}, \citenamefont {Degroote},
  \citenamefont {Heimonen}, \citenamefont {Kottmann}, \citenamefont {Menke},
  \citenamefont {Mok}, \citenamefont {Sim}, \citenamefont {Kwek},\ and\
  \citenamefont {Aspuru-Guzik}}]{bharti2021noisy}%
  \BibitemOpen
  \bibfield  {author} {\bibinfo {author} {\bibfnamefont {K.}~\bibnamefont
  {Bharti}}, \bibinfo {author} {\bibfnamefont {A.}~\bibnamefont
  {Cervera-Lierta}}, \bibinfo {author} {\bibfnamefont {T.~H.}\ \bibnamefont
  {Kyaw}}, \bibinfo {author} {\bibfnamefont {T.}~\bibnamefont {Haug}}, \bibinfo
  {author} {\bibfnamefont {S.}~\bibnamefont {Alperin-Lea}}, \bibinfo {author}
  {\bibfnamefont {A.}~\bibnamefont {Anand}}, \bibinfo {author} {\bibfnamefont
  {M.}~\bibnamefont {Degroote}}, \bibinfo {author} {\bibfnamefont
  {H.}~\bibnamefont {Heimonen}}, \bibinfo {author} {\bibfnamefont {J.~S.}\
  \bibnamefont {Kottmann}}, \bibinfo {author} {\bibfnamefont {T.}~\bibnamefont
  {Menke}}, \bibinfo {author} {\bibfnamefont {W.-K.}\ \bibnamefont {Mok}},
  \bibinfo {author} {\bibfnamefont {S.}~\bibnamefont {Sim}}, \bibinfo {author}
  {\bibfnamefont {L.-C.}\ \bibnamefont {Kwek}},\ and\ \bibinfo {author}
  {\bibfnamefont {A.}~\bibnamefont {Aspuru-Guzik}},\ }\bibfield  {title}
  {\bibinfo {title} {Noisy intermediate-scale quantum (nisq) algorithms},\
  }\href@noop {} {\bibfield  {journal} {\bibinfo  {journal} {arXiv:2101.08448}\
  } (\bibinfo {year} {2021})}\BibitemShut {NoStop}%
\bibitem [{\citenamefont {Peruzzo}\ \emph {et~al.}(2014)\citenamefont
  {Peruzzo}, \citenamefont {McClean}, \citenamefont {Shadbolt}, \citenamefont
  {Yung}, \citenamefont {Zhou}, \citenamefont {Love}, \citenamefont
  {Aspuru-Guzik},\ and\ \citenamefont {Obrien}}]{peruzzo2014variational}%
  \BibitemOpen
  \bibfield  {author} {\bibinfo {author} {\bibfnamefont {A.}~\bibnamefont
  {Peruzzo}}, \bibinfo {author} {\bibfnamefont {J.}~\bibnamefont {McClean}},
  \bibinfo {author} {\bibfnamefont {P.}~\bibnamefont {Shadbolt}}, \bibinfo
  {author} {\bibfnamefont {M.-H.}\ \bibnamefont {Yung}}, \bibinfo {author}
  {\bibfnamefont {X.-Q.}\ \bibnamefont {Zhou}}, \bibinfo {author}
  {\bibfnamefont {P.~J.}\ \bibnamefont {Love}}, \bibinfo {author}
  {\bibfnamefont {A.}~\bibnamefont {Aspuru-Guzik}},\ and\ \bibinfo {author}
  {\bibfnamefont {J.~L.}\ \bibnamefont {Obrien}},\ }\bibfield  {title}
  {\bibinfo {title} {A variational eigenvalue solver on a photonic quantum
  processor},\ }\href@noop {} {\bibfield  {journal} {\bibinfo  {journal}
  {Nature communications}\ }\textbf {\bibinfo {volume} {5}},\ \bibinfo {pages}
  {4213} (\bibinfo {year} {2014})}\BibitemShut {NoStop}%
\bibitem [{\citenamefont {Kandala}\ \emph {et~al.}(2017)\citenamefont
  {Kandala}, \citenamefont {Mezzacapo}, \citenamefont {Temme}, \citenamefont
  {Takita}, \citenamefont {Brink}, \citenamefont {Chow},\ and\ \citenamefont
  {Gambetta}}]{kandala2017hardware}%
  \BibitemOpen
  \bibfield  {author} {\bibinfo {author} {\bibfnamefont {A.}~\bibnamefont
  {Kandala}}, \bibinfo {author} {\bibfnamefont {A.}~\bibnamefont {Mezzacapo}},
  \bibinfo {author} {\bibfnamefont {K.}~\bibnamefont {Temme}}, \bibinfo
  {author} {\bibfnamefont {M.}~\bibnamefont {Takita}}, \bibinfo {author}
  {\bibfnamefont {M.}~\bibnamefont {Brink}}, \bibinfo {author} {\bibfnamefont
  {J.~M.}\ \bibnamefont {Chow}},\ and\ \bibinfo {author} {\bibfnamefont
  {J.~M.}\ \bibnamefont {Gambetta}},\ }\bibfield  {title} {\bibinfo {title}
  {Hardware-efficient variational quantum eigensolver for small molecules and
  quantum magnets},\ }\href@noop {} {\bibfield  {journal} {\bibinfo  {journal}
  {Nature}\ }\textbf {\bibinfo {volume} {549}},\ \bibinfo {pages} {242}
  (\bibinfo {year} {2017})}\BibitemShut {NoStop}%
\bibitem [{\citenamefont {McClean}\ \emph {et~al.}(2016)\citenamefont
  {McClean}, \citenamefont {Romero}, \citenamefont {Babbush},\ and\
  \citenamefont {Aspuru-Guzik}}]{mcclean2016theory}%
  \BibitemOpen
  \bibfield  {author} {\bibinfo {author} {\bibfnamefont {J.~R.}\ \bibnamefont
  {McClean}}, \bibinfo {author} {\bibfnamefont {J.}~\bibnamefont {Romero}},
  \bibinfo {author} {\bibfnamefont {R.}~\bibnamefont {Babbush}},\ and\ \bibinfo
  {author} {\bibfnamefont {A.}~\bibnamefont {Aspuru-Guzik}},\ }\bibfield
  {title} {\bibinfo {title} {The theory of variational hybrid quantum-classical
  algorithms},\ }\href@noop {} {\bibfield  {journal} {\bibinfo  {journal} {New
  Journal of Physics}\ }\textbf {\bibinfo {volume} {18}},\ \bibinfo {pages}
  {023023} (\bibinfo {year} {2016})}\BibitemShut {NoStop}%
\bibitem [{\citenamefont {Cerezo}\ \emph {et~al.}(2020)\citenamefont {Cerezo},
  \citenamefont {Arrasmith}, \citenamefont {Babbush}, \citenamefont {Benjamin},
  \citenamefont {Endo}, \citenamefont {Fujii}, \citenamefont {McClean},
  \citenamefont {Mitarai}, \citenamefont {Yuan}, \citenamefont {Cincio} \emph
  {et~al.}}]{cerezo2020variational}%
  \BibitemOpen
  \bibfield  {author} {\bibinfo {author} {\bibfnamefont {M.}~\bibnamefont
  {Cerezo}}, \bibinfo {author} {\bibfnamefont {A.}~\bibnamefont {Arrasmith}},
  \bibinfo {author} {\bibfnamefont {R.}~\bibnamefont {Babbush}}, \bibinfo
  {author} {\bibfnamefont {S.~C.}\ \bibnamefont {Benjamin}}, \bibinfo {author}
  {\bibfnamefont {S.}~\bibnamefont {Endo}}, \bibinfo {author} {\bibfnamefont
  {K.}~\bibnamefont {Fujii}}, \bibinfo {author} {\bibfnamefont {J.~R.}\
  \bibnamefont {McClean}}, \bibinfo {author} {\bibfnamefont {K.}~\bibnamefont
  {Mitarai}}, \bibinfo {author} {\bibfnamefont {X.}~\bibnamefont {Yuan}},
  \bibinfo {author} {\bibfnamefont {L.}~\bibnamefont {Cincio}}, \emph
  {et~al.},\ }\bibfield  {title} {\bibinfo {title} {Variational quantum
  algorithms},\ }\href@noop {} {\bibfield  {journal} {\bibinfo  {journal}
  {arXiv preprint arXiv:2012.09265}\ } (\bibinfo {year} {2020})}\BibitemShut
  {NoStop}%
\bibitem [{\citenamefont {McClean}\ \emph {et~al.}(2018)\citenamefont
  {McClean}, \citenamefont {Boixo}, \citenamefont {Smelyanskiy}, \citenamefont
  {Babbush},\ and\ \citenamefont {Neven}}]{mcclean2018barren}%
  \BibitemOpen
  \bibfield  {author} {\bibinfo {author} {\bibfnamefont {J.~R.}\ \bibnamefont
  {McClean}}, \bibinfo {author} {\bibfnamefont {S.}~\bibnamefont {Boixo}},
  \bibinfo {author} {\bibfnamefont {V.~N.}\ \bibnamefont {Smelyanskiy}},
  \bibinfo {author} {\bibfnamefont {R.}~\bibnamefont {Babbush}},\ and\ \bibinfo
  {author} {\bibfnamefont {H.}~\bibnamefont {Neven}},\ }\bibfield  {title}
  {\bibinfo {title} {Barren plateaus in quantum neural network training
  landscapes},\ }\href@noop {} {\bibfield  {journal} {\bibinfo  {journal}
  {Nature communications}\ }\textbf {\bibinfo {volume} {9}},\ \bibinfo {pages}
  {4812} (\bibinfo {year} {2018})}\BibitemShut {NoStop}%
\bibitem [{\citenamefont {Cerezo}\ \emph
  {et~al.}(2021{\natexlab{a}})\citenamefont {Cerezo}, \citenamefont {Sone},
  \citenamefont {Volkoff}, \citenamefont {Cincio},\ and\ \citenamefont
  {Coles}}]{cerezo2021cost}%
  \BibitemOpen
  \bibfield  {author} {\bibinfo {author} {\bibfnamefont {M.}~\bibnamefont
  {Cerezo}}, \bibinfo {author} {\bibfnamefont {A.}~\bibnamefont {Sone}},
  \bibinfo {author} {\bibfnamefont {T.}~\bibnamefont {Volkoff}}, \bibinfo
  {author} {\bibfnamefont {L.}~\bibnamefont {Cincio}},\ and\ \bibinfo {author}
  {\bibfnamefont {P.~J.}\ \bibnamefont {Coles}},\ }\bibfield  {title} {\bibinfo
  {title} {Cost function dependent barren plateaus in shallow parametrized
  quantum circuits},\ }\href@noop {} {\bibfield  {journal} {\bibinfo  {journal}
  {Nature Communications}\ }\textbf {\bibinfo {volume} {12}},\ \bibinfo {pages}
  {1} (\bibinfo {year} {2021}{\natexlab{a}})}\BibitemShut {NoStop}%
\bibitem [{\citenamefont {Marrero}\ \emph {et~al.}(2020)\citenamefont
  {Marrero}, \citenamefont {Kieferov{\'a}},\ and\ \citenamefont
  {Wiebe}}]{marrero2020entanglement}%
  \BibitemOpen
  \bibfield  {author} {\bibinfo {author} {\bibfnamefont {C.~O.}\ \bibnamefont
  {Marrero}}, \bibinfo {author} {\bibfnamefont {M.}~\bibnamefont
  {Kieferov{\'a}}},\ and\ \bibinfo {author} {\bibfnamefont {N.}~\bibnamefont
  {Wiebe}},\ }\bibfield  {title} {\bibinfo {title} {Entanglement induced barren
  plateaus},\ }\href@noop {} {\bibfield  {journal} {\bibinfo  {journal}
  {arXiv:2010.15968}\ } (\bibinfo {year} {2020})}\BibitemShut {NoStop}%
\bibitem [{\citenamefont {Wang}\ \emph {et~al.}(2020)\citenamefont {Wang},
  \citenamefont {Fontana}, \citenamefont {Cerezo}, \citenamefont {Sharma},
  \citenamefont {Sone}, \citenamefont {Cincio},\ and\ \citenamefont
  {Coles}}]{wang2020noise}%
  \BibitemOpen
  \bibfield  {author} {\bibinfo {author} {\bibfnamefont {S.}~\bibnamefont
  {Wang}}, \bibinfo {author} {\bibfnamefont {E.}~\bibnamefont {Fontana}},
  \bibinfo {author} {\bibfnamefont {M.}~\bibnamefont {Cerezo}}, \bibinfo
  {author} {\bibfnamefont {K.}~\bibnamefont {Sharma}}, \bibinfo {author}
  {\bibfnamefont {A.}~\bibnamefont {Sone}}, \bibinfo {author} {\bibfnamefont
  {L.}~\bibnamefont {Cincio}},\ and\ \bibinfo {author} {\bibfnamefont {P.~J.}\
  \bibnamefont {Coles}},\ }\bibfield  {title} {\bibinfo {title} {Noise-induced
  barren plateaus in variational quantum algorithms},\ }\href@noop {}
  {\bibfield  {journal} {\bibinfo  {journal} {arXiv:2007.14384}\ } (\bibinfo
  {year} {2020})}\BibitemShut {NoStop}%
\bibitem [{\citenamefont {Bittel}\ and\ \citenamefont
  {Kliesch}(2021)}]{bittel2021training}%
  \BibitemOpen
  \bibfield  {author} {\bibinfo {author} {\bibfnamefont {L.}~\bibnamefont
  {Bittel}}\ and\ \bibinfo {author} {\bibfnamefont {M.}~\bibnamefont
  {Kliesch}},\ }\bibfield  {title} {\bibinfo {title} {Training variational
  quantum algorithms is np-hard -- even for logarithmically many qubits and
  free fermionic systems},\ }\href@noop {} {\bibfield  {journal} {\bibinfo
  {journal} {arXiv:2101.07267}\ } (\bibinfo {year} {2021})}\BibitemShut
  {NoStop}%
\bibitem [{\citenamefont {Bharti}(2020)}]{bharti2020quantum}%
  \BibitemOpen
  \bibfield  {author} {\bibinfo {author} {\bibfnamefont {K.}~\bibnamefont
  {Bharti}},\ }\bibfield  {title} {\bibinfo {title} {Quantum assisted
  eigensolver},\ }\href@noop {} {\bibfield  {journal} {\bibinfo  {journal}
  {arXiv:2009.11001}\ } (\bibinfo {year} {2020})}\BibitemShut {NoStop}%
\bibitem [{\citenamefont {Bharti}\ and\ \citenamefont
  {Haug}(2020{\natexlab{a}})}]{bharti2020quantum2}%
  \BibitemOpen
  \bibfield  {author} {\bibinfo {author} {\bibfnamefont {K.}~\bibnamefont
  {Bharti}}\ and\ \bibinfo {author} {\bibfnamefont {T.}~\bibnamefont {Haug}},\
  }\bibfield  {title} {\bibinfo {title} {Quantum assisted simulator},\
  }\href@noop {} {\bibfield  {journal} {\bibinfo  {journal} {arXiv:2011.06911}\
  } (\bibinfo {year} {2020}{\natexlab{a}})}\BibitemShut {NoStop}%
\bibitem [{\citenamefont {Bharti}\ and\ \citenamefont
  {Haug}(2020{\natexlab{b}})}]{bharti2020iterative}%
  \BibitemOpen
  \bibfield  {author} {\bibinfo {author} {\bibfnamefont {K.}~\bibnamefont
  {Bharti}}\ and\ \bibinfo {author} {\bibfnamefont {T.}~\bibnamefont {Haug}},\
  }\bibfield  {title} {\bibinfo {title} {Iterative quantum assisted
  eigensolver},\ }\href@noop {} {\bibfield  {journal} {\bibinfo  {journal}
  {arXiv:2010.05638}\ } (\bibinfo {year} {2020}{\natexlab{b}})}\BibitemShut
  {NoStop}%
\bibitem [{\citenamefont {Haug}\ and\ \citenamefont
  {Bharti}(2020)}]{haug2020generalized}%
  \BibitemOpen
  \bibfield  {author} {\bibinfo {author} {\bibfnamefont {T.}~\bibnamefont
  {Haug}}\ and\ \bibinfo {author} {\bibfnamefont {K.}~\bibnamefont {Bharti}},\
  }\bibfield  {title} {\bibinfo {title} {Generalized quantum assisted
  simulator},\ }\href {https://arxiv.org/abs/2011.14737} {\bibfield  {journal}
  {\bibinfo  {journal} {arXiv:2011.14737}\ } (\bibinfo {year}
  {2020})}\BibitemShut {NoStop}%
\bibitem [{\citenamefont {Lau}\ \emph {et~al.}(2021)\citenamefont {Lau},
  \citenamefont {Haug}, \citenamefont {Kwek},\ and\ \citenamefont
  {Bharti}}]{lau2021nisq}%
  \BibitemOpen
  \bibfield  {author} {\bibinfo {author} {\bibfnamefont {J.~W.~Z.}\
  \bibnamefont {Lau}}, \bibinfo {author} {\bibfnamefont {T.}~\bibnamefont
  {Haug}}, \bibinfo {author} {\bibfnamefont {L.~C.}\ \bibnamefont {Kwek}},\
  and\ \bibinfo {author} {\bibfnamefont {K.}~\bibnamefont {Bharti}},\
  }\bibfield  {title} {\bibinfo {title} {Nisq algorithm for hamiltonian
  simulation via truncated taylor series},\ }\href@noop {} {\bibfield
  {journal} {\bibinfo  {journal} {arXiv:2103.05500}\ } (\bibinfo {year}
  {2021})}\BibitemShut {NoStop}%
\bibitem [{\citenamefont {Lim}\ \emph {et~al.}(2021)\citenamefont {Lim},
  \citenamefont {Haug}, \citenamefont {Kwek},\ and\ \citenamefont
  {Bharti}}]{lim2021fastforwarding}%
  \BibitemOpen
  \bibfield  {author} {\bibinfo {author} {\bibfnamefont {K.~H.}\ \bibnamefont
  {Lim}}, \bibinfo {author} {\bibfnamefont {T.}~\bibnamefont {Haug}}, \bibinfo
  {author} {\bibfnamefont {L.~C.}\ \bibnamefont {Kwek}},\ and\ \bibinfo
  {author} {\bibfnamefont {K.}~\bibnamefont {Bharti}},\ }\bibfield  {title}
  {\bibinfo {title} {Fast-forwarding with nisq processors without feedback
  loop},\ }\href@noop {} {\bibfield  {journal} {\bibinfo  {journal}
  {arXiv:2104.01931}\ } (\bibinfo {year} {2021})}\BibitemShut {NoStop}%
\bibitem [{\citenamefont {Stokes}\ \emph {et~al.}(2020)\citenamefont {Stokes},
  \citenamefont {Izaac}, \citenamefont {Killoran},\ and\ \citenamefont
  {Carleo}}]{stokes2020quantum}%
  \BibitemOpen
  \bibfield  {author} {\bibinfo {author} {\bibfnamefont {J.}~\bibnamefont
  {Stokes}}, \bibinfo {author} {\bibfnamefont {J.}~\bibnamefont {Izaac}},
  \bibinfo {author} {\bibfnamefont {N.}~\bibnamefont {Killoran}},\ and\
  \bibinfo {author} {\bibfnamefont {G.}~\bibnamefont {Carleo}},\ }\bibfield
  {title} {\bibinfo {title} {Quantum natural gradient},\ }\href@noop {}
  {\bibfield  {journal} {\bibinfo  {journal} {Quantum}\ }\textbf {\bibinfo
  {volume} {4}},\ \bibinfo {pages} {269} (\bibinfo {year} {2020})}\BibitemShut
  {NoStop}%
\bibitem [{\citenamefont {Koczor}\ and\ \citenamefont
  {Benjamin}(2019)}]{koczor2019quantum}%
  \BibitemOpen
  \bibfield  {author} {\bibinfo {author} {\bibfnamefont {B.}~\bibnamefont
  {Koczor}}\ and\ \bibinfo {author} {\bibfnamefont {S.~C.}\ \bibnamefont
  {Benjamin}},\ }\bibfield  {title} {\bibinfo {title} {Quantum natural gradient
  generalised to non-unitary circuits},\ }\href@noop {} {\bibfield  {journal}
  {\bibinfo  {journal} {arXiv preprint arXiv:1912.08660}\ } (\bibinfo {year}
  {2019})}\BibitemShut {NoStop}%
\bibitem [{\citenamefont {Yamamoto}(2019)}]{yamamoto2019natural}%
  \BibitemOpen
  \bibfield  {author} {\bibinfo {author} {\bibfnamefont {N.}~\bibnamefont
  {Yamamoto}},\ }\bibfield  {title} {\bibinfo {title} {On the natural gradient
  for variational quantum eigensolver},\ }\href@noop {} {\bibfield  {journal}
  {\bibinfo  {journal} {arXiv:1909.05074}\ } (\bibinfo {year}
  {2019})}\BibitemShut {NoStop}%
\bibitem [{\citenamefont {van Straaten}\ and\ \citenamefont
  {Koczor}(2020)}]{van2020measurement}%
  \BibitemOpen
  \bibfield  {author} {\bibinfo {author} {\bibfnamefont {B.}~\bibnamefont {van
  Straaten}}\ and\ \bibinfo {author} {\bibfnamefont {B.}~\bibnamefont
  {Koczor}},\ }\bibfield  {title} {\bibinfo {title} {Measurement cost of
  metric-aware variational quantum algorithms},\ }\href@noop {} {\bibfield
  {journal} {\bibinfo  {journal} {arXiv:2005.05172}\ } (\bibinfo {year}
  {2020})}\BibitemShut {NoStop}%
\bibitem [{\citenamefont {Wierichs}\ \emph {et~al.}(2020)\citenamefont
  {Wierichs}, \citenamefont {Gogolin},\ and\ \citenamefont
  {Kastoryano}}]{wierichs2020avoiding}%
  \BibitemOpen
  \bibfield  {author} {\bibinfo {author} {\bibfnamefont {D.}~\bibnamefont
  {Wierichs}}, \bibinfo {author} {\bibfnamefont {C.}~\bibnamefont {Gogolin}},\
  and\ \bibinfo {author} {\bibfnamefont {M.}~\bibnamefont {Kastoryano}},\
  }\bibfield  {title} {\bibinfo {title} {Avoiding local minima in variational
  quantum eigensolvers with the natural gradient optimizer},\ }\href@noop {}
  {\bibfield  {journal} {\bibinfo  {journal} {arXiv:2004.14666}\ } (\bibinfo
  {year} {2020})}\BibitemShut {NoStop}%
\bibitem [{\citenamefont {Gacon}\ \emph {et~al.}(2021)\citenamefont {Gacon},
  \citenamefont {Zoufal}, \citenamefont {Carleo},\ and\ \citenamefont
  {Woerner}}]{gacon2021simultaneous}%
  \BibitemOpen
  \bibfield  {author} {\bibinfo {author} {\bibfnamefont {J.}~\bibnamefont
  {Gacon}}, \bibinfo {author} {\bibfnamefont {C.}~\bibnamefont {Zoufal}},
  \bibinfo {author} {\bibfnamefont {G.}~\bibnamefont {Carleo}},\ and\ \bibinfo
  {author} {\bibfnamefont {S.}~\bibnamefont {Woerner}},\ }\bibfield  {title}
  {\bibinfo {title} {Simultaneous perturbation stochastic approximation of the
  quantum fisher information},\ }\href@noop {} {\bibfield  {journal} {\bibinfo
  {journal} {arXiv:2103.09232}\ } (\bibinfo {year} {2021})}\BibitemShut
  {NoStop}%
\bibitem [{\citenamefont {Haug}\ \emph {et~al.}(2021)\citenamefont {Haug},
  \citenamefont {Bharti},\ and\ \citenamefont {Kim}}]{haug2021capacity}%
  \BibitemOpen
  \bibfield  {author} {\bibinfo {author} {\bibfnamefont {T.}~\bibnamefont
  {Haug}}, \bibinfo {author} {\bibfnamefont {K.}~\bibnamefont {Bharti}},\ and\
  \bibinfo {author} {\bibfnamefont {M.}~\bibnamefont {Kim}},\ }\bibfield
  {title} {\bibinfo {title} {Capacity and quantum geometry of parametrized
  quantum circuits},\ }\href@noop {} {\bibfield  {journal} {\bibinfo  {journal}
  {arXiv:2102.01659}\ } (\bibinfo {year} {2021})}\BibitemShut {NoStop}%
\bibitem [{\citenamefont {d'Alessandro}(2007)}]{d2007introduction}%
  \BibitemOpen
  \bibfield  {author} {\bibinfo {author} {\bibfnamefont {D.}~\bibnamefont
  {d'Alessandro}},\ }\href@noop {} {\emph {\bibinfo {title} {Introduction to
  quantum control and dynamics}}}\ (\bibinfo  {publisher} {CRC press},\
  \bibinfo {year} {2007})\BibitemShut {NoStop}%
\bibitem [{\citenamefont {Machnes}\ \emph {et~al.}(2011)\citenamefont
  {Machnes}, \citenamefont {Sander}, \citenamefont {Glaser}, \citenamefont
  {de~Fouquieres}, \citenamefont {Gruslys}, \citenamefont {Schirmer},\ and\
  \citenamefont {Schulte-Herbr{\"u}ggen}}]{machnes2011comparing}%
  \BibitemOpen
  \bibfield  {author} {\bibinfo {author} {\bibfnamefont {S.}~\bibnamefont
  {Machnes}}, \bibinfo {author} {\bibfnamefont {U.}~\bibnamefont {Sander}},
  \bibinfo {author} {\bibfnamefont {S.~J.}\ \bibnamefont {Glaser}}, \bibinfo
  {author} {\bibfnamefont {P.}~\bibnamefont {de~Fouquieres}}, \bibinfo {author}
  {\bibfnamefont {A.}~\bibnamefont {Gruslys}}, \bibinfo {author} {\bibfnamefont
  {S.}~\bibnamefont {Schirmer}},\ and\ \bibinfo {author} {\bibfnamefont
  {T.}~\bibnamefont {Schulte-Herbr{\"u}ggen}},\ }\bibfield  {title} {\bibinfo
  {title} {Comparing, optimizing, and benchmarking quantum-control algorithms
  in a unifying programming framework},\ }\href@noop {} {\bibfield  {journal}
  {\bibinfo  {journal} {Physical Review A}\ }\textbf {\bibinfo {volume} {84}},\
  \bibinfo {pages} {022305} (\bibinfo {year} {2011})}\BibitemShut {NoStop}%
\bibitem [{\citenamefont {Otten}\ \emph {et~al.}(2019)\citenamefont {Otten},
  \citenamefont {Cortes},\ and\ \citenamefont {Gray}}]{otten2019noise}%
  \BibitemOpen
  \bibfield  {author} {\bibinfo {author} {\bibfnamefont {M.}~\bibnamefont
  {Otten}}, \bibinfo {author} {\bibfnamefont {C.~L.}\ \bibnamefont {Cortes}},\
  and\ \bibinfo {author} {\bibfnamefont {S.~K.}\ \bibnamefont {Gray}},\
  }\bibfield  {title} {\bibinfo {title} {Noise-resilient quantum dynamics using
  symmetry-preserving ansatzes},\ }\href@noop {} {\bibfield  {journal}
  {\bibinfo  {journal} {arXiv preprint arXiv:1910.06284}\ } (\bibinfo {year}
  {2019})}\BibitemShut {NoStop}%
\bibitem [{\citenamefont {Barison}\ \emph {et~al.}(2021)\citenamefont
  {Barison}, \citenamefont {Vicentini},\ and\ \citenamefont
  {Carleo}}]{barison2021efficient}%
  \BibitemOpen
  \bibfield  {author} {\bibinfo {author} {\bibfnamefont {S.}~\bibnamefont
  {Barison}}, \bibinfo {author} {\bibfnamefont {F.}~\bibnamefont {Vicentini}},\
  and\ \bibinfo {author} {\bibfnamefont {G.}~\bibnamefont {Carleo}},\
  }\bibfield  {title} {\bibinfo {title} {An efficient quantum algorithm for the
  time evolution of parameterized circuits},\ }\href@noop {} {\bibfield
  {journal} {\bibinfo  {journal} {arXiv:2101.04579}\ } (\bibinfo {year}
  {2021})}\BibitemShut {NoStop}%
\bibitem [{\citenamefont {Gibbs}\ \emph {et~al.}(2021)\citenamefont {Gibbs},
  \citenamefont {Gili}, \citenamefont {Holmes}, \citenamefont {Commeau},
  \citenamefont {Arrasmith}, \citenamefont {Cincio}, \citenamefont {Coles},\
  and\ \citenamefont {Sornborger}}]{gibbs2021long}%
  \BibitemOpen
  \bibfield  {author} {\bibinfo {author} {\bibfnamefont {J.}~\bibnamefont
  {Gibbs}}, \bibinfo {author} {\bibfnamefont {K.}~\bibnamefont {Gili}},
  \bibinfo {author} {\bibfnamefont {Z.}~\bibnamefont {Holmes}}, \bibinfo
  {author} {\bibfnamefont {B.}~\bibnamefont {Commeau}}, \bibinfo {author}
  {\bibfnamefont {A.}~\bibnamefont {Arrasmith}}, \bibinfo {author}
  {\bibfnamefont {L.}~\bibnamefont {Cincio}}, \bibinfo {author} {\bibfnamefont
  {P.~J.}\ \bibnamefont {Coles}},\ and\ \bibinfo {author} {\bibfnamefont
  {A.}~\bibnamefont {Sornborger}},\ }\bibfield  {title} {\bibinfo {title}
  {Long-time simulations with high fidelity on quantum hardware},\ }\href@noop
  {} {\bibfield  {journal} {\bibinfo  {journal} {arXiv:2102.04313}\ } (\bibinfo
  {year} {2021})}\BibitemShut {NoStop}%
\bibitem [{\citenamefont {Holmes}\ \emph {et~al.}(2020)\citenamefont {Holmes},
  \citenamefont {Arrasmith}, \citenamefont {Yan}, \citenamefont {Coles},
  \citenamefont {Albrecht},\ and\ \citenamefont
  {Sornborger}}]{holmes2020barren}%
  \BibitemOpen
  \bibfield  {author} {\bibinfo {author} {\bibfnamefont {Z.}~\bibnamefont
  {Holmes}}, \bibinfo {author} {\bibfnamefont {A.}~\bibnamefont {Arrasmith}},
  \bibinfo {author} {\bibfnamefont {B.}~\bibnamefont {Yan}}, \bibinfo {author}
  {\bibfnamefont {P.~J.}\ \bibnamefont {Coles}}, \bibinfo {author}
  {\bibfnamefont {A.}~\bibnamefont {Albrecht}},\ and\ \bibinfo {author}
  {\bibfnamefont {A.~T.}\ \bibnamefont {Sornborger}},\ }\bibfield  {title}
  {\bibinfo {title} {Barren plateaus preclude learning scramblers},\
  }\href@noop {} {\bibfield  {journal} {\bibinfo  {journal} {arXiv:2009.14808}\
  } (\bibinfo {year} {2020})}\BibitemShut {NoStop}%
\bibitem [{\citenamefont {Benedetti}\ \emph {et~al.}(2019)\citenamefont
  {Benedetti}, \citenamefont {Garcia-Pintos}, \citenamefont {Perdomo},
  \citenamefont {Leyton-Ortega}, \citenamefont {Nam},\ and\ \citenamefont
  {Perdomo-Ortiz}}]{benedetti2019generative}%
  \BibitemOpen
  \bibfield  {author} {\bibinfo {author} {\bibfnamefont {M.}~\bibnamefont
  {Benedetti}}, \bibinfo {author} {\bibfnamefont {D.}~\bibnamefont
  {Garcia-Pintos}}, \bibinfo {author} {\bibfnamefont {O.}~\bibnamefont
  {Perdomo}}, \bibinfo {author} {\bibfnamefont {V.}~\bibnamefont
  {Leyton-Ortega}}, \bibinfo {author} {\bibfnamefont {Y.}~\bibnamefont {Nam}},\
  and\ \bibinfo {author} {\bibfnamefont {A.}~\bibnamefont {Perdomo-Ortiz}},\
  }\bibfield  {title} {\bibinfo {title} {A generative modeling approach for
  benchmarking and training shallow quantum circuits},\ }\href@noop {}
  {\bibfield  {journal} {\bibinfo  {journal} {npj Quantum Information}\
  }\textbf {\bibinfo {volume} {5}},\ \bibinfo {pages} {1} (\bibinfo {year}
  {2019})}\BibitemShut {NoStop}%
\bibitem [{\citenamefont {Huang}\ \emph
  {et~al.}(2020{\natexlab{a}})\citenamefont {Huang}, \citenamefont {Du},
  \citenamefont {Gong}, \citenamefont {Zhao}, \citenamefont {Wu}, \citenamefont
  {Wang}, \citenamefont {Li}, \citenamefont {Liang}, \citenamefont {Lin},
  \citenamefont {Xu} \emph {et~al.}}]{huang2020experimental}%
  \BibitemOpen
  \bibfield  {author} {\bibinfo {author} {\bibfnamefont {H.-L.}\ \bibnamefont
  {Huang}}, \bibinfo {author} {\bibfnamefont {Y.}~\bibnamefont {Du}}, \bibinfo
  {author} {\bibfnamefont {M.}~\bibnamefont {Gong}}, \bibinfo {author}
  {\bibfnamefont {Y.}~\bibnamefont {Zhao}}, \bibinfo {author} {\bibfnamefont
  {Y.}~\bibnamefont {Wu}}, \bibinfo {author} {\bibfnamefont {C.}~\bibnamefont
  {Wang}}, \bibinfo {author} {\bibfnamefont {S.}~\bibnamefont {Li}}, \bibinfo
  {author} {\bibfnamefont {F.}~\bibnamefont {Liang}}, \bibinfo {author}
  {\bibfnamefont {J.}~\bibnamefont {Lin}}, \bibinfo {author} {\bibfnamefont
  {Y.}~\bibnamefont {Xu}}, \emph {et~al.},\ }\bibfield  {title} {\bibinfo
  {title} {Experimental quantum generative adversarial networks for image
  generation},\ }\href@noop {} {\bibfield  {journal} {\bibinfo  {journal}
  {arXiv:2010.06201}\ } (\bibinfo {year} {2020}{\natexlab{a}})}\BibitemShut
  {NoStop}%
\bibitem [{\citenamefont {Romero}\ \emph {et~al.}(2017)\citenamefont {Romero},
  \citenamefont {Olson},\ and\ \citenamefont
  {Aspuru-Guzik}}]{romero2017quantum}%
  \BibitemOpen
  \bibfield  {author} {\bibinfo {author} {\bibfnamefont {J.}~\bibnamefont
  {Romero}}, \bibinfo {author} {\bibfnamefont {J.~P.}\ \bibnamefont {Olson}},\
  and\ \bibinfo {author} {\bibfnamefont {A.}~\bibnamefont {Aspuru-Guzik}},\
  }\bibfield  {title} {\bibinfo {title} {Quantum autoencoders for efficient
  compression of quantum data},\ }\href
  {https://doi.org/10.1088/2058-9565/aa8072} {\bibfield  {journal} {\bibinfo
  {journal} {Quantum Sci. Technol.}\ }\textbf {\bibinfo {volume} {2}},\
  \bibinfo {pages} {045001} (\bibinfo {year} {2017})}\BibitemShut {NoStop}%
\bibitem [{\citenamefont {Higgott}\ \emph {et~al.}(2019)\citenamefont
  {Higgott}, \citenamefont {Wang},\ and\ \citenamefont
  {Brierley}}]{higgott2019variational}%
  \BibitemOpen
  \bibfield  {author} {\bibinfo {author} {\bibfnamefont {O.}~\bibnamefont
  {Higgott}}, \bibinfo {author} {\bibfnamefont {D.}~\bibnamefont {Wang}},\ and\
  \bibinfo {author} {\bibfnamefont {S.}~\bibnamefont {Brierley}},\ }\bibfield
  {title} {\bibinfo {title} {Variational quantum computation of excited
  states},\ }\href@noop {} {\bibfield  {journal} {\bibinfo  {journal}
  {Quantum}\ }\textbf {\bibinfo {volume} {3}},\ \bibinfo {pages} {156}
  (\bibinfo {year} {2019})}\BibitemShut {NoStop}%
\bibitem [{\citenamefont {Meyer}(2021)}]{meyer2021fisher}%
  \BibitemOpen
  \bibfield  {author} {\bibinfo {author} {\bibfnamefont {J.~J.}\ \bibnamefont
  {Meyer}},\ }\bibfield  {title} {\bibinfo {title} {Fisher information in noisy
  intermediate-scale quantum applications},\ }\href@noop {} {\bibfield
  {journal} {\bibinfo  {journal} {arXiv:2103.15191}\ } (\bibinfo {year}
  {2021})}\BibitemShut {NoStop}%
\bibitem [{\citenamefont {Liu}\ \emph {et~al.}(2019)\citenamefont {Liu},
  \citenamefont {Yuan}, \citenamefont {Lu},\ and\ \citenamefont
  {Wang}}]{liu2019quantum}%
  \BibitemOpen
  \bibfield  {author} {\bibinfo {author} {\bibfnamefont {J.}~\bibnamefont
  {Liu}}, \bibinfo {author} {\bibfnamefont {H.}~\bibnamefont {Yuan}}, \bibinfo
  {author} {\bibfnamefont {X.-M.}\ \bibnamefont {Lu}},\ and\ \bibinfo {author}
  {\bibfnamefont {X.}~\bibnamefont {Wang}},\ }\bibfield  {title} {\bibinfo
  {title} {Quantum fisher information matrix and multiparameter estimation},\
  }\href@noop {} {\bibfield  {journal} {\bibinfo  {journal} {Journal of Physics
  A: Mathematical and Theoretical}\ }\textbf {\bibinfo {volume} {53}},\
  \bibinfo {pages} {023001} (\bibinfo {year} {2019})}\BibitemShut {NoStop}%
\bibitem [{\citenamefont {Brown}\ and\ \citenamefont
  {Viola}(2010)}]{brown2010convergence}%
  \BibitemOpen
  \bibfield  {author} {\bibinfo {author} {\bibfnamefont {W.~G.}\ \bibnamefont
  {Brown}}\ and\ \bibinfo {author} {\bibfnamefont {L.}~\bibnamefont {Viola}},\
  }\bibfield  {title} {\bibinfo {title} {Convergence rates for arbitrary
  statistical moments of random quantum circuits},\ }\href@noop {} {\bibfield
  {journal} {\bibinfo  {journal} {Physical review letters}\ }\textbf {\bibinfo
  {volume} {104}},\ \bibinfo {pages} {250501} (\bibinfo {year}
  {2010})}\BibitemShut {NoStop}%
\bibitem [{\citenamefont {Luo}\ \emph {et~al.}(2020)\citenamefont {Luo},
  \citenamefont {Liu}, \citenamefont {Zhang},\ and\ \citenamefont
  {Wang}}]{yao}%
  \BibitemOpen
  \bibfield  {author} {\bibinfo {author} {\bibfnamefont {X.-Z.}\ \bibnamefont
  {Luo}}, \bibinfo {author} {\bibfnamefont {J.-G.}\ \bibnamefont {Liu}},
  \bibinfo {author} {\bibfnamefont {P.}~\bibnamefont {Zhang}},\ and\ \bibinfo
  {author} {\bibfnamefont {L.}~\bibnamefont {Wang}},\ }\bibfield  {title}
  {\bibinfo {title} {Yao. jl: Extensible, efficient framework for quantum
  algorithm design},\ }\href {https://doi.org/10.22331/q-2020-10-11-341}
  {\bibfield  {journal} {\bibinfo  {journal} {Quantum}\ }\textbf {\bibinfo
  {volume} {4}},\ \bibinfo {pages} {341} (\bibinfo {year} {2020})}\BibitemShut
  {NoStop}%
\bibitem [{\citenamefont {Johansson}\ \emph {et~al.}(2012)\citenamefont
  {Johansson}, \citenamefont {Nation},\ and\ \citenamefont
  {Nori}}]{johansson2012qutip}%
  \BibitemOpen
  \bibfield  {author} {\bibinfo {author} {\bibfnamefont {J.~R.}\ \bibnamefont
  {Johansson}}, \bibinfo {author} {\bibfnamefont {P.~D.}\ \bibnamefont
  {Nation}},\ and\ \bibinfo {author} {\bibfnamefont {F.}~\bibnamefont {Nori}},\
  }\bibfield  {title} {\bibinfo {title} {Qutip: An open-source python framework
  for the dynamics of open quantum systems},\ }\href@noop {} {\bibfield
  {journal} {\bibinfo  {journal} {Computer Physics Communications}\ }\textbf
  {\bibinfo {volume} {183}},\ \bibinfo {pages} {1760} (\bibinfo {year}
  {2012})}\BibitemShut {NoStop}%
\bibitem [{\citenamefont {Sim}\ \emph {et~al.}(2019)\citenamefont {Sim},
  \citenamefont {Johnson},\ and\ \citenamefont
  {Aspuru-Guzik}}]{sim2019expressibility}%
  \BibitemOpen
  \bibfield  {author} {\bibinfo {author} {\bibfnamefont {S.}~\bibnamefont
  {Sim}}, \bibinfo {author} {\bibfnamefont {P.~D.}\ \bibnamefont {Johnson}},\
  and\ \bibinfo {author} {\bibfnamefont {A.}~\bibnamefont {Aspuru-Guzik}},\
  }\bibfield  {title} {\bibinfo {title} {Expressibility and entangling
  capability of parameterized quantum circuits for hybrid quantum-classical
  algorithms},\ }\href@noop {} {\bibfield  {journal} {\bibinfo  {journal}
  {Advanced Quantum Technologies}\ }\textbf {\bibinfo {volume} {2}},\ \bibinfo
  {pages} {1900070} (\bibinfo {year} {2019})}\BibitemShut {NoStop}%
\bibitem [{\citenamefont {Jones}\ \emph {et~al.}(2019)\citenamefont {Jones},
  \citenamefont {Endo}, \citenamefont {McArdle}, \citenamefont {Yuan},\ and\
  \citenamefont {Benjamin}}]{jones2019variational}%
  \BibitemOpen
  \bibfield  {author} {\bibinfo {author} {\bibfnamefont {T.}~\bibnamefont
  {Jones}}, \bibinfo {author} {\bibfnamefont {S.}~\bibnamefont {Endo}},
  \bibinfo {author} {\bibfnamefont {S.}~\bibnamefont {McArdle}}, \bibinfo
  {author} {\bibfnamefont {X.}~\bibnamefont {Yuan}},\ and\ \bibinfo {author}
  {\bibfnamefont {S.~C.}\ \bibnamefont {Benjamin}},\ }\bibfield  {title}
  {\bibinfo {title} {Variational quantum algorithms for discovering hamiltonian
  spectra},\ }\href@noop {} {\bibfield  {journal} {\bibinfo  {journal}
  {Physical Review A}\ }\textbf {\bibinfo {volume} {99}},\ \bibinfo {pages}
  {062304} (\bibinfo {year} {2019})}\BibitemShut {NoStop}%
\bibitem [{\citenamefont {Khatri}\ \emph {et~al.}(2019)\citenamefont {Khatri},
  \citenamefont {LaRose}, \citenamefont {Poremba}, \citenamefont {Cincio},
  \citenamefont {Sornborger},\ and\ \citenamefont {Coles}}]{khatri2019quantum}%
  \BibitemOpen
  \bibfield  {author} {\bibinfo {author} {\bibfnamefont {S.}~\bibnamefont
  {Khatri}}, \bibinfo {author} {\bibfnamefont {R.}~\bibnamefont {LaRose}},
  \bibinfo {author} {\bibfnamefont {A.}~\bibnamefont {Poremba}}, \bibinfo
  {author} {\bibfnamefont {L.}~\bibnamefont {Cincio}}, \bibinfo {author}
  {\bibfnamefont {A.~T.}\ \bibnamefont {Sornborger}},\ and\ \bibinfo {author}
  {\bibfnamefont {P.~J.}\ \bibnamefont {Coles}},\ }\bibfield  {title} {\bibinfo
  {title} {Quantum-assisted quantum compiling},\ }\href@noop {} {\bibfield
  {journal} {\bibinfo  {journal} {Quantum}\ }\textbf {\bibinfo {volume} {3}},\
  \bibinfo {pages} {140} (\bibinfo {year} {2019})}\BibitemShut {NoStop}%
\bibitem [{\citenamefont {Mari}\ \emph {et~al.}(2021)\citenamefont {Mari},
  \citenamefont {Bromley},\ and\ \citenamefont
  {Killoran}}]{mari2021estimating}%
  \BibitemOpen
  \bibfield  {author} {\bibinfo {author} {\bibfnamefont {A.}~\bibnamefont
  {Mari}}, \bibinfo {author} {\bibfnamefont {T.~R.}\ \bibnamefont {Bromley}},\
  and\ \bibinfo {author} {\bibfnamefont {N.}~\bibnamefont {Killoran}},\
  }\bibfield  {title} {\bibinfo {title} {Estimating the gradient and
  higher-order derivatives on quantum hardware},\ }\href@noop {} {\bibfield
  {journal} {\bibinfo  {journal} {Physical Review A}\ }\textbf {\bibinfo
  {volume} {103}},\ \bibinfo {pages} {012405} (\bibinfo {year}
  {2021})}\BibitemShut {NoStop}%
\bibitem [{\citenamefont {Cerezo}\ \emph
  {et~al.}(2021{\natexlab{b}})\citenamefont {Cerezo}, \citenamefont {Sone},
  \citenamefont {Beckey},\ and\ \citenamefont {Coles}}]{cerezo2021sub}%
  \BibitemOpen
  \bibfield  {author} {\bibinfo {author} {\bibfnamefont {M.}~\bibnamefont
  {Cerezo}}, \bibinfo {author} {\bibfnamefont {A.}~\bibnamefont {Sone}},
  \bibinfo {author} {\bibfnamefont {J.~L.}\ \bibnamefont {Beckey}},\ and\
  \bibinfo {author} {\bibfnamefont {P.~J.}\ \bibnamefont {Coles}},\ }\bibfield
  {title} {\bibinfo {title} {Sub-quantum fisher information},\ }\href@noop {}
  {\bibfield  {journal} {\bibinfo  {journal} {arXiv:2101.10144}\ } (\bibinfo
  {year} {2021}{\natexlab{b}})}\BibitemShut {NoStop}%
\bibitem [{\citenamefont {Beckey}\ \emph {et~al.}(2020)\citenamefont {Beckey},
  \citenamefont {Cerezo}, \citenamefont {Sone},\ and\ \citenamefont
  {Coles}}]{beckey2020variational}%
  \BibitemOpen
  \bibfield  {author} {\bibinfo {author} {\bibfnamefont {J.~L.}\ \bibnamefont
  {Beckey}}, \bibinfo {author} {\bibfnamefont {M.}~\bibnamefont {Cerezo}},
  \bibinfo {author} {\bibfnamefont {A.}~\bibnamefont {Sone}},\ and\ \bibinfo
  {author} {\bibfnamefont {P.~J.}\ \bibnamefont {Coles}},\ }\bibfield  {title}
  {\bibinfo {title} {Variational quantum algorithm for estimating the quantum
  fisher information},\ }\href@noop {} {\bibfield  {journal} {\bibinfo
  {journal} {arXiv:2010.10488}\ } (\bibinfo {year} {2020})}\BibitemShut
  {NoStop}%
\bibitem [{\citenamefont {Jones}(2020)}]{jones2020efficient}%
  \BibitemOpen
  \bibfield  {author} {\bibinfo {author} {\bibfnamefont {T.}~\bibnamefont
  {Jones}},\ }\bibfield  {title} {\bibinfo {title} {Efficient classical
  calculation of the quantum natural gradient},\ }\href@noop {} {\bibfield
  {journal} {\bibinfo  {journal} {arXiv preprint arXiv:2011.02991}\ } (\bibinfo
  {year} {2020})}\BibitemShut {NoStop}%
\bibitem [{\citenamefont {Glaser}\ \emph {et~al.}(2015)\citenamefont {Glaser},
  \citenamefont {Boscain}, \citenamefont {Calarco}, \citenamefont {Koch},
  \citenamefont {K{\"o}ckenberger}, \citenamefont {Kosloff}, \citenamefont
  {Kuprov}, \citenamefont {Luy}, \citenamefont {Schirmer}, \citenamefont
  {Schulte-Herbr{\"u}ggen} \emph {et~al.}}]{glaser2015training}%
  \BibitemOpen
  \bibfield  {author} {\bibinfo {author} {\bibfnamefont {S.~J.}\ \bibnamefont
  {Glaser}}, \bibinfo {author} {\bibfnamefont {U.}~\bibnamefont {Boscain}},
  \bibinfo {author} {\bibfnamefont {T.}~\bibnamefont {Calarco}}, \bibinfo
  {author} {\bibfnamefont {C.~P.}\ \bibnamefont {Koch}}, \bibinfo {author}
  {\bibfnamefont {W.}~\bibnamefont {K{\"o}ckenberger}}, \bibinfo {author}
  {\bibfnamefont {R.}~\bibnamefont {Kosloff}}, \bibinfo {author} {\bibfnamefont
  {I.}~\bibnamefont {Kuprov}}, \bibinfo {author} {\bibfnamefont
  {B.}~\bibnamefont {Luy}}, \bibinfo {author} {\bibfnamefont {S.}~\bibnamefont
  {Schirmer}}, \bibinfo {author} {\bibfnamefont {T.}~\bibnamefont
  {Schulte-Herbr{\"u}ggen}}, \emph {et~al.},\ }\bibfield  {title} {\bibinfo
  {title} {Training schr{\"o}dinger’s cat: quantum optimal control},\
  }\href@noop {} {\bibfield  {journal} {\bibinfo  {journal} {The European
  Physical Journal D}\ }\textbf {\bibinfo {volume} {69}},\ \bibinfo {pages} {1}
  (\bibinfo {year} {2015})}\BibitemShut {NoStop}%
\bibitem [{\citenamefont {Bastidas}\ \emph {et~al.}(2020)\citenamefont
  {Bastidas}, \citenamefont {Haug}, \citenamefont {Gravel}, \citenamefont
  {Kwek}, \citenamefont {Munro},\ and\ \citenamefont
  {Nemoto}}]{bastidas2020fully}%
  \BibitemOpen
  \bibfield  {author} {\bibinfo {author} {\bibfnamefont {V.}~\bibnamefont
  {Bastidas}}, \bibinfo {author} {\bibfnamefont {T.}~\bibnamefont {Haug}},
  \bibinfo {author} {\bibfnamefont {C.}~\bibnamefont {Gravel}}, \bibinfo
  {author} {\bibfnamefont {L.-C.}\ \bibnamefont {Kwek}}, \bibinfo {author}
  {\bibfnamefont {W.}~\bibnamefont {Munro}},\ and\ \bibinfo {author}
  {\bibfnamefont {K.}~\bibnamefont {Nemoto}},\ }\bibfield  {title} {\bibinfo
  {title} {Fully-programmable universal quantum simulator with a
  one-dimensional quantum processor},\ }\href
  {https://arxiv.org/abs/2009.00823} {\bibfield  {journal} {\bibinfo  {journal}
  {arXiv:2009.00823}\ } (\bibinfo {year} {2020})}\BibitemShut {NoStop}%
\bibitem [{\citenamefont {Khaneja}\ \emph {et~al.}(2005)\citenamefont
  {Khaneja}, \citenamefont {Reiss}, \citenamefont {Kehlet}, \citenamefont
  {Schulte-Herbr{\"u}ggen},\ and\ \citenamefont {Glaser}}]{khaneja2005optimal}%
  \BibitemOpen
  \bibfield  {author} {\bibinfo {author} {\bibfnamefont {N.}~\bibnamefont
  {Khaneja}}, \bibinfo {author} {\bibfnamefont {T.}~\bibnamefont {Reiss}},
  \bibinfo {author} {\bibfnamefont {C.}~\bibnamefont {Kehlet}}, \bibinfo
  {author} {\bibfnamefont {T.}~\bibnamefont {Schulte-Herbr{\"u}ggen}},\ and\
  \bibinfo {author} {\bibfnamefont {S.~J.}\ \bibnamefont {Glaser}},\ }\bibfield
   {title} {\bibinfo {title} {Optimal control of coupled spin dynamics: design
  of nmr pulse sequences by gradient ascent algorithms},\ }\href@noop {}
  {\bibfield  {journal} {\bibinfo  {journal} {Journal of magnetic resonance}\
  }\textbf {\bibinfo {volume} {172}},\ \bibinfo {pages} {296} (\bibinfo {year}
  {2005})}\BibitemShut {NoStop}%
\bibitem [{\citenamefont {Arrasmith}\ \emph {et~al.}(2021)\citenamefont
  {Arrasmith}, \citenamefont {Holmes}, \citenamefont {Cerezo},\ and\
  \citenamefont {Coles}}]{arrasmith2021equivalence}%
  \BibitemOpen
  \bibfield  {author} {\bibinfo {author} {\bibfnamefont {A.}~\bibnamefont
  {Arrasmith}}, \bibinfo {author} {\bibfnamefont {Z.}~\bibnamefont {Holmes}},
  \bibinfo {author} {\bibfnamefont {M.}~\bibnamefont {Cerezo}},\ and\ \bibinfo
  {author} {\bibfnamefont {P.~J.}\ \bibnamefont {Coles}},\ }\bibfield  {title}
  {\bibinfo {title} {Equivalence of quantum barren plateaus to cost
  concentration and narrow gorges},\ }\href@noop {} {\bibfield  {journal}
  {\bibinfo  {journal} {arXiv:2104.05868}\ } (\bibinfo {year}
  {2021})}\BibitemShut {NoStop}%
\bibitem [{\citenamefont {Anandan}\ and\ \citenamefont
  {Aharonov}(1990)}]{anandan1990geometry}%
  \BibitemOpen
  \bibfield  {author} {\bibinfo {author} {\bibfnamefont {J.}~\bibnamefont
  {Anandan}}\ and\ \bibinfo {author} {\bibfnamefont {Y.}~\bibnamefont
  {Aharonov}},\ }\bibfield  {title} {\bibinfo {title} {Geometry of quantum
  evolution},\ }\href@noop {} {\bibfield  {journal} {\bibinfo  {journal}
  {Physical review letters}\ }\textbf {\bibinfo {volume} {65}},\ \bibinfo
  {pages} {1697} (\bibinfo {year} {1990})}\BibitemShut {NoStop}%
\bibitem [{\citenamefont {Grimsley}\ \emph {et~al.}(2019)\citenamefont
  {Grimsley}, \citenamefont {Economou}, \citenamefont {Barnes},\ and\
  \citenamefont {Mayhall}}]{grimsley2019adaptive}%
  \BibitemOpen
  \bibfield  {author} {\bibinfo {author} {\bibfnamefont {H.~R.}\ \bibnamefont
  {Grimsley}}, \bibinfo {author} {\bibfnamefont {S.~E.}\ \bibnamefont
  {Economou}}, \bibinfo {author} {\bibfnamefont {E.}~\bibnamefont {Barnes}},\
  and\ \bibinfo {author} {\bibfnamefont {N.~J.}\ \bibnamefont {Mayhall}},\
  }\bibfield  {title} {\bibinfo {title} {An adaptive variational algorithm for
  exact molecular simulations on a quantum computer},\ }\href@noop {}
  {\bibfield  {journal} {\bibinfo  {journal} {Nature communications}\ }\textbf
  {\bibinfo {volume} {10}},\ \bibinfo {pages} {1} (\bibinfo {year}
  {2019})}\BibitemShut {NoStop}%
\bibitem [{\citenamefont {Zhang}\ \emph {et~al.}(2021)\citenamefont {Zhang},
  \citenamefont {Kyaw}, \citenamefont {Kottmann}, \citenamefont {Degroote},\
  and\ \citenamefont {Aspuru-Guzik}}]{zhang2021mutual}%
  \BibitemOpen
  \bibfield  {author} {\bibinfo {author} {\bibfnamefont {Z.-J.}\ \bibnamefont
  {Zhang}}, \bibinfo {author} {\bibfnamefont {T.~H.}\ \bibnamefont {Kyaw}},
  \bibinfo {author} {\bibfnamefont {J.}~\bibnamefont {Kottmann}}, \bibinfo
  {author} {\bibfnamefont {M.}~\bibnamefont {Degroote}},\ and\ \bibinfo
  {author} {\bibfnamefont {A.}~\bibnamefont {Aspuru-Guzik}},\ }\bibfield
  {title} {\bibinfo {title} {Mutual information-assisted adaptive variational
  quantum eigensolver},\ }\href@noop {} {\bibfield  {journal} {\bibinfo
  {journal} {Quantum Science and Technology}\ } (\bibinfo {year}
  {2021})}\BibitemShut {NoStop}%
\bibitem [{\citenamefont {Biamonte}\ \emph {et~al.}(2017)\citenamefont
  {Biamonte}, \citenamefont {Wittek}, \citenamefont {Pancotti}, \citenamefont
  {Rebentrost}, \citenamefont {Wiebe},\ and\ \citenamefont
  {Lloyd}}]{biamonte2017quantum}%
  \BibitemOpen
  \bibfield  {author} {\bibinfo {author} {\bibfnamefont {J.}~\bibnamefont
  {Biamonte}}, \bibinfo {author} {\bibfnamefont {P.}~\bibnamefont {Wittek}},
  \bibinfo {author} {\bibfnamefont {N.}~\bibnamefont {Pancotti}}, \bibinfo
  {author} {\bibfnamefont {P.}~\bibnamefont {Rebentrost}}, \bibinfo {author}
  {\bibfnamefont {N.}~\bibnamefont {Wiebe}},\ and\ \bibinfo {author}
  {\bibfnamefont {S.}~\bibnamefont {Lloyd}},\ }\bibfield  {title} {\bibinfo
  {title} {Quantum machine learning},\ }\href@noop {} {\bibfield  {journal}
  {\bibinfo  {journal} {Nature}\ }\textbf {\bibinfo {volume} {549}},\ \bibinfo
  {pages} {195} (\bibinfo {year} {2017})}\BibitemShut {NoStop}%
\bibitem [{\citenamefont {Schuld}\ and\ \citenamefont
  {Killoran}(2019)}]{schuld2019quantum}%
  \BibitemOpen
  \bibfield  {author} {\bibinfo {author} {\bibfnamefont {M.}~\bibnamefont
  {Schuld}}\ and\ \bibinfo {author} {\bibfnamefont {N.}~\bibnamefont
  {Killoran}},\ }\bibfield  {title} {\bibinfo {title} {Quantum machine learning
  in feature hilbert spaces},\ }\href@noop {} {\bibfield  {journal} {\bibinfo
  {journal} {Physical review letters}\ }\textbf {\bibinfo {volume} {122}},\
  \bibinfo {pages} {040504} (\bibinfo {year} {2019})}\BibitemShut {NoStop}%
\bibitem [{\citenamefont {Schuld}\ \emph {et~al.}(2021)\citenamefont {Schuld},
  \citenamefont {Sweke},\ and\ \citenamefont {Meyer}}]{schuld2021effect}%
  \BibitemOpen
  \bibfield  {author} {\bibinfo {author} {\bibfnamefont {M.}~\bibnamefont
  {Schuld}}, \bibinfo {author} {\bibfnamefont {R.}~\bibnamefont {Sweke}},\ and\
  \bibinfo {author} {\bibfnamefont {J.~J.}\ \bibnamefont {Meyer}},\ }\bibfield
  {title} {\bibinfo {title} {Effect of data encoding on the expressive power of
  variational quantum-machine-learning models},\ }\href@noop {} {\bibfield
  {journal} {\bibinfo  {journal} {Physical Review A}\ }\textbf {\bibinfo
  {volume} {103}},\ \bibinfo {pages} {032430} (\bibinfo {year}
  {2021})}\BibitemShut {NoStop}%
\bibitem [{\citenamefont {Goodfellow}\ \emph {et~al.}(2016)\citenamefont
  {Goodfellow}, \citenamefont {Bengio}, \citenamefont {Courville},\ and\
  \citenamefont {Bengio}}]{goodfellow2016deep}%
  \BibitemOpen
  \bibfield  {author} {\bibinfo {author} {\bibfnamefont {I.}~\bibnamefont
  {Goodfellow}}, \bibinfo {author} {\bibfnamefont {Y.}~\bibnamefont {Bengio}},
  \bibinfo {author} {\bibfnamefont {A.}~\bibnamefont {Courville}},\ and\
  \bibinfo {author} {\bibfnamefont {Y.}~\bibnamefont {Bengio}},\ }\href@noop {}
  {\emph {\bibinfo {title} {Deep learning}}},\ Vol.~\bibinfo {volume} {1}\
  (\bibinfo  {publisher} {MIT press Cambridge},\ \bibinfo {year}
  {2016})\BibitemShut {NoStop}%
\bibitem [{\citenamefont {Chatterjee}\ and\ \citenamefont
  {Yu}(2016)}]{chatterjee2016generalized}%
  \BibitemOpen
  \bibfield  {author} {\bibinfo {author} {\bibfnamefont {R.}~\bibnamefont
  {Chatterjee}}\ and\ \bibinfo {author} {\bibfnamefont {T.}~\bibnamefont
  {Yu}},\ }\bibfield  {title} {\bibinfo {title} {Generalized coherent states,
  reproducing kernels, and quantum support vector machines},\ }\href@noop {}
  {\bibfield  {journal} {\bibinfo  {journal} {arXiv:1612.03713}\ } (\bibinfo
  {year} {2016})}\BibitemShut {NoStop}%
\bibitem [{\citenamefont {Schuld}(2021)}]{schuld2021quantum}%
  \BibitemOpen
  \bibfield  {author} {\bibinfo {author} {\bibfnamefont {M.}~\bibnamefont
  {Schuld}},\ }\bibfield  {title} {\bibinfo {title} {Quantum machine learning
  models are kernel methods},\ }\href@noop {} {\bibfield  {journal} {\bibinfo
  {journal} {arXiv:2101.11020}\ } (\bibinfo {year} {2021})}\BibitemShut
  {NoStop}%
\bibitem [{\citenamefont {Otten}\ \emph {et~al.}(2020)\citenamefont {Otten},
  \citenamefont {Goumiri}, \citenamefont {Priest}, \citenamefont {Chapline},\
  and\ \citenamefont {Schneider}}]{otten2020quantum}%
  \BibitemOpen
  \bibfield  {author} {\bibinfo {author} {\bibfnamefont {M.}~\bibnamefont
  {Otten}}, \bibinfo {author} {\bibfnamefont {I.~R.}\ \bibnamefont {Goumiri}},
  \bibinfo {author} {\bibfnamefont {B.~W.}\ \bibnamefont {Priest}}, \bibinfo
  {author} {\bibfnamefont {G.~F.}\ \bibnamefont {Chapline}},\ and\ \bibinfo
  {author} {\bibfnamefont {M.~D.}\ \bibnamefont {Schneider}},\ }\bibfield
  {title} {\bibinfo {title} {Quantum machine learning using gaussian processes
  with performant quantum kernels},\ }\href@noop {} {\bibfield  {journal}
  {\bibinfo  {journal} {arXiv preprint arXiv:2004.11280}\ } (\bibinfo {year}
  {2020})}\BibitemShut {NoStop}%
\bibitem [{\citenamefont {Huang}\ \emph
  {et~al.}(2020{\natexlab{b}})\citenamefont {Huang}, \citenamefont {Broughton},
  \citenamefont {Mohseni}, \citenamefont {Babbush}, \citenamefont {Boixo},
  \citenamefont {Neven},\ and\ \citenamefont {McClean}}]{huang2020power}%
  \BibitemOpen
  \bibfield  {author} {\bibinfo {author} {\bibfnamefont {H.-Y.}\ \bibnamefont
  {Huang}}, \bibinfo {author} {\bibfnamefont {M.}~\bibnamefont {Broughton}},
  \bibinfo {author} {\bibfnamefont {M.}~\bibnamefont {Mohseni}}, \bibinfo
  {author} {\bibfnamefont {R.}~\bibnamefont {Babbush}}, \bibinfo {author}
  {\bibfnamefont {S.}~\bibnamefont {Boixo}}, \bibinfo {author} {\bibfnamefont
  {H.}~\bibnamefont {Neven}},\ and\ \bibinfo {author} {\bibfnamefont {J.~R.}\
  \bibnamefont {McClean}},\ }\bibfield  {title} {\bibinfo {title} {Power of
  data in quantum machine learning},\ }\href@noop {} {\bibfield  {journal}
  {\bibinfo  {journal} {arXiv:2011.01938}\ } (\bibinfo {year}
  {2020}{\natexlab{b}})}\BibitemShut {NoStop}%
\bibitem [{\citenamefont {Havl{\'\i}{\v{c}}ek}\ \emph
  {et~al.}(2019)\citenamefont {Havl{\'\i}{\v{c}}ek}, \citenamefont
  {C{\'o}rcoles}, \citenamefont {Temme}, \citenamefont {Harrow}, \citenamefont
  {Kandala}, \citenamefont {Chow},\ and\ \citenamefont
  {Gambetta}}]{havlivcek2019supervised}%
  \BibitemOpen
  \bibfield  {author} {\bibinfo {author} {\bibfnamefont {V.}~\bibnamefont
  {Havl{\'\i}{\v{c}}ek}}, \bibinfo {author} {\bibfnamefont {A.~D.}\
  \bibnamefont {C{\'o}rcoles}}, \bibinfo {author} {\bibfnamefont
  {K.}~\bibnamefont {Temme}}, \bibinfo {author} {\bibfnamefont {A.~W.}\
  \bibnamefont {Harrow}}, \bibinfo {author} {\bibfnamefont {A.}~\bibnamefont
  {Kandala}}, \bibinfo {author} {\bibfnamefont {J.~M.}\ \bibnamefont {Chow}},\
  and\ \bibinfo {author} {\bibfnamefont {J.~M.}\ \bibnamefont {Gambetta}},\
  }\bibfield  {title} {\bibinfo {title} {Supervised learning with
  quantum-enhanced feature spaces},\ }\href@noop {} {\bibfield  {journal}
  {\bibinfo  {journal} {Nature}\ }\textbf {\bibinfo {volume} {567}},\ \bibinfo
  {pages} {209} (\bibinfo {year} {2019})}\BibitemShut {NoStop}%
\bibitem [{\citenamefont {Blank}\ \emph {et~al.}(2020)\citenamefont {Blank},
  \citenamefont {Park}, \citenamefont {Rhee},\ and\ \citenamefont
  {Petruccione}}]{blank2020quantum}%
  \BibitemOpen
  \bibfield  {author} {\bibinfo {author} {\bibfnamefont {C.}~\bibnamefont
  {Blank}}, \bibinfo {author} {\bibfnamefont {D.~K.}\ \bibnamefont {Park}},
  \bibinfo {author} {\bibfnamefont {J.-K.~K.}\ \bibnamefont {Rhee}},\ and\
  \bibinfo {author} {\bibfnamefont {F.}~\bibnamefont {Petruccione}},\
  }\bibfield  {title} {\bibinfo {title} {Quantum classifier with tailored
  quantum kernel},\ }\href@noop {} {\bibfield  {journal} {\bibinfo  {journal}
  {npj Quantum Information}\ }\textbf {\bibinfo {volume} {6}},\ \bibinfo
  {pages} {1} (\bibinfo {year} {2020})}\BibitemShut {NoStop}%
\bibitem [{\citenamefont {Rasmussen}\ and\ \citenamefont
  {Williams}(2005)}]{rasmussen2005gaussian}%
  \BibitemOpen
  \bibfield  {author} {\bibinfo {author} {\bibfnamefont {C.~E.}\ \bibnamefont
  {Rasmussen}}\ and\ \bibinfo {author} {\bibfnamefont {C.~K.~I.}\ \bibnamefont
  {Williams}},\ }\href@noop {} {\emph {\bibinfo {title} {Gaussian Processes for
  Machine Learning (Adaptive Computation and Machine Learning)}}}\ (\bibinfo
  {publisher} {The MIT Press},\ \bibinfo {year} {2005})\BibitemShut {NoStop}%
\bibitem [{\citenamefont {Haug}()}]{haug2021optimalquantumlearning}%
  \BibitemOpen
  \bibfield  {author} {\bibinfo {author} {\bibfnamefont {T.}~\bibnamefont
  {Haug}},\ }\href@noop {} {\bibinfo {title} {Optimal quantum learning}},\
  \bibinfo {howpublished}
  {\url{https://github.com/txhaug/optimal-quantum-learning}}\BibitemShut
  {NoStop}%
\end{thebibliography}%

\appendix 

\section{Parameterized quantum circuits}\label{app:pqc}
\begin{figure*}[htbp]
	\centering
	\includegraphics[width=0.8\textwidth]{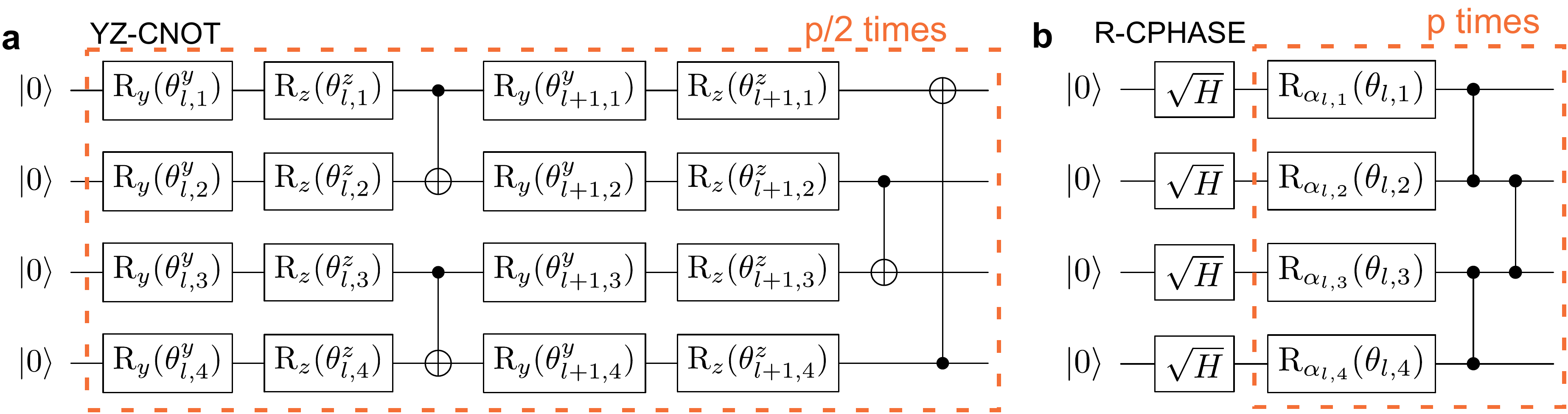}
	\caption{Description of PQCs used in the main text.
	\idg{a} YZ-CNOT consists of $p$ layers, with each layer having Y and Z rotations applied on each qubit, and CNOT gates between neighboring qubits applied in an alternating fashion. For odd layer number $l$, CNOT gate is applied between qubit $2n-1$ and $2n$, for even $l$ CNOT is applied on $2n$ and $2n+1$. For YZ-$\sqrt{i\text{SWAP}}$, the CNOT gates are replaced with $\sqrt{i\text{SWAP}}$ gates. 
	\idg{b} R-CPHASE PQC, which consists of an initial layer of $\pi/2$ rotation around the $y$-axis, followed by $p$ layers. Each layer consists of randomly chosen rotations with $\sigma^x$, $\sigma^y$ or $\sigma^z$, followed by CPHASE gates applied as a chain of nearest-neighbors.
	}
	\label{fig:circuits}
\end{figure*}

The PQCs used in the main text are described in Fig.\ref{fig:circuits}. The PQCs are given by $\psi(\boldsymbol{\theta})=U(\theta)\ket{0}^{\otimes N}=\prod_{l=p}^1 \left[  W_l V_l(\theta_l)\right]\ket{0}^{\otimes N}$, which consist of $p$ layers of entangling gates $W_l$ and parameterized rotations $V_l(\theta_l)$. The parameters $\boldsymbol{\theta}$ are chosen randomly. 

\section{Further data on variance of gradient}\label{app:data_gradient}
\begin{figure}[htbp]
	\centering	
	\subfigimg[width=0.28\textwidth]{}{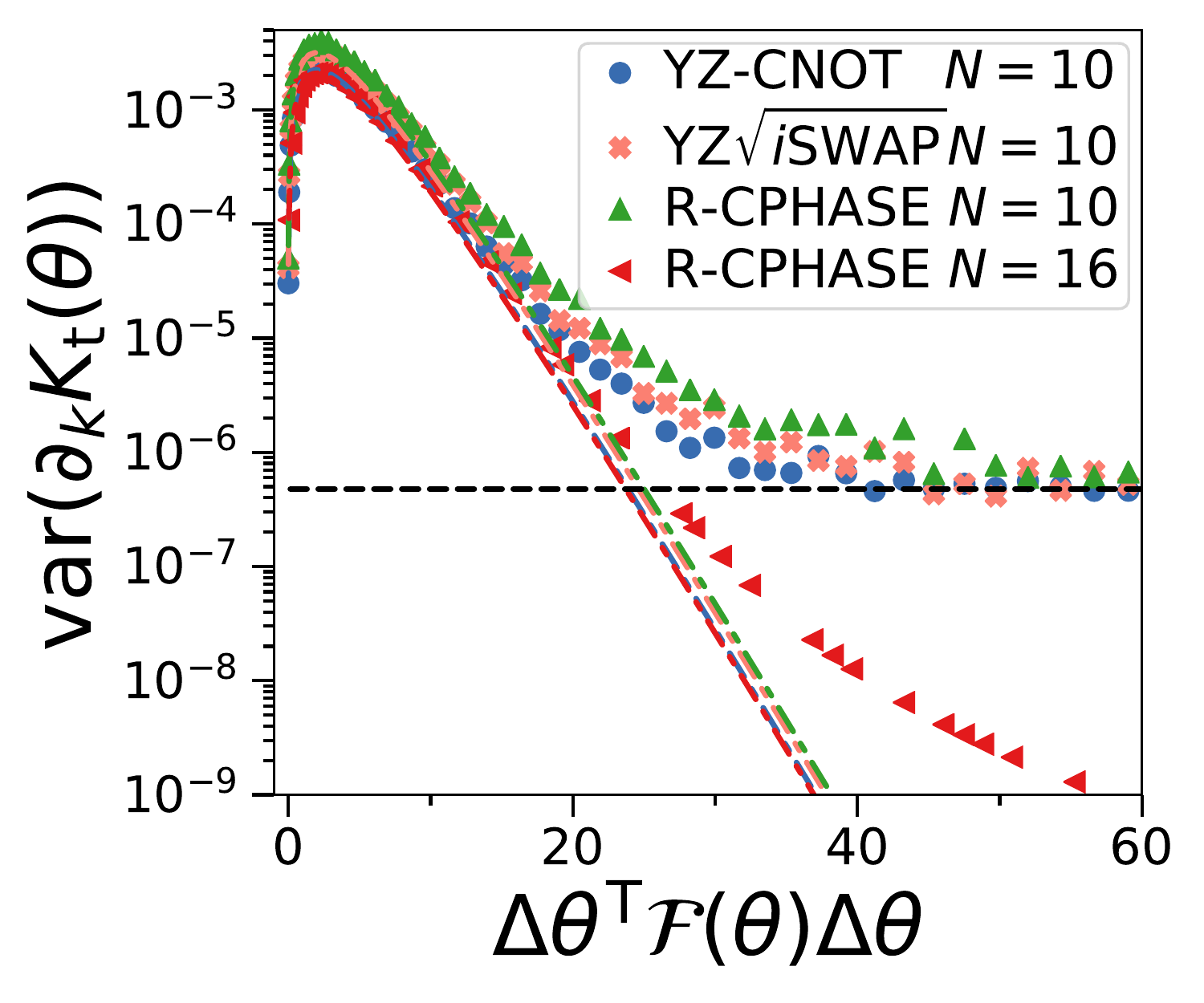}
	\caption{ Variance of gradient $\text{var}(\partial_k K_\text{t}(\boldsymbol{\theta}))$ against parameter norm $\Delta\boldsymbol{\theta}^\text{T}\mathcal{F}(\boldsymbol{\theta})\Delta\boldsymbol{\theta}$ for different types of PQCs. Dashed line is the analytic formula~\eqref{eq:var_grad_sup} for the variance. The dashed horizontal line is the variance of the gradient for states of a random PQC $\text{var}(\nabla K_\text{t}(\boldsymbol{\theta}_\text{rand}))=\frac{1}{2^{2N+1}}$~\cite{mcclean2018barren}.
    Number of layers $p=20$ for R-CPHASE $N=10$, $p=16$ for $N=16$, else $p=10$. Average over 50 random instances of $\boldsymbol{\theta}_\text{t}$.
	}
	\label{fig:vargrad_sup}
\end{figure}

In Fig.\ref{fig:vargrad_sup}, we show the variance of the gradient as function of parameter norm $\boldsymbol{\theta}^\text{T}\mathcal{F}\boldsymbol{\theta}$. We see a good match with the analytic formula. For large $\boldsymbol{\theta}^\text{T}\mathcal{F}\boldsymbol{\theta}$, we see deviation from the analytic result, as the variance of the gradient approaches the value given by a random state sampled from a sufficiently deep PQC with $\text{var}(\nabla K_\text{t}(\boldsymbol{\theta}_\text{rand}))=\frac{1}{2^{2N+1}}$~\cite{mcclean2018barren}.

\section{Further data on adaptive learning rate}\label{app:data_learning}
\begin{figure*}[htbp]
	\centering	
	\subfigimg[width=0.24\textwidth]{a}{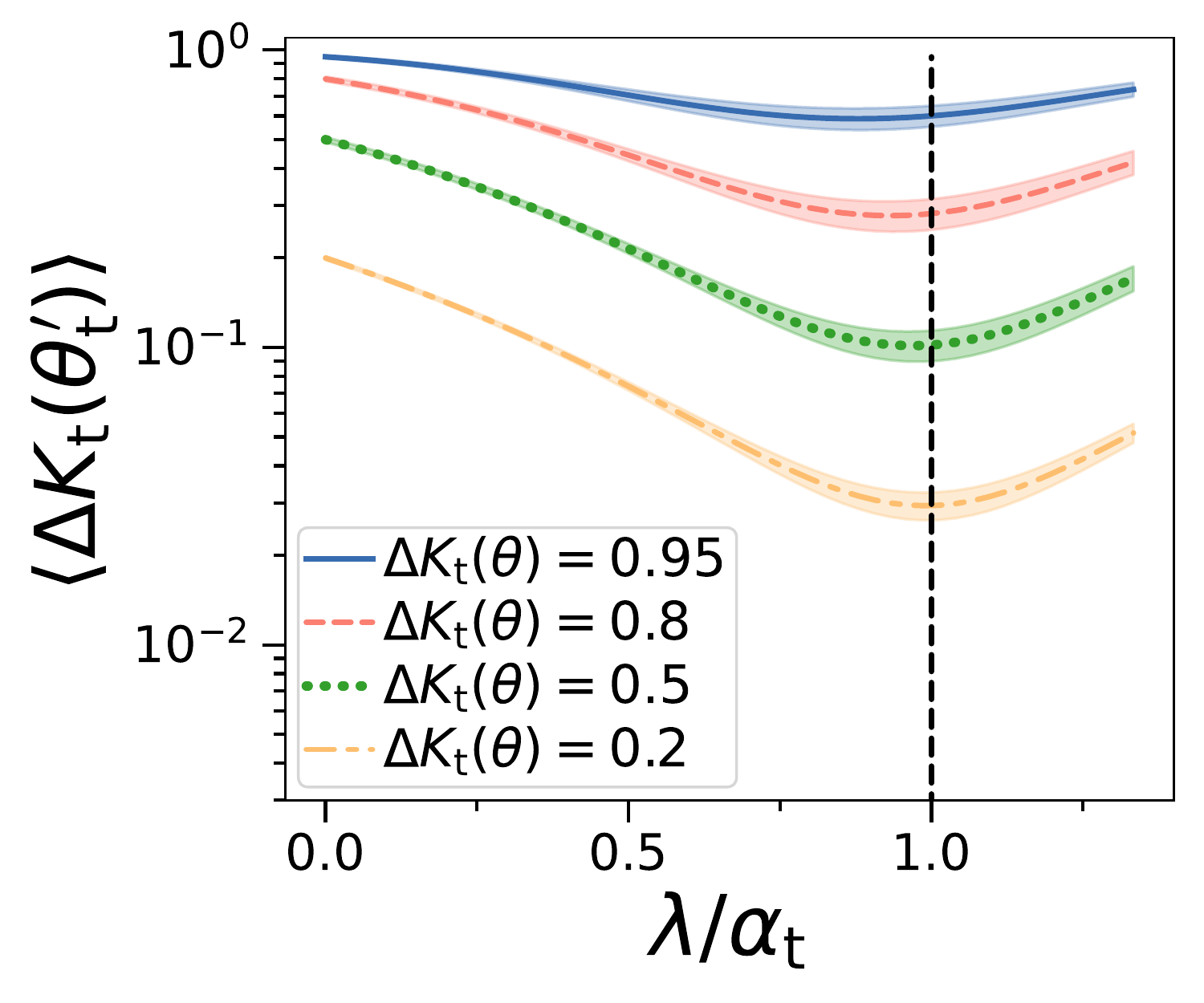}
	\subfigimg[width=0.24\textwidth]{b}{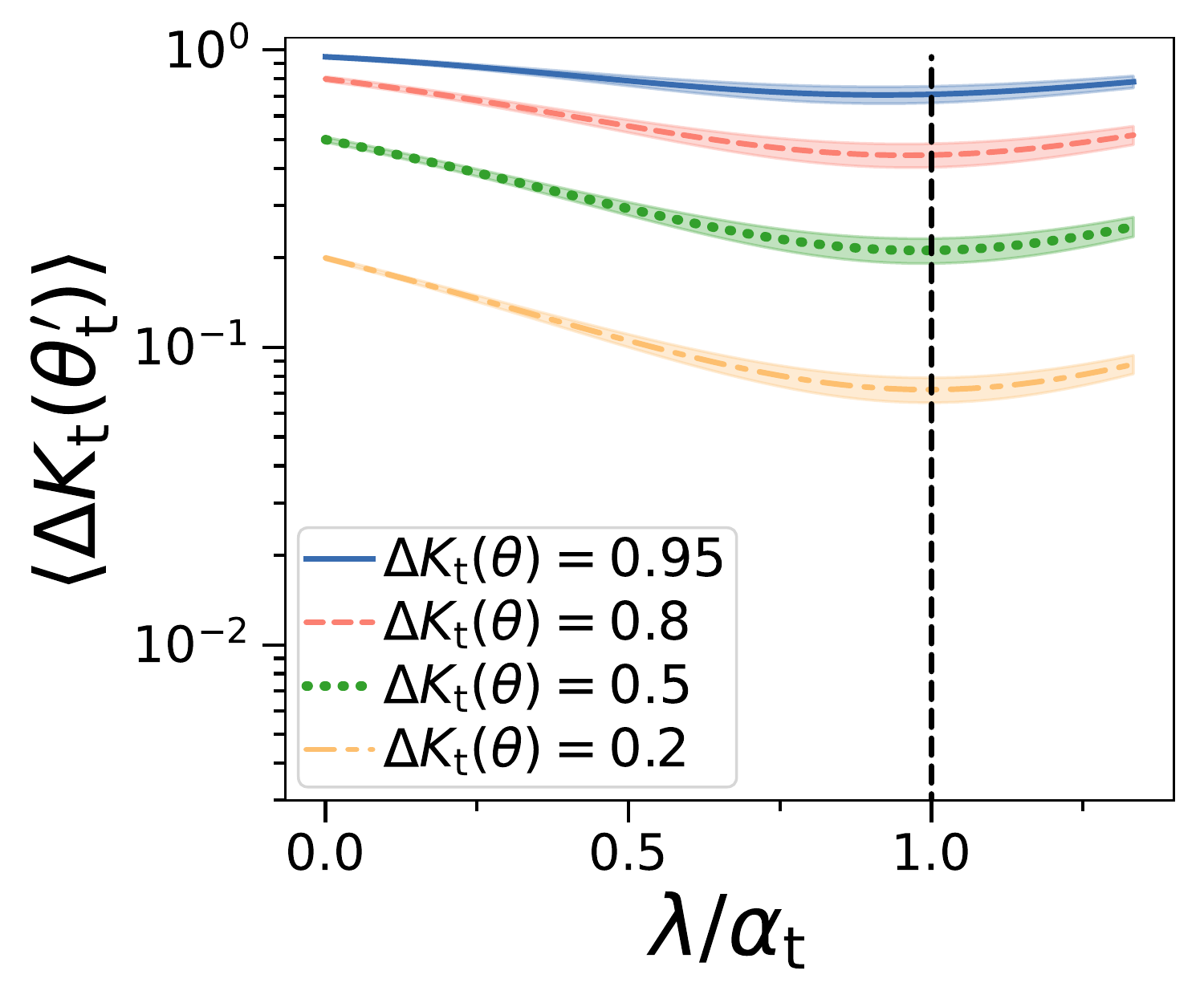}
	\subfigimg[width=0.24\textwidth]{c}{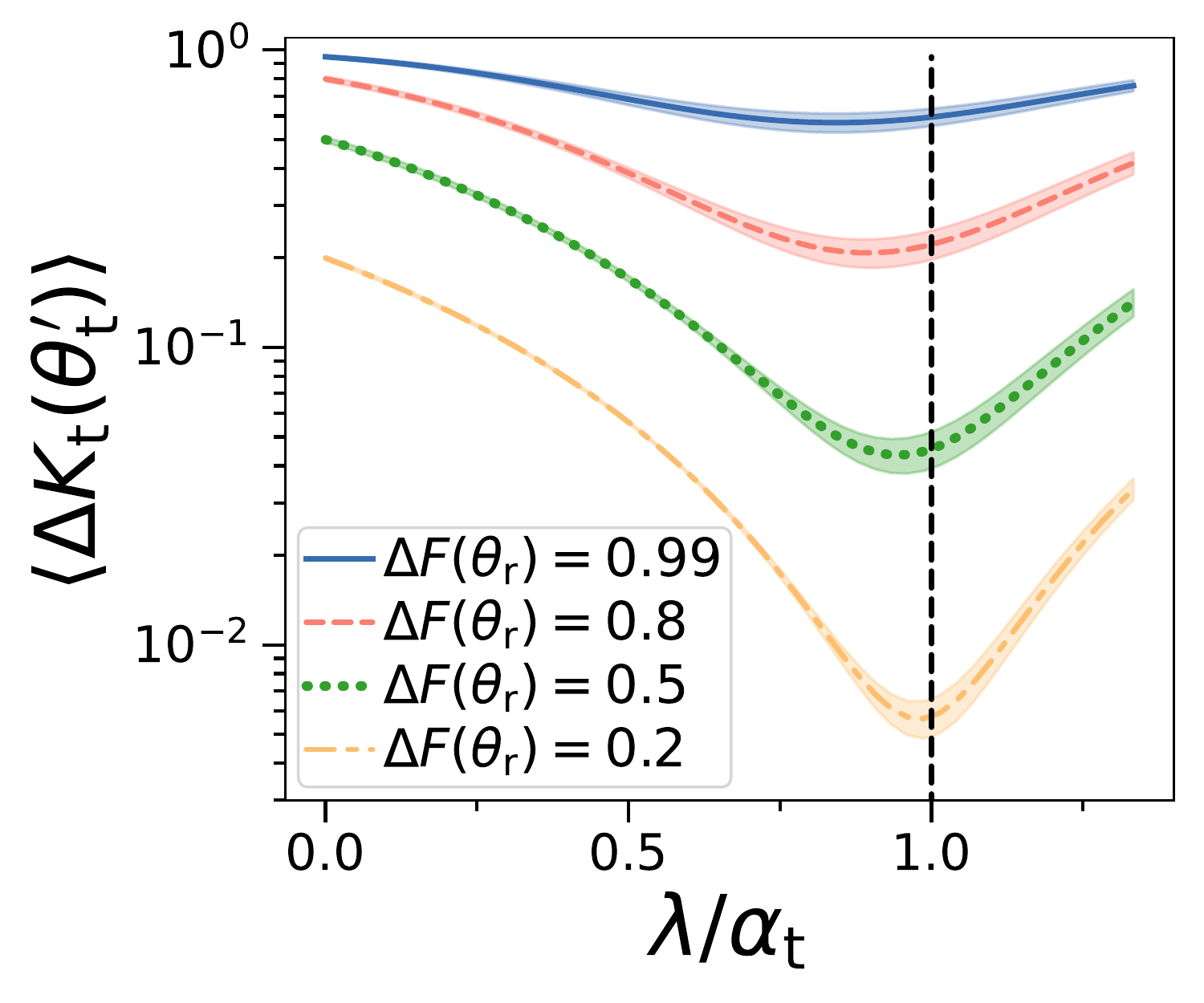}
	\caption{Average infidelity after one iteration of gradient ascent $\langle\Delta K_\text{t}(\boldsymbol{\theta}_\text{t}')\rangle$  plotted against learning rate $\lambda$, for gradient ascent update $\boldsymbol{\theta}_\text{t}'=\boldsymbol{\theta}+\lambda G_\beta(\boldsymbol{\theta})$. $\lambda$ is normalized in respect to analytically calculated learning rate $\alpha_\text{t}$ (\eqref{eq:update_add_sup}), shown as vertical dashed line. Curves show various initial infidelities $\Delta K_\text{t}(\boldsymbol{\theta})$, with shaded area being the standard deviation of $\Delta K_\text{t}(\boldsymbol{\theta}_\text{t}')$. Infidelity averaged over 50 random instances of YZ-CNOT PQC. 
	We show \idg{a} the GQNG $\beta=\frac{1}{2}$, \idg{b} the regular gradient $\beta=0$ and \idg{c} QNG $\beta=1$ with regularization $\epsilon_\text{R}=10^{-1}$.
	}
	\label{fig:scale_comp_GQNG}
\end{figure*}

Here, we show further results on the adaptive learning rate.
We compare the adaptive gradient ascent using the GQNG, regular gradient and QNG in Fig.\ref{fig:scale_comp_GQNG}. We find for all types of gradients, the adaptive gradient ascent finds the nearly optimal learning rate.

\begin{figure*}[htbp]
	\centering	
	\subfigimg[width=0.24\textwidth]{a}{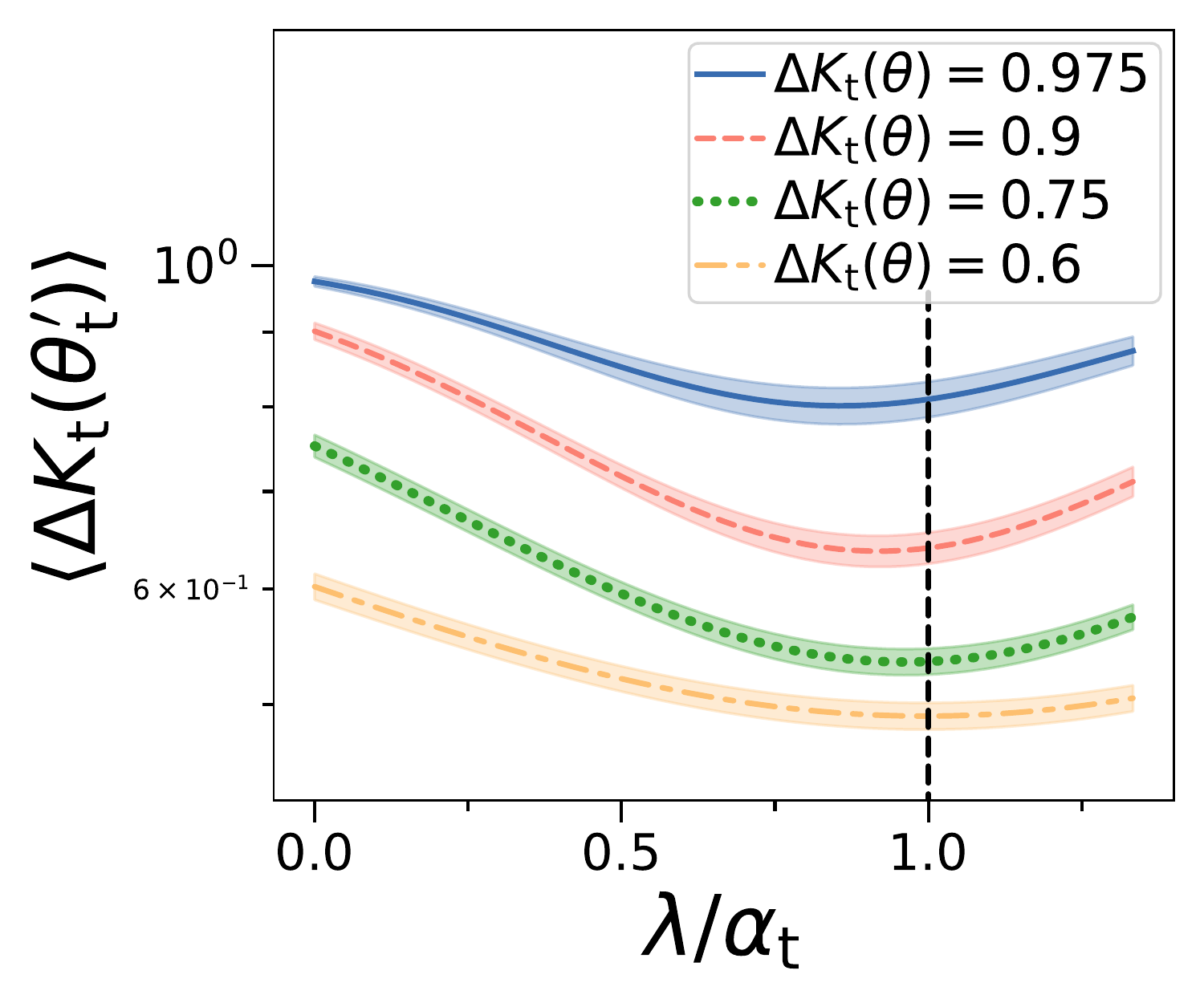}
	\subfigimg[width=0.24\textwidth]{b}{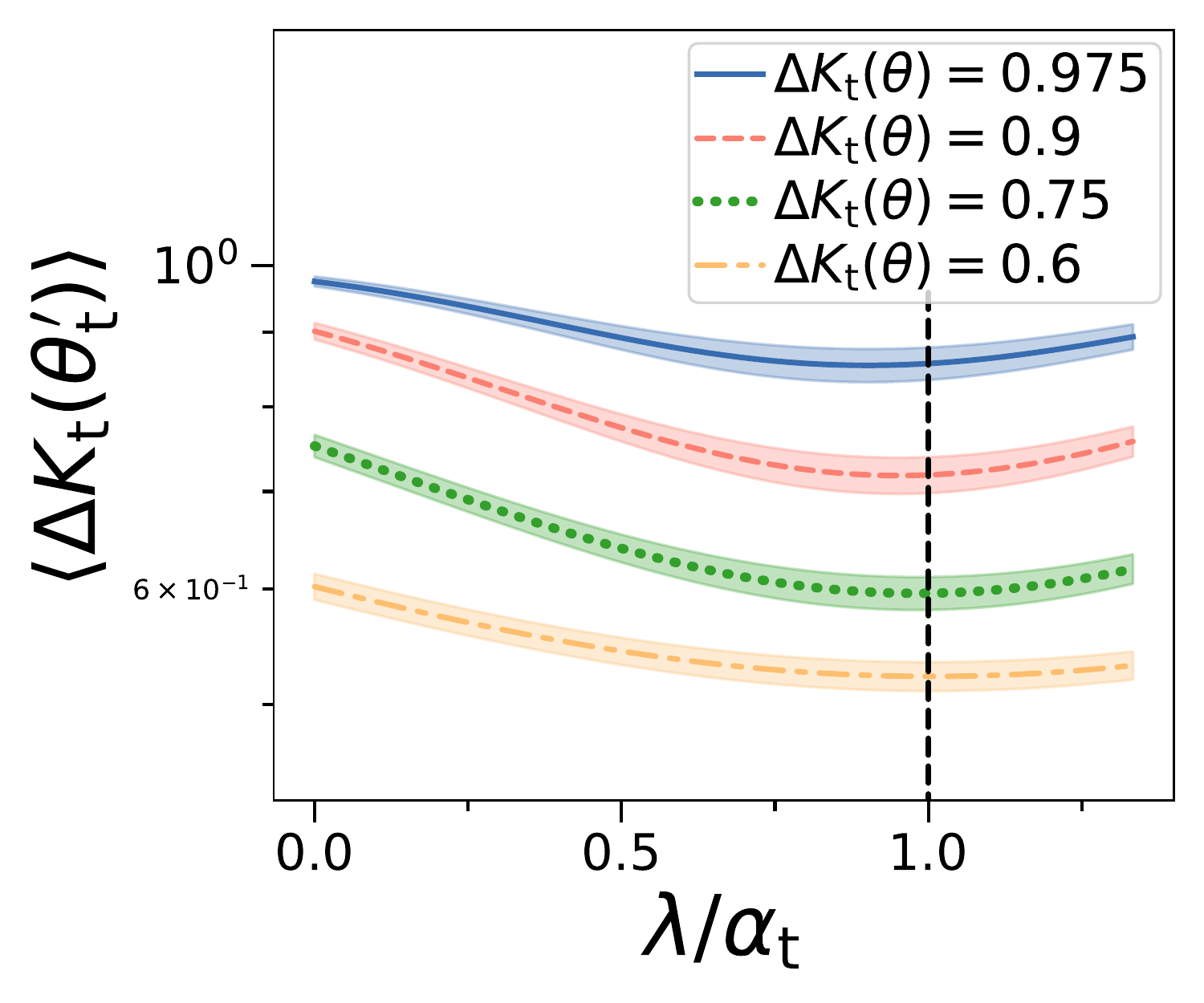}
	\subfigimg[width=0.24\textwidth]{c}{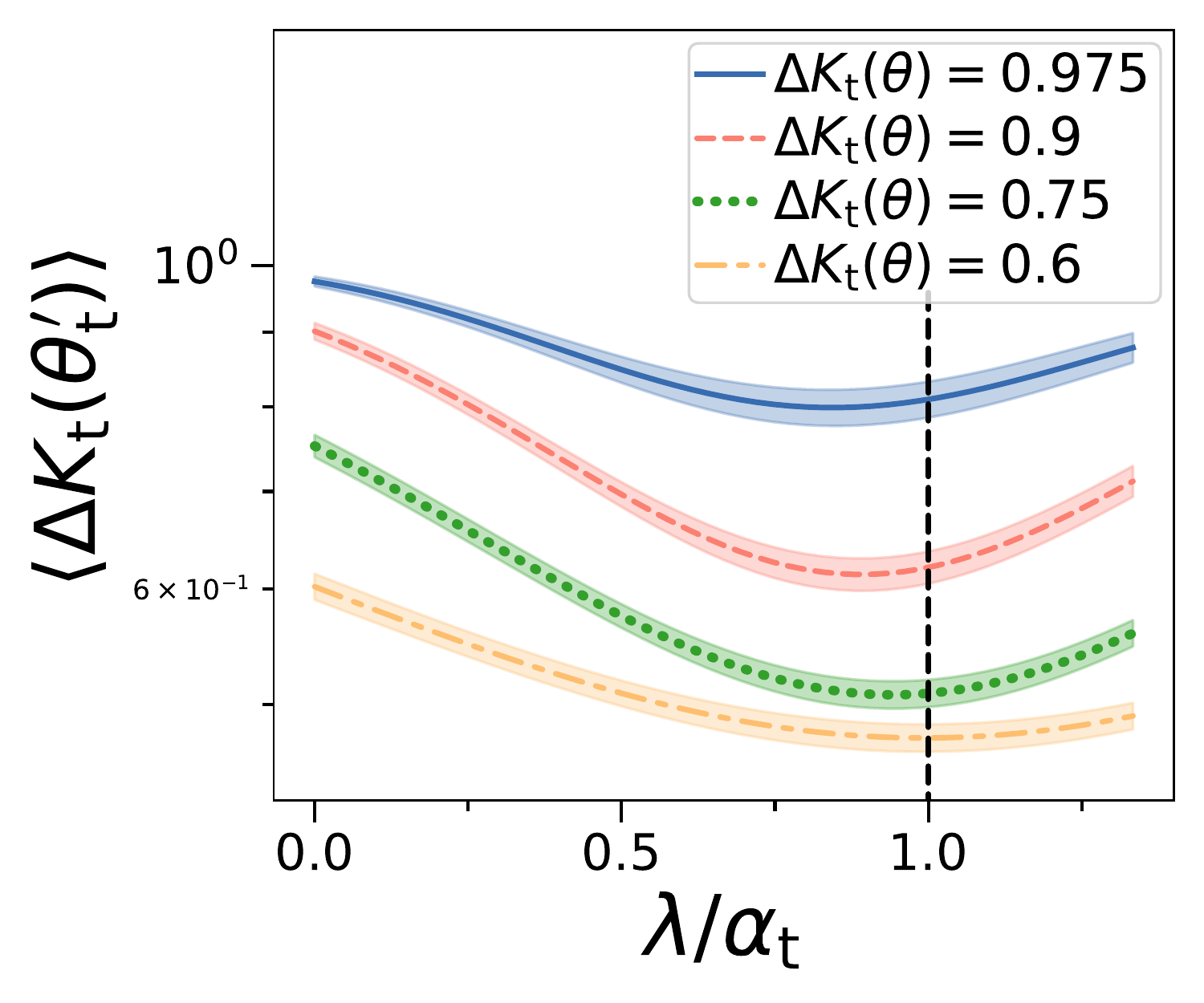}
	\caption{Target state to be learned cannot be perfectly represented by PQC, i.e. $K_0=\text{max}_{\boldsymbol{\theta}}K_\text{t}(\boldsymbol{\theta})\approx0.5$.
	Average infidelity after one iteration of gradient ascent $\langle\Delta K_\text{t}(\boldsymbol{\theta}_\text{t}')\rangle$ plotted against learning rate $\lambda$, for gradient ascent update $\boldsymbol{\theta}_\text{t}'=\boldsymbol{\theta}+\lambda G_\beta(\boldsymbol{\theta})$. $\lambda$ is normalized in respect to analytically calculated learning rate $\alpha_\text{t}$, shown as vertical dashed line. Curves show various initial infidelities $\Delta K_\text{t}(\boldsymbol{\theta})$, with shaded area being the standard deviation of $\Delta K_\text{t}(\boldsymbol{\theta}_\text{t}')$. Infidelity averaged over 50 random instances of YZ-CNOT PQC. 
	We show \idg{a} the GQNG $\beta=\frac{1}{2}$, \idg{b} the regular gradient $\beta=0$ and \idg{c} QNG $\beta=1$ with regularization $\epsilon_\text{R}=10^{-1}$.
	}
	\label{fig:scale_comp_GQNG_stateshift}
\end{figure*}

Now, we choose a target state that cannot be perfectly represented by the PQC, i.e. $K_0=\text{max}_{\boldsymbol{\theta}}K_\text{t}(\boldsymbol{\theta}<1$. In Fig.\ref{fig:scale_comp_GQNG_stateshift}, we investigate the learning rate $\lambda$ normalized to the optimal learning rate $\alpha_\text{t}$ and choose $K_0=0.5$. We find for all types of gradients, adaptive gradient ascent finds the nearly optimal learning rate.

\section{Further data on training VQAs}\label{app:vqa_data}
\begin{figure}[htbp]
	\centering
	\subfigimg[width=0.24\textwidth]{a}{trainStdQutipEvalEvalQGN10d10e0r50i1r2c4a2n0s0_02t11H3U0S1n20o7g1b1q0r0_1a0_9s0C12.pdf}\hfill
	\subfigimg[width=0.24\textwidth]{b}{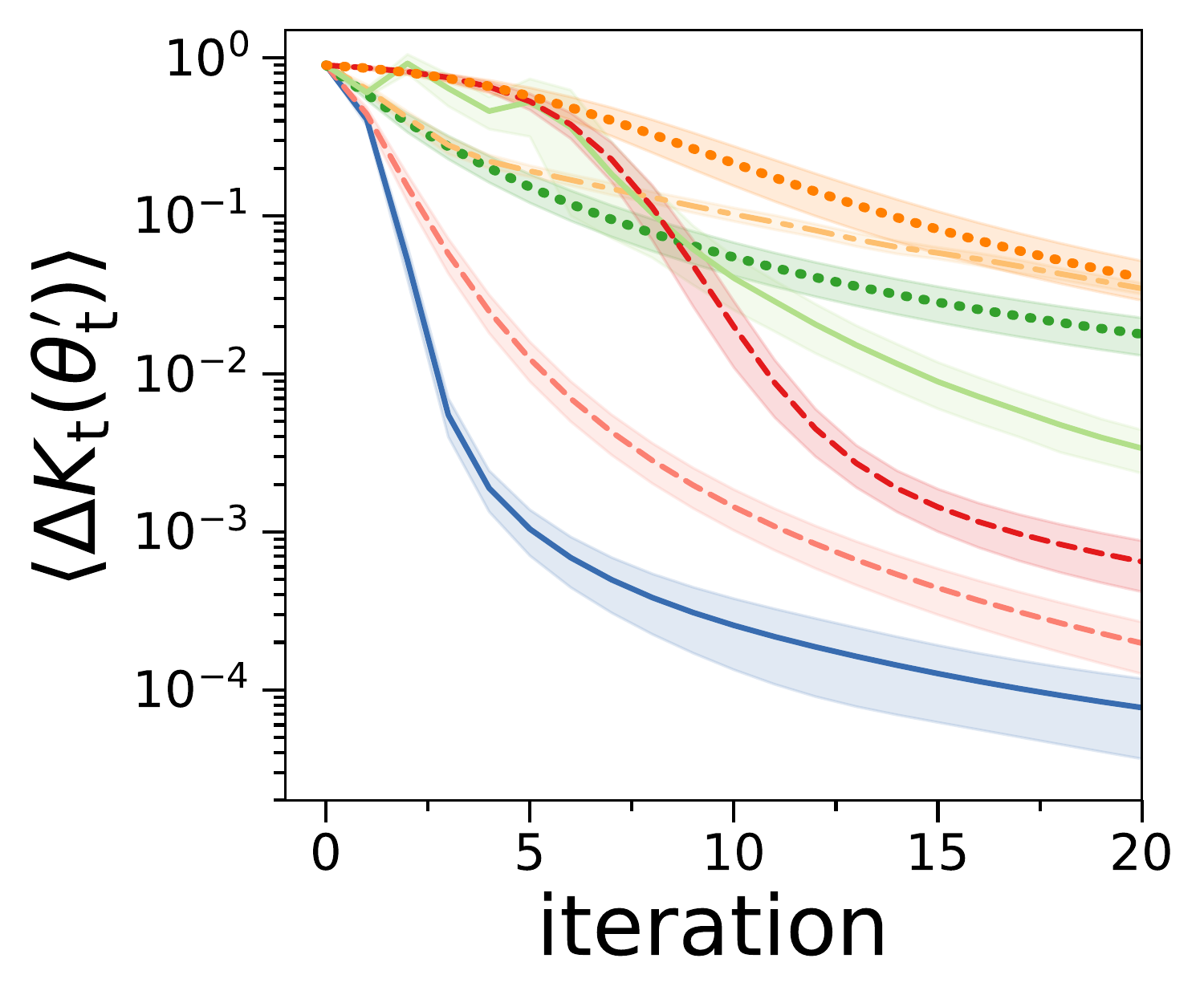}
	\caption{Training with YZ-CNOT PQC.
	\idg{a} Linear Plot \idg{b} Logarithmic plot. 
	Average infidelity $\langle \Delta K_\text{t}(\boldsymbol{\theta}_\text{t}') \rangle$ against number of iterations of gradient ascent. Shaded area is the standard deviation over 50 instances of training. We compare different optimization methods against each other. We find that adaptive gradient ascent with QNG (A-QNG, $\beta=1$, $\epsilon_\text{R}=10^{-1}$) or GQNG (A-GQNG, $\beta=\frac{1}{2}$, $\epsilon_\text{R}=0$) performs best by a large margin, followed by adaptive method with regular gradient (A-G). Standard optimization methods such as Adam and LBFGS perform comparable to A-G. Initial infidelity is $\Delta K_\text{t}(\boldsymbol{\theta})=0.9$, $N=10$ and $p=10$.
	}
	\label{fig:training_sup}
\end{figure}

In this section we show further data for the training of VQAs.
In Fig.\ref{fig:training_sup} we show both linear and logarithmic plots for training the YZ-CNOT PQC. We see that QNG and GQNG with adaptive learning rates provides about one order of magnitude smaller infidelities compared to other methods.

\begin{figure}[htbp]
	\centering
	\subfigimg[width=0.24\textwidth]{a}{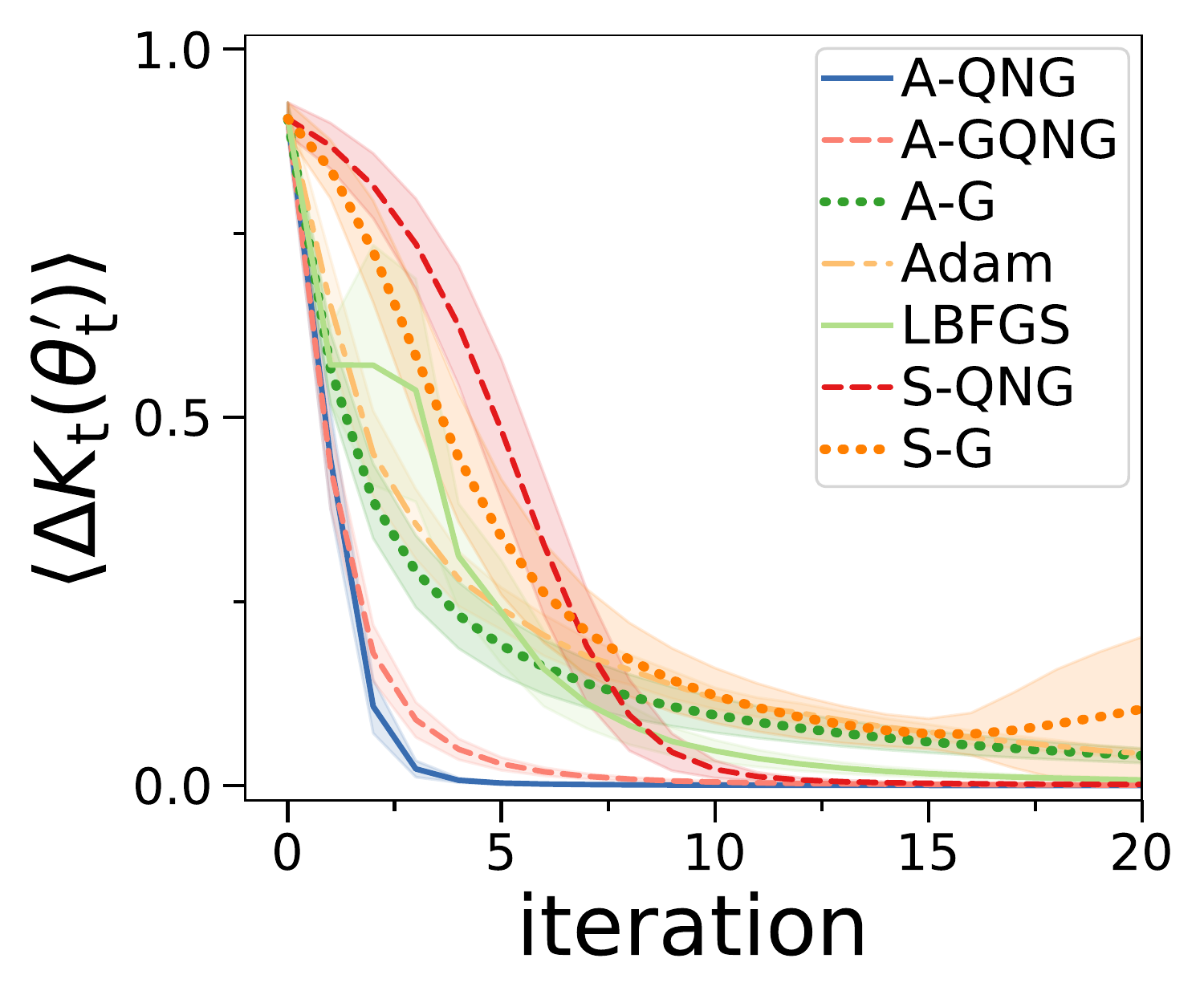}\hfill
	\subfigimg[width=0.24\textwidth]{b}{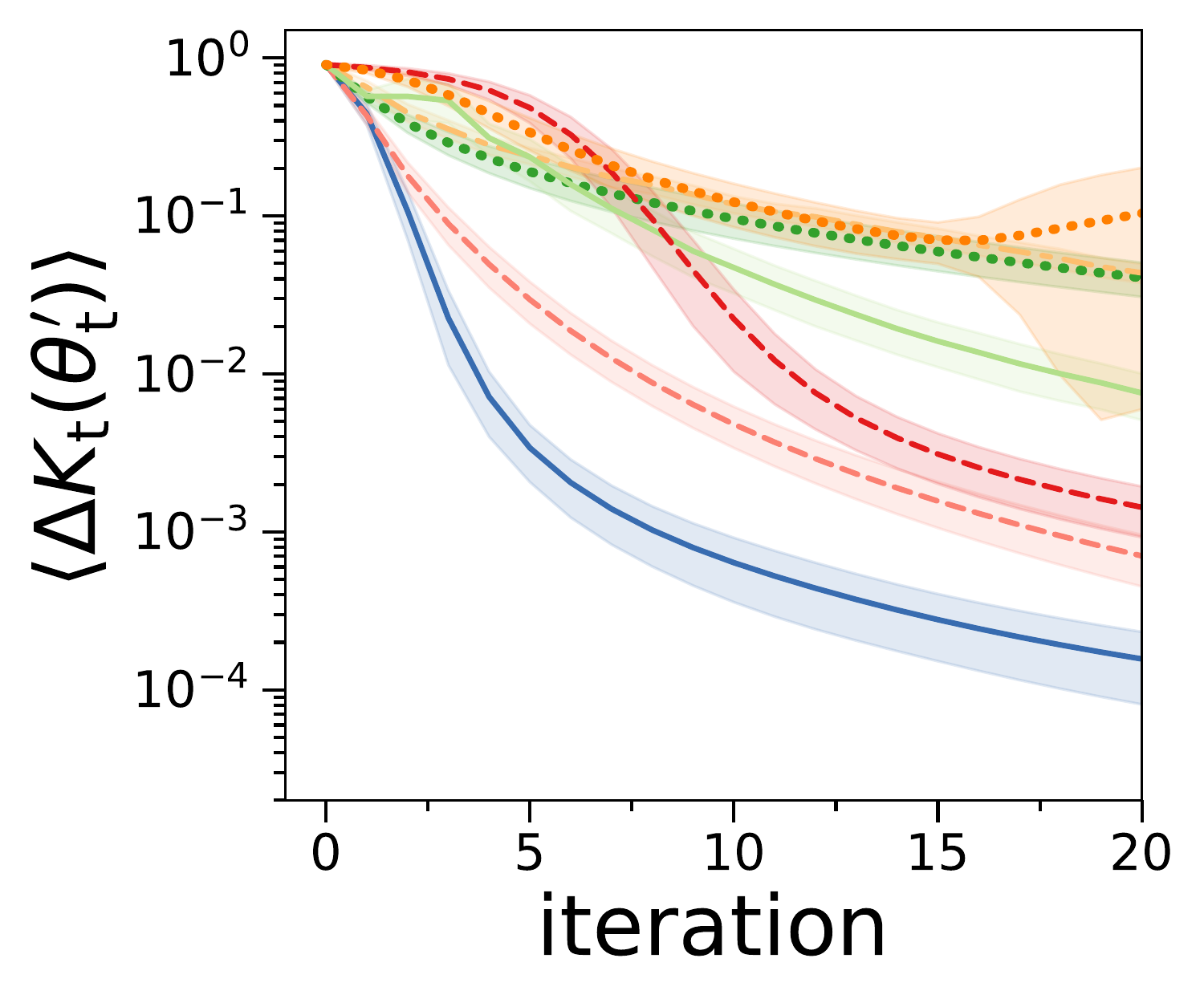}
	\caption{Training with R-CPHASE PQC.
	\idg{a} Linear Plot \idg{b} Logarithmic plot. 
	Average infidelity $\langle \Delta K_\text{t}(\boldsymbol{\theta}_\text{t}') \rangle$ against number of iterations of gradient ascent. 
	Initial infidelity is $\Delta K_\text{t}(\boldsymbol{\theta})=0.9$,  $N=10$ and $p=20$.
	}
	\label{fig:training_RCPHASE_sup}
\end{figure}

We show further training of our VQA in Fig.\ref{fig:training_RCPHASE_sup} for for the R-CPHASE PQC. We find similar trajectories as before, however in general the infidelities are a bit higher. This may be related to the fact that the R-CPHASE PQC has a QFIM with more smaller eigenvalues compared to the YZ-CNOT PQC. This may affect training adversely here.

\begin{figure}[htbp]
	\centering
	\subfigimg[width=0.24\textwidth]{a}{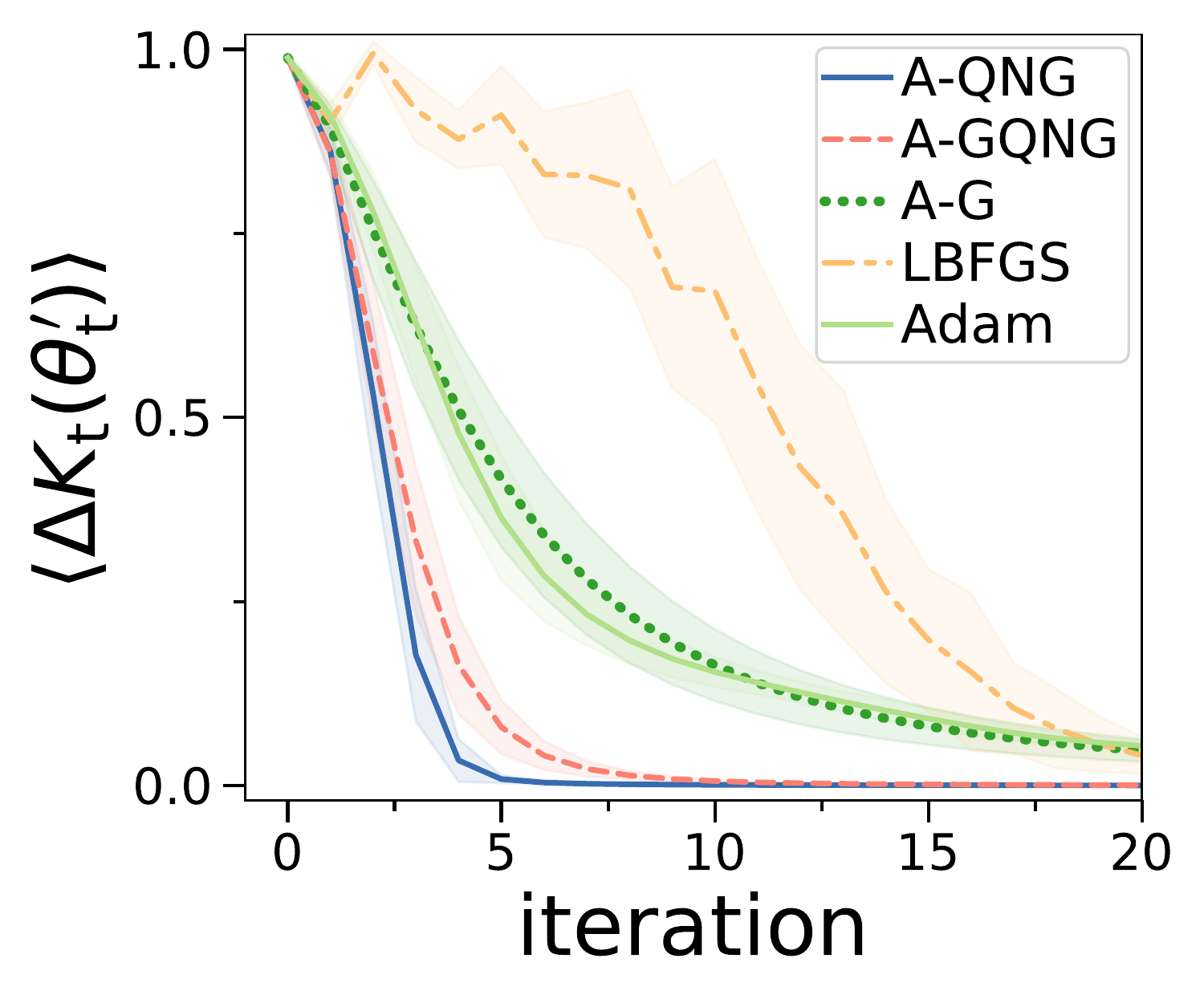}\hfill
	\subfigimg[width=0.24\textwidth]{b}{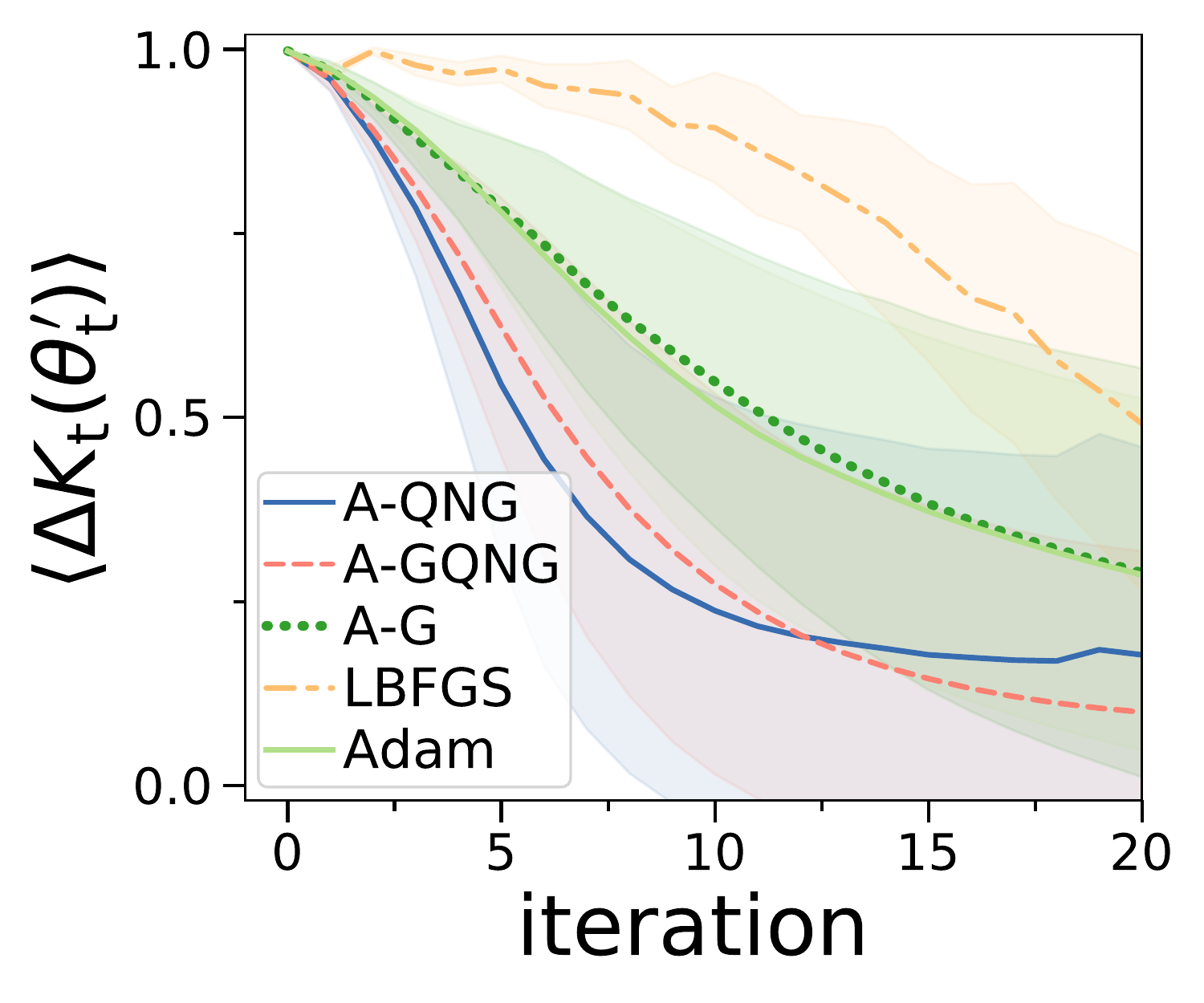}
	\caption{Average infidelity $\langle \Delta K_\text{t}(\boldsymbol{\theta}_\text{t}') \rangle$ against number of iterations of gradient ascent. Shaded area is the standard deviation over 50 instances of training. We compare QNG, GQNG and standard gradient with adaptive learning rate against Adam and LBFGS.
	\idg{a} Initial infidelity $\Delta K_\text{t}(\boldsymbol{\theta})=0.99$ \idg{b} $\Delta K_\text{t}(\boldsymbol{\theta})=0.999$.
    The PQC used is YZ-CNOT with $N=10$ and $p=10$. The regularization for A-QNG is $\epsilon_\text{R}=10^{-1}$.
	}
	\label{fig:training_comp_methods_sup}
\end{figure}

In Fig.\ref{fig:training_comp_methods_sup}, we compare QNG (A-QNG), GQNG (A-GQNG) and standard gradient (A-G) with adaptive learning rate against Adam and LBFGS for different values of initial infidelity. We find superior performance of A-QNG and A-GQNG for higher infidelities as well. A-G performs comparable to Adam.

\begin{figure}[htbp]
	\centering
	\subfigimg[width=0.35\textwidth]{a}{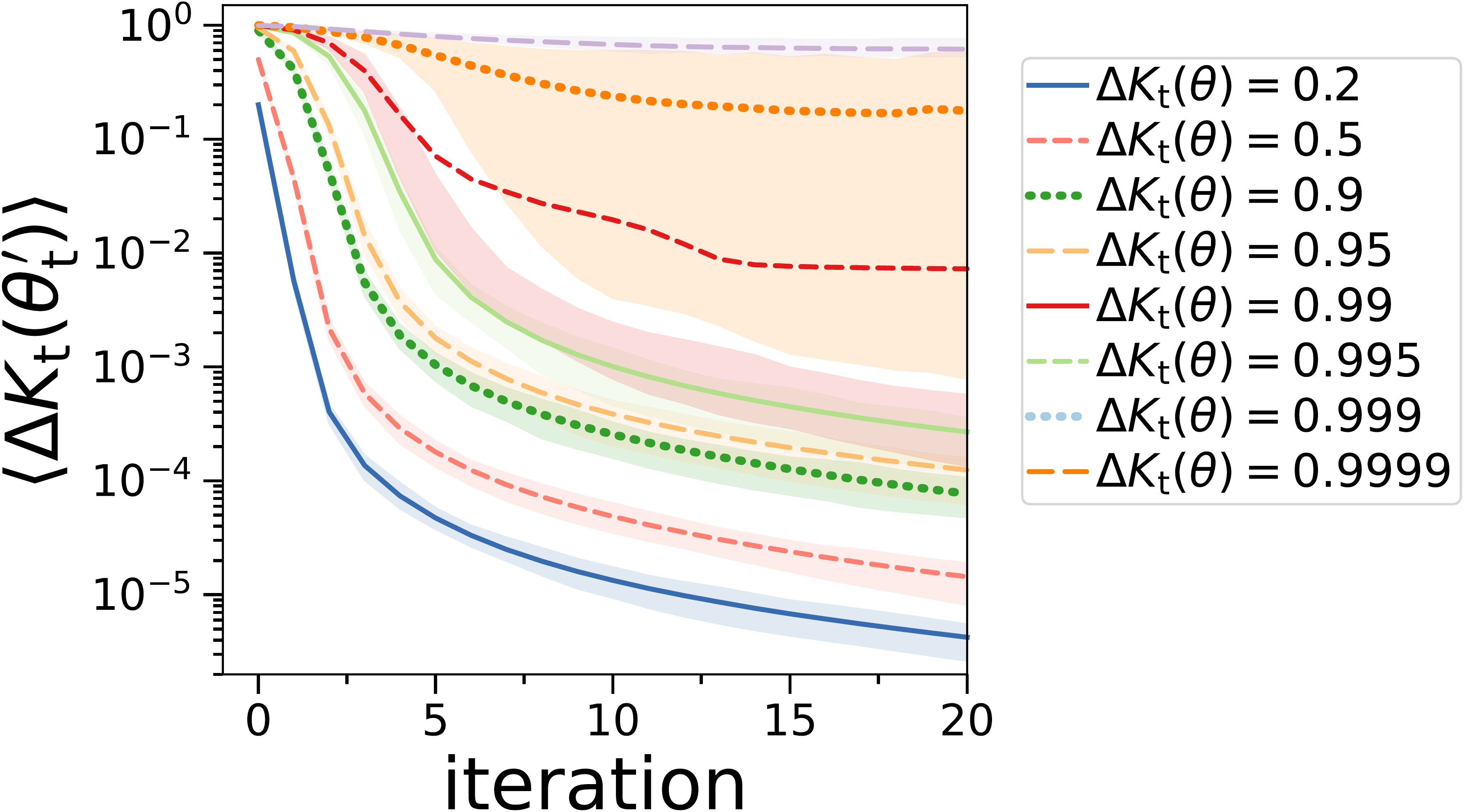}
	\subfigimg[width=0.35\textwidth]{b}{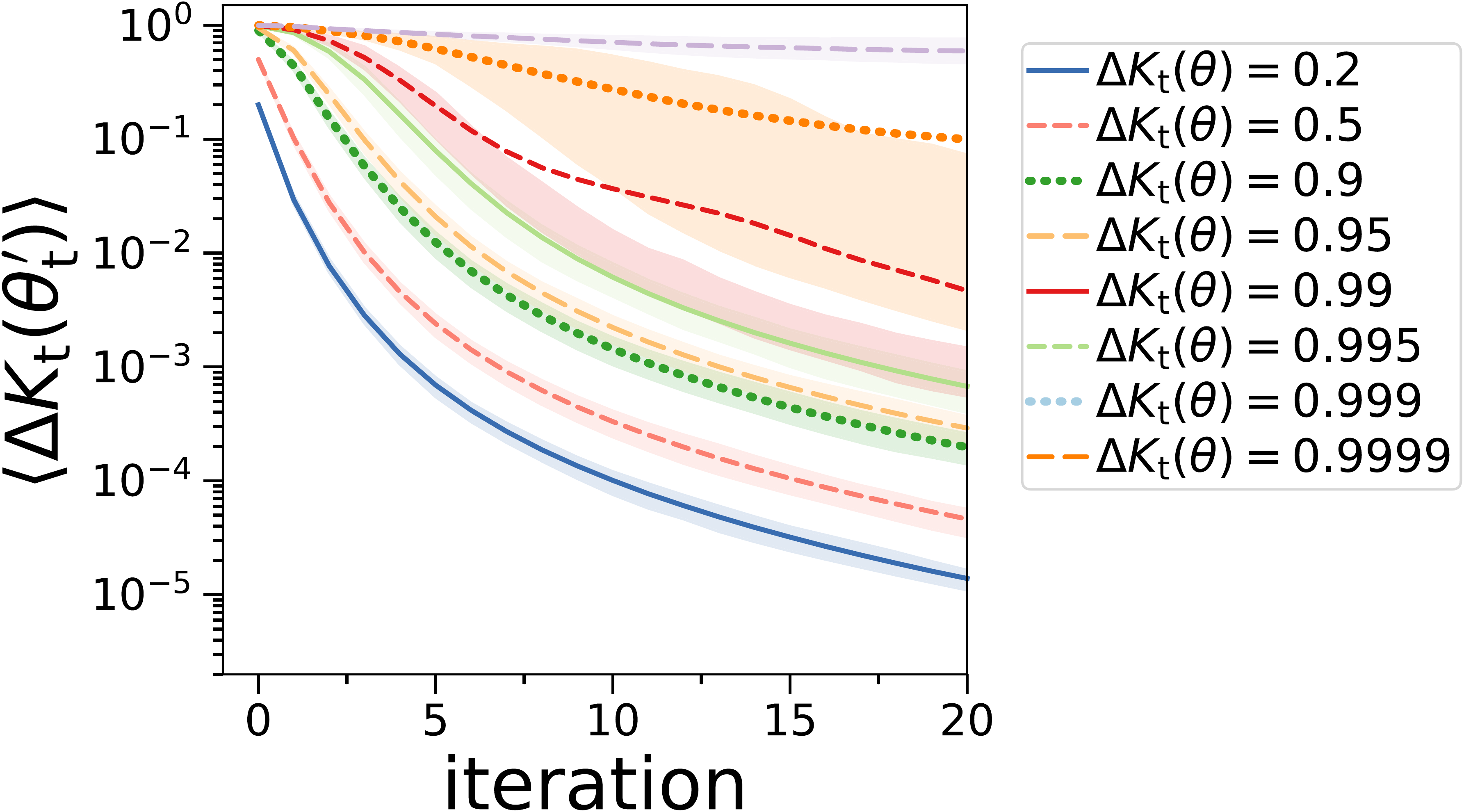}
	\caption{Different initial infidelities $\Delta K_\text{t}(\boldsymbol{\theta})$ for \idg{a} A-QNG ($\epsilon_\text{R}=10^{-1}$) and \idg{b} A-GQNG ($\beta=\frac{1}{2}$). Average infidelity $\langle \Delta K_\text{t}(\boldsymbol{\theta}_\text{t}') \rangle$ against number of iterations of gradient ascent. Shaded area is the 20 and 80 percentile over 50 instances of training. 
    The PQC is YZ-CNOT with $N=10$ and $p=10$.
	}
	\label{fig:training_comp_fidelity}
\end{figure}

In Fig.\ref{fig:training_comp_fidelity}, we compare training with A-QNG and A-GQNG for different initial infidelities $\Delta K_\text{t}(\boldsymbol{\theta})$. We can train the PQC even for very high infidelities. Note that the Gaussian kernel approximation breaks down for $\Delta K_\text{t}(\boldsymbol{\theta})>1-\frac{1}{2^N}\approx0.999$. However, training is still possible then, although at a slower rate.

\begin{figure}[htbp]
	\centering
	\subfigimg[width=0.35\textwidth]{a}{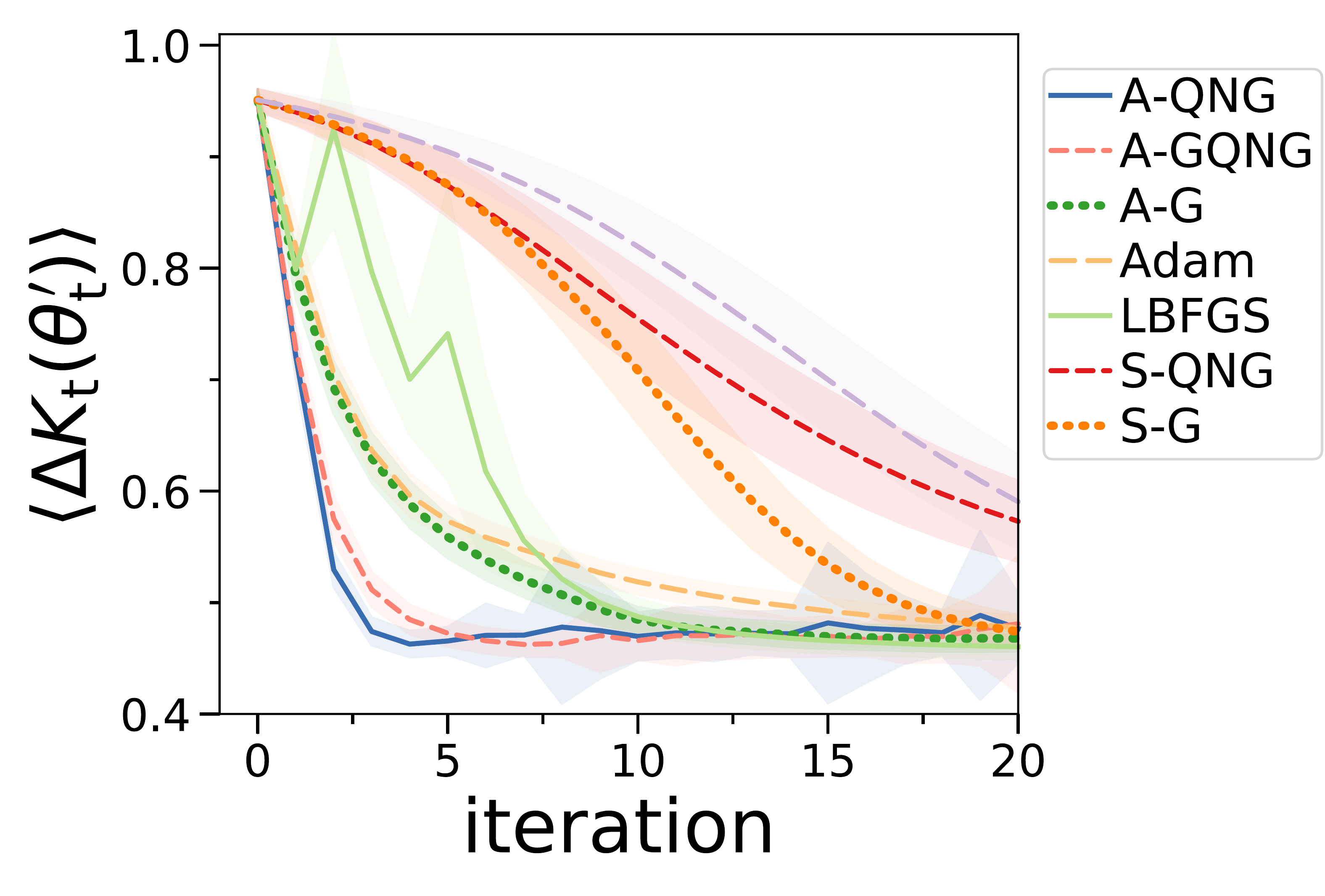}
	\caption{Training with YZ-CNOT PQC where target state cannot be learned perfectly, i.e. $K_0=\text{max}_{\boldsymbol{\theta}}K_\text{t}(\boldsymbol{\theta})\approx0.5$.
	Average infidelity $\langle \Delta K_\text{t}(\boldsymbol{\theta}_\text{t}') \rangle$ against number of iterations of gradient ascent for different optimization routines. 
	Initial infidelity is $\Delta K_\text{t}(\boldsymbol{\theta})=0.95$,  $N=10$ and $p=10$.
	}
	\label{fig:training_stateshift_sup}
\end{figure}

Now, again the target state to be learned cannot be perfectly represented by PQC, i.e. $K_0=\text{max}_{\boldsymbol{\theta}}K_\text{t}(\boldsymbol{\theta})\approx0.5$. We show the trajectories for the YZ-CNOT PQC in Fig.\ref{fig:training_stateshift_sup}. We note that A-QNG and A-GQNG find the solution faster compared to other algorithms.

\section{Further data on training quantum control problems}\label{app:control_data}
Here, we show further data on training quantum control problems.
The one-dimensional Ising model with a transverse field $h$ and longitudinal field $g$ is given by
\begin{equation}
H_0=\sum_{i=1}^N\label{eq:ising_sup} \sigma^x_n\sigma^x_{n+1}+h\sigma^z_n+g\sigma^x_n\,.
\end{equation}
The driving Hamiltonian $H(t)$ with arbitrary local control over the transverse field $h_n(t)$ is given by
\begin{equation}\label{eq:control_sup}
H(t)=\sum_{i=1}^N \sigma^x_n\sigma^x_{n+1}+h_n(t)\sigma^z_n+g\sigma^x_n\,.
\end{equation}
We initially start from the all zero state $\ket{\psi(0)}=\ket{0}$ and evolve the quantum state in time with the time-dependent Schr\"odinger equation $\partial_t\ket{\psi(t)}=H(t)\ket{\psi(t)}$.

\begin{figure}[htbp]
	\centering
	\subfigimg[width=0.24\textwidth]{b}{trainStdQutipEvalEvalQCN6d16r20n0s0_02H10S1n60o7g1b1q0r0_1h1g1.pdf}\hfill
	\subfigimg[width=0.24\textwidth]{b}{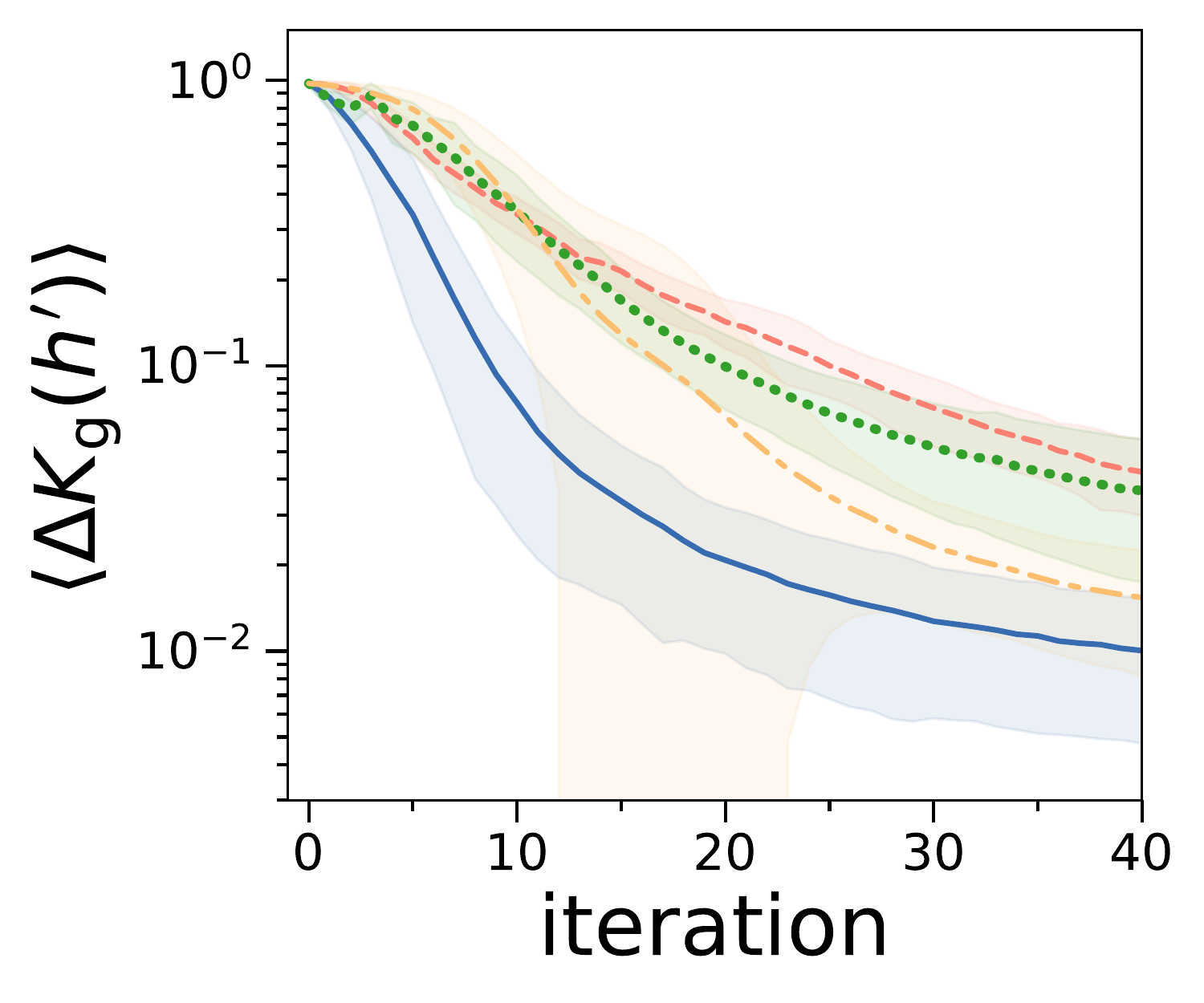}
	\caption{\idg{a} Average infidelity $\langle \Delta K_\text{g}(h') \rangle$ against number of iterations for finding quantum control protocol for driving parameters $h'$. We use driving Hamiltonian \eqref{eq:control_sup} with $g=1$, $N=6$, $\Delta t=1$ and $T=d=16$. The initial state is the zero state and the target state is the ground state of \eqref{eq:ising_sup} with $g=1$ and $h=1$. We average training over 20 instances of initially random protocols. Shaded area is the standard deviation of the infidelity.
	\idg{b} Log plot of the same training data.
	}
	\label{fig:training_control_sup}
\end{figure}

In Fig.\ref{fig:training_control_sup}, we show both linear and logarithmic plots for the quantum control problem investigated in the main text. 

\begin{figure}[htbp]
	\centering
	\subfigimg[width=0.24\textwidth]{b}{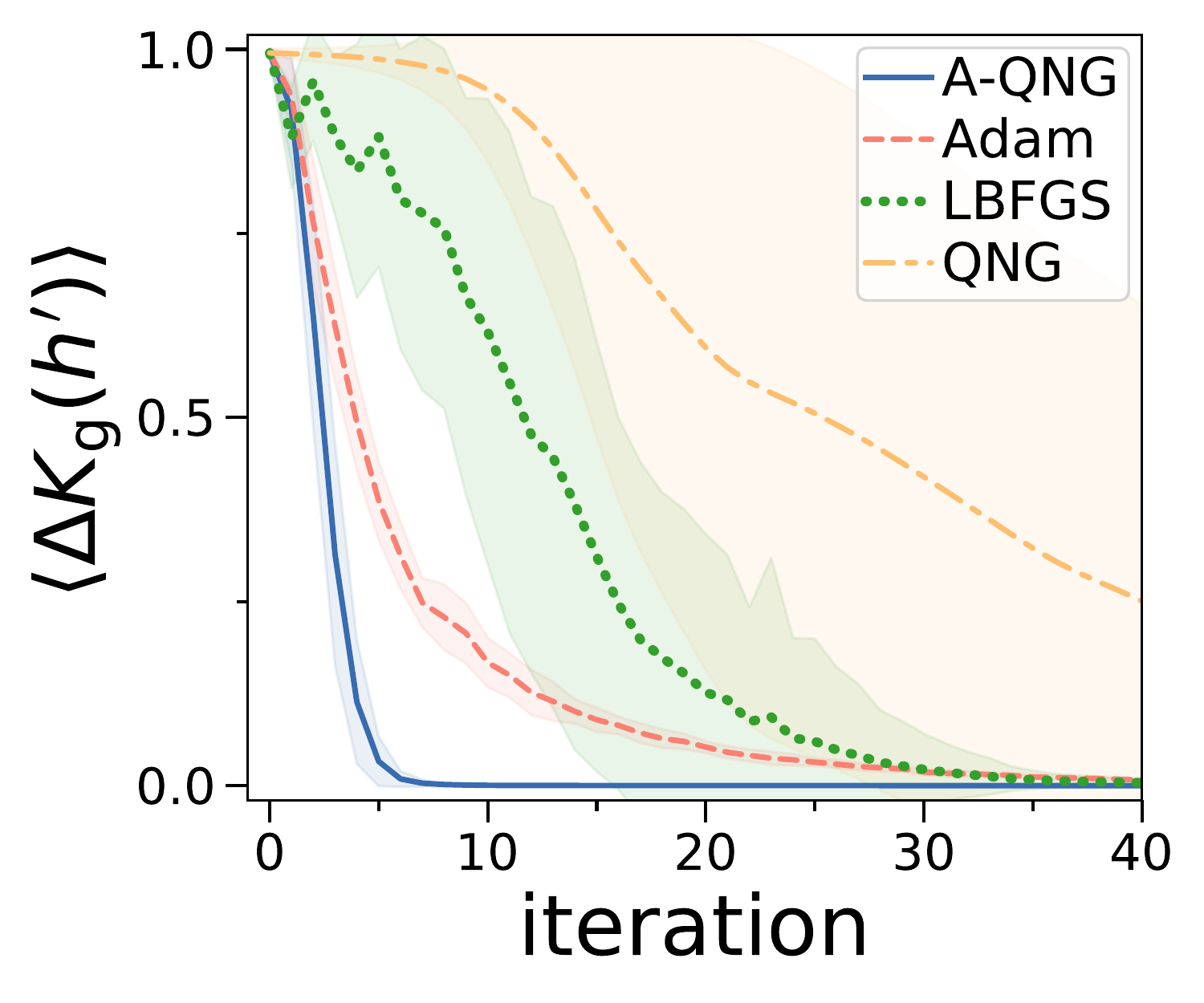}\hfill
	\subfigimg[width=0.24\textwidth]{b}{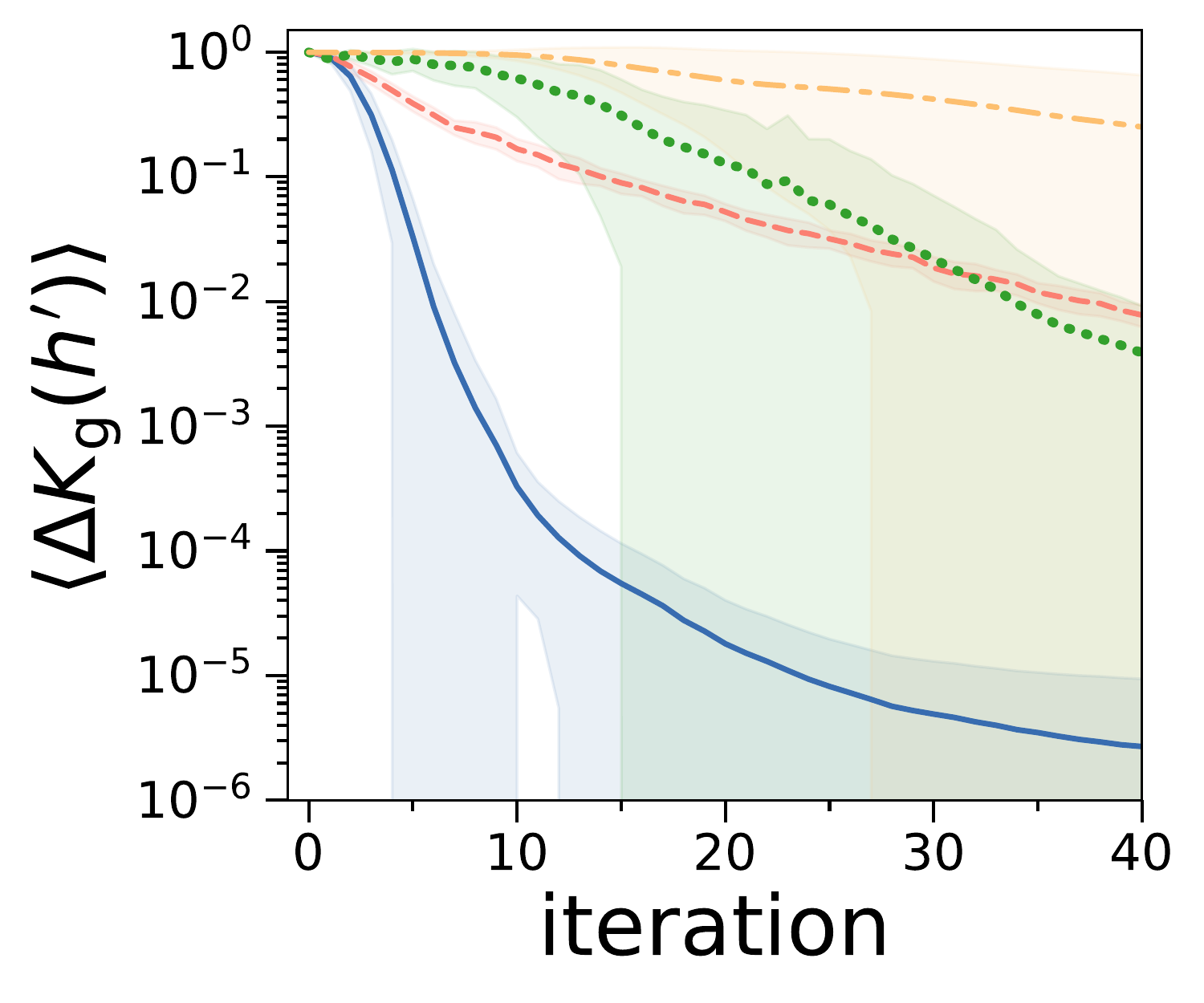}
	\caption{
	\idg{a} Average infidelity $\langle \Delta K_\text{g}(h') \rangle$ against number of iterations for finding quantum control protocol for driving parameters $h'$. We use driving Hamiltonian \eqref{eq:control_sup} with $g=0$, $N=8$, $\Delta t=1$ and $T=d=10$. The target state is the ground state of \eqref{eq:ising_sup} with $g=0$ and $h=1$. We average training over 20 instances of initially random protocols. Shaded area is the standard deviation of the infidelity.
	\idg{b} Log plot of the same training data.
	}
	\label{fig:training_control_ising_sup}
\end{figure}

Next, we study the case of finding a protocol for the ground state of the transverse Ising model with $g=0$. The result is shown in Fig.\ref{fig:training_control_ising_sup}. We observe that A-QNG converges to a good solution with low infidelity much faster than the other investigated algorithms.

\section{PQC with Gaussian kernel}\label{app:gaussian}
We show a PQC with Gaussian kernel that can be analytically calculated with the following $N$ qubit quantum state 
\begin{equation}
\ket{\psi(\boldsymbol{\theta})}=\bigotimes_{n=1}^N(\cos(\frac{\theta_n}{2})\ket{0}+\sin(\frac{\theta_n}{2})\ket{1})\,.
\end{equation}
The fidelity of two states parameterized by $\boldsymbol{\theta}$, $\boldsymbol{\theta}'$ is given by
\begin{equation}
K(\boldsymbol{\theta},\boldsymbol{\theta}')=\abs{\braket{\psi(\boldsymbol{\theta})}{\psi(\boldsymbol{\theta}')}}^2=\prod_{n=1}^N(1+\frac{1}{2}\cos(\Delta\theta_n))
\end{equation}
where we define $\Delta\boldsymbol{\theta}=\boldsymbol{\theta}-\boldsymbol{\theta}'$ as the difference between the two parameter sets. We now assume $\abs{\Delta\theta_n}\ll 1$ and that all the differences of the parameters are equal $\Delta\theta_1=\dots=\Delta\theta_N$. We then find in the limit of many qubits $N$
\begin{equation}\label{eq:expfidelity}
K(\boldsymbol{\theta},\boldsymbol{\theta}')\approx\prod_{n=1}^N(1-\frac{1}{4}\Delta\theta_n^2)\xrightarrow[N \to \infty]{}\text{exp}(-\frac{1}{4}\Delta\boldsymbol{\theta}^\text{T}\Delta\boldsymbol{\theta})\,.
\end{equation}
The QFIM of above model is $\mathcal{F}(\boldsymbol{\theta})=I$, $I$ being the identity matrix.
We can now modify the model to describe a generic PQC with the QFIM $\mathcal{F}$. 
We now assume that the parameters of the PQC $\boldsymbol{\theta}$ are not independent, but are related via $\boldsymbol{\theta}=\mathcal{F}^{\frac{1}{2}}\boldsymbol{\mu}$, where $\boldsymbol{\mu}$ is some $N$ dimensional parameter. We find for the transformed parameters $\boldsymbol{\mu}$
\begin{equation}
K(\boldsymbol{\mu},\boldsymbol{\mu}')\approx\text{exp}(-\frac{1}{4}\Delta\boldsymbol{\mu}^{\text{T}} \mathcal{F}\Delta\boldsymbol{\mu})\,,
\end{equation}
where $\Delta\boldsymbol{\mu}=\boldsymbol{\mu}-\boldsymbol{\mu}'$ and we used $\vert\mathcal{F}^{\frac{1}{2}}\boldsymbol{\mu}\vert^2=\boldsymbol{\mu}^{\text{T}} \mathcal{F}\boldsymbol{\mu}$. A first order Taylor expansion gives us
\begin{equation}
K(\boldsymbol{\mu},\boldsymbol{\mu}')\approx\text{exp}(-\frac{1}{4}\Delta\boldsymbol{\mu}^{\text{T}} \mathcal{F}\Delta\boldsymbol{\mu})\approx 1-\frac{1}{4}\Delta\boldsymbol{\mu}^{\text{T}} \mathcal{F}\Delta\boldsymbol{\mu}\,,
\end{equation}
which is the relation between fidelity and parameter distance for the QFIM~\cite{meyer2021fisher,liu2019quantum}.

\section{Generalized quantum natural gradient}\label{app:gqng}
The generalized quantum natural gradient (GQNG) is given by
\begin{equation}\label{eq:general_gradient_sup}
G_\beta(\boldsymbol{\theta})=\mathcal{F}^{-\beta}(\boldsymbol{\theta})\nabla K_\text{t}(\boldsymbol{\theta})\,,
\end{equation}
with $\beta\in[0,1]$, the derivative of the fidelity $\nabla K_\text{t}(\boldsymbol{\theta})$ and the QFIM $\mathcal{F}(\boldsymbol{\theta})$.
Standard gradient ascent is $G_0(\boldsymbol{\theta})=\nabla K_\text{t}(\boldsymbol{\theta})$, whereas the QNG is $G_1(\boldsymbol{\theta})=\mathcal{F}^{-1}(\boldsymbol{\theta})\nabla K_\text{t}(\boldsymbol{\theta})$~\cite{stokes2020quantum,yamamoto2019natural}. The standard gradient is reliable, however it can require many iterations before convergence. The QNG gives the optimal gradient in parameter space, however it is unstable without regularization. We trace this instability back to the inverse of the QFIM $\mathcal{F}^{-1}(\boldsymbol{\theta})$, which when ill conditioned is highly sensitive to small changes in parameter $\boldsymbol{\theta}$. 
A common approach is to take care of the ill-conditioned QFIM is by adding the identity matrix to the QFIM before inversion $\mathcal{F}'=\mathcal{F}+\epsilon_\text{R}I$, where $I$ is the identity matrix and $\epsilon_\text{R}$ a small hyperparameter.
By choosing appropriate $\epsilon_\text{R}>0$ the QNG can become stable.
We now want to investigate when regularization is necessary as a function of $\beta$ in the first place.
$\beta$ allows us to tune the trade-off between stability and optimal updates.
With the Gaussian kernel, a gradient ascent update $\boldsymbol{\theta}'=\boldsymbol{\theta}+\alpha G_\beta(\boldsymbol{\theta})$ gives us the fidelity
\begin{equation}
    K_\text{t}(\boldsymbol{\theta}')=K_0\exp[-\frac{1}{4}(\Delta\boldsymbol{\theta}+\alpha G_\beta(\boldsymbol{\theta}))^\text{T}\mathcal{F}(\boldsymbol{\theta})(\Delta\boldsymbol{\theta}+\alpha G_\beta(\boldsymbol{\theta})]\,,
\end{equation}
where $K_0$ is the maximal possible fidelity and $\Delta\boldsymbol{\theta}=\boldsymbol{\theta}-\boldsymbol{\theta}_\text{t}$.
For convenience, in the following we do not explicitly write the $\boldsymbol{\theta}$ parameter for $\mathcal{F}(\boldsymbol{\theta})$ and $\nabla K_\text{t}(\boldsymbol{\theta})$ and we set $\epsilon_\text{R}=0$
\begin{align*}
&K_\text{t}(\boldsymbol{\theta}')=K_\text{t}(\boldsymbol{\theta})\exp[-\frac{\alpha}{4}(\alpha G_\beta^\text{T}\mathcal{F} G_\beta+G_\beta^\text{T}\mathcal{F}\Delta\boldsymbol{\theta}+\Delta\boldsymbol{\theta}^\text{T}\mathcal{F} G_\beta)]\\
&=K_\text{t}(\boldsymbol{\theta})\exp[-\frac{\alpha}{4}(\alpha \nabla K_\text{t}^\text{T}\mathcal{F}^{1-2\beta} \nabla K_\text{t}+2\Delta\boldsymbol{\theta}^\text{T}\mathcal{F}^{1-\beta} \nabla K_\text{t})]\,,
\end{align*}
where we used $\mathcal{F}^\text{T}=\mathcal{F}$.
For choice $\beta\le\frac{1}{2}$, the updated fidelity does not contain any terims with $\mathcal{F}^{-\delta}$ having a negative exponent, which could cause instabilities. Thus, gradient ascent with $\beta=\frac{1}{2}$ and $G_\frac{1}{2}(\boldsymbol{\theta})=\mathcal{F}^{-\frac{1}{2}}\nabla K_\text{t}(\boldsymbol{\theta})$ is stable without the need of regularization. For $\beta>\frac{1}{2}$, $\epsilon_\text{R}>0$ has to be chosen to take care of the ill-conditioned inverse of the QFIM.

\section{Adaptive learning rate}\label{app:learning}
Here, we derive the adaptive learning rates for gradient ascent.
We would like to solve the optimisation problem $\boldsymbol{\theta}_\text{t}=\text{argmax}_{\boldsymbol{\theta}}\abs{\braket{\psi(\boldsymbol{\theta})}{\psi_\text{t}}}^2$. 
First, we assume that $\text{max}_{\boldsymbol{\theta}}\abs{\braket{\psi(\boldsymbol{\theta})}{\psi_\text{t}}}^2=K_0=1$, i.e. the PQC is able to perfectly represent the state. We relax $K_0<1$ further below.
For the initial parameter $\boldsymbol{\theta}$ we have
a fidelity $K_\text{t}(\boldsymbol{\theta})$.
Gradient ascent with the GQNG uses the update rule for the next parameter $\boldsymbol{\theta}_1$
\begin{equation}\label{eq:grad_update}
\boldsymbol{\theta}_1=\boldsymbol{\theta}+\alpha_1 G_\beta(\boldsymbol{\theta})\,,
\end{equation}
with learning rate $\alpha_1$ and GQNG $G_\beta(\boldsymbol{\theta})=\mathcal{F}^{-\beta}(\boldsymbol{\theta})\nabla K_\text{t}(\boldsymbol{\theta})$.
We assume now that the fidelity follows a Gaussian kernel, and we would like to choose $\alpha_1$ such that this update is as close as possible to the correct solution $\boldsymbol{\theta}_1\approx\boldsymbol{\theta}_\text{t}$.
Given the Gaussian kernel, we have
\begin{equation}
K_\text{t}(\boldsymbol{\theta})=e^{-\frac{1}{4}\Delta\boldsymbol{\theta}^{\text{T}}\mathcal{F}(\boldsymbol{\theta})\Delta\boldsymbol{\theta}}\,.
\end{equation}
where we defined the distance between target parameter and initial parameter $\Delta \boldsymbol{\theta}=\boldsymbol{\theta}_\text{t}-\boldsymbol{\theta}$. We then find by applying the logarithm
\begin{equation}\label{eq:log_kernel}
-4\log(K_\text{t}(\boldsymbol{\theta}))=\Delta\boldsymbol{\theta}^{\text{T}}\mathcal{F}(\boldsymbol{\theta})\Delta\boldsymbol{\theta}\,.
\end{equation}
A reordering  of \eqref{eq:grad_update} gives us
\begin{equation}
\Delta\boldsymbol{\theta}=\alpha_1 G_\beta(\boldsymbol{\theta})\,
\end{equation}
We now multiply both sides with $\mathcal{F}^{\frac{1}{2}}(\boldsymbol{\theta})$ and get
\begin{equation}
\mathcal{F}^{\frac{1}{2}}(\boldsymbol{\theta})\Delta\boldsymbol{\theta}=\alpha_1\mathcal{F}^{\frac{1}{2}}(\boldsymbol{\theta}) G_\beta(\boldsymbol{\theta})\,,
\end{equation}
followed by taking square on both sides
\begin{equation}
\Delta\boldsymbol{\theta}^{\text{T}}\mathcal{F}(\boldsymbol{\theta})\Delta\boldsymbol{\theta}=\alpha_1^2 G_\beta(\boldsymbol{\theta})^{\text{T}}\mathcal{F}(\boldsymbol{\theta})\nabla G_\beta(\boldsymbol{\theta})\,,
\end{equation}
where we used $\vert\mathcal{F}^{\frac{1}{2}}\boldsymbol{\mu}\vert^2=\boldsymbol{\mu}^{\text{T}} \mathcal{F}\boldsymbol{\mu}$.
We insert \eqref{eq:log_kernel} and get
\begin{equation}
\alpha_1=\frac{2\sqrt{-\log(K_\text{t}(\boldsymbol{\theta}))}}{\sqrt{G_\beta(\boldsymbol{\theta})^{\text{T}}\mathcal{F}(\boldsymbol{\theta}) G_\beta(\boldsymbol{\theta})}}\,,
\end{equation}
with the first update rule
\begin{equation}
\boldsymbol{\theta}_1=\boldsymbol{\theta}+\alpha_1 G_\beta(\boldsymbol{\theta})\,.
\end{equation}
We assumed for above calculations that the PQC is able to represent the target quantum state perfectly, i.e. $\text{max}_{\boldsymbol{\theta}}\abs{\braket{\psi(\boldsymbol{\theta})}{\psi_\text{t}}}^2=1$. If this is not the case, we have to adjust the learning rate to take this into account.

Assume the target state is given by 
\begin{equation}
\ket{\psi_\text{t}}=\sqrt{K_0}\ket{\psi(\boldsymbol{\theta}_\text{t})}+\sqrt{1-K_0}\ket{\psi_\text{o}}\,,
\end{equation}
where $\ket{\psi_\text{o}}$ is some state that is orthogonal to any other state that can be represented by the PQC, i.e. $\abs{\braket{\psi_\text{o}}{\psi(\boldsymbol{\theta})}}^2=0\,\,\forall\boldsymbol{\theta}$ and $K_0$ is the maximal possible fidelity of the PQC.
Then, the first update rule is moving in the correct direction, however overshoots the target parameters. We now calculate the update rule that takes this into account.
We find
\begin{equation}
K_\text{t}(\boldsymbol{\theta})=K_0e^{-\frac{1}{4}\Delta\boldsymbol{\theta}^{\text{T}}\mathcal{F}(\boldsymbol{\theta})\Delta\boldsymbol{\theta}}
\end{equation}
\begin{equation}
K_\text{t}(\boldsymbol{\theta}_1)=K_0e^{-\frac{1}{4}(\boldsymbol{\theta}_1-\boldsymbol{\theta}_\text{t})^{\text{T}}\mathcal{F}(\boldsymbol{\theta})(\boldsymbol{\theta}_1-\boldsymbol{\theta}_\text{t})}
\end{equation}
where $K_\text{t}(\boldsymbol{\theta}_1)$ is the fidelity after applying the first update rule.
Now our goal is to find the corrected update rule
\begin{equation}
\boldsymbol{\theta}_\text{t}=\boldsymbol{\theta}+\alpha_\text{t}G_\beta(\boldsymbol{\theta})\,,
\end{equation}
with updated learning rate $\alpha_\text{t}$. By subtracting the two update rules we get
\begin{equation}
\boldsymbol{\theta}_1-\boldsymbol{\theta}_\text{t}=(\alpha_1-\alpha_\text{t})G_\beta(\boldsymbol{\theta})\,.
\end{equation}
We insert above equations into the fidelities and get
\begin{equation}
K_\text{t}(\boldsymbol{\theta})=K_0 e^{-\frac{1}{4}\alpha_\text{t}^2 G_\beta^{\text{T}}\mathcal{F}G_\beta}
\end{equation}
\begin{equation}
K_\text{t}(\boldsymbol{\theta}_1)=K_0 e^{-\frac{1}{4}(\alpha_1-\alpha_\text{t})^2G_\beta^{\text{T}}\mathcal{F}G_\beta}
\end{equation}
By dividing above equations, we can solve for $\alpha_\text{t}$ and find
\begin{equation}\label{eq:update_add_sup}
\alpha_\text{t}=\frac{1}{2}\left(\frac{4}{\alpha_1 G_\beta(\boldsymbol{\theta})^{\text{T}}\mathcal{F}(\boldsymbol{\theta})G_\beta(\boldsymbol{\theta})}\log\left(\frac{K_\text{t}(\boldsymbol{\theta}_1)}{K_\text{t}(\boldsymbol{\theta})}\right)+\alpha_1\right)
\end{equation}
with final update rule
\begin{equation}
\boldsymbol{\theta}_\text{t}'=\boldsymbol{\theta}+\alpha_\text{t}G_\beta(\boldsymbol{\theta})
\end{equation}
where $\boldsymbol{\theta}_\text{t}'$ is the parameter for the PQC after one iteration of gradient ascent.

\section{Variance of gradient}\label{app:variance}
We now show how to calculate the variance of the gradient $\nabla K_\text{t}(\boldsymbol{\theta})$ using the Gaussian kernel.
We assume a PQC with $M$ parameters, QFIM $\mathcal{F}(\boldsymbol{\theta})$ and a target state $\ket{\psi_\text{t}}=\sqrt{K_0}\ket{\psi(\boldsymbol{\theta}_\text{t})}+\sqrt{1-K_0}\ket{\psi_0}$, where $\ket{\psi_\text{o}}$ is some state that is orthogonal to any other state that can be represented by the PQC, i.e. $\abs{\braket{\psi_\text{o}}{\psi(\boldsymbol{\theta})}}^2=0\,\,\forall\boldsymbol{\theta}$.
The initial state of the PQC is $\ket{\psi(\boldsymbol{\theta})}$ with random parameter $\boldsymbol{\theta}$.
We define the vector to the correct solution as $\Delta\boldsymbol{\theta}=\boldsymbol{\theta}_\text{t}-\boldsymbol{\theta}$. 
We now assume that the entries of the vector $\Delta\boldsymbol{\theta}$ are sampled from the uniform distribution $\Delta\boldsymbol{\theta}^{(n)}\sim\text{uniform}(-\sqrt{3}\frac{\sqrt{-4\log(A_0)}}{\sqrt{\text{Tr}(\mathcal{F})}},\sqrt{3}\frac{\sqrt{-4\log(A_0)}}{\sqrt{\text{Tr}(\mathcal{F})}})$, where $A_0$ is identified later on.
The mean is $\langle\Delta\boldsymbol{\theta}^{(n)}\rangle=0$ and variance $\langle(\Delta\boldsymbol{\theta}^{(n)})^2\rangle=\frac{-4\log(A_0)}{\text{Tr}(\mathcal{F})}$. Further, the average of the product of index $n$ and $m$ of the parameter $\Delta\boldsymbol{\theta}$ is given by $\langle(\Delta\boldsymbol{\theta}^{(n)}\Delta\boldsymbol{\theta}^{(m)})\rangle=\delta_{nm}\frac{-4\log(A_0)}{\text{Tr}(\mathcal{F})}$, where $\delta_{nm}$ is the Kronecker delta.
First, we calculate
\begin{align*}
\langle\Delta\boldsymbol{\theta}^\text{T}\mathcal{F}\Delta\boldsymbol{\theta}\rangle&=\sum_{n,m}\mathcal{F}_{nm}\langle\Delta\boldsymbol{\theta}^{(n)}\Delta\boldsymbol{\theta}^{(m)}\rangle\\
&=\sum_{n}\mathcal{F}_{nn}\langle\Delta\boldsymbol{\theta}^{(n)}\Delta\boldsymbol{\theta}^{(n)}\rangle=-4\log(A_0)
\end{align*}
Now, we assume that $M$ is large, such that we can apply the central limit theorem $\langle e^{-\frac{1}{4}\Delta\boldsymbol{\theta}^\text{T}\mathcal{F}\Delta\boldsymbol{\theta}}\rangle\approx e^{-\frac{1}{4}\langle\Delta\boldsymbol{\theta}^\text{T}\mathcal{F}\Delta\boldsymbol{\theta}\rangle}$.
Then, we find
\begin{equation}
A_0=e^{-\frac{1}{4}\langle \Delta\boldsymbol{\theta}^\text{T}\mathcal{F}\Delta\boldsymbol{\theta}\rangle}\,.
\end{equation}
We can now identify with the fidelity
\begin{equation}
 K_\text{t}(\boldsymbol{\theta})=K_0 e^{-\frac{1}{4}\langle \Delta\boldsymbol{\theta}^\text{T}\mathcal{F}\Delta\boldsymbol{\theta}\rangle}=A_0K_0\,.
\end{equation}
The gradient of the fidelity is given by
\begin{equation}
\nabla K_\text{t}(\boldsymbol{\theta})=-\frac{1}{2}\mathcal{F}\Delta\boldsymbol{\theta}K_0 e^{-\frac{1}{4}\Delta\boldsymbol{\theta}^\text{T}\mathcal{F}\Delta\boldsymbol{\theta}}\,.
\end{equation}
We now want to calculate the variance of the $n$-th element of the gradient vector. The mean of the gradient $\langle \nabla K_\text{t}(\boldsymbol{\theta})\rangle=0$.
For the square of the gradient, we find
\begin{equation}
\langle (\nabla K_\text{t}(\boldsymbol{\theta})^{(n)})^2\rangle=\frac{1}{4}\langle((\mathcal{F}\Delta\boldsymbol{\theta})^{(n)})^2\rangle K_0^2 e^{-\frac{1}{8}\langle\Delta\boldsymbol{\theta}^\text{T}\mathcal{F}\Delta\boldsymbol{\theta}\rangle}\,.
\end{equation}
We now calculate the element
\begin{align*}
&\langle((\mathcal{F}\Delta\boldsymbol{\theta})^{(n)})^2\rangle=\sum_{k,m}\langle  \mathcal{F}_{nm}\Delta\boldsymbol{\theta}^{(m)}\mathcal{F}_{nk}\Delta\boldsymbol{\theta}^{(k)}\rangle\\
&=\sum_{m}\mathcal{F}_{nm}\mathcal{F}_{nm}\langle  (\Delta\boldsymbol{\theta}^{(m)})^2\rangle=\sum_{m}\mathcal{F}_{nm}\mathcal{F}_{mn}\frac{-4\log(A_0)}{\text{Tr}(\mathcal{F})}\\
&=\frac{-4\log(A_0)(\mathcal{F}^2)_{nn}}{\text{Tr}(\mathcal{F})}
\end{align*}
where we used $\mathcal{F}_{mn}=\mathcal{F}_{nm}$ and $\sum_{m}\mathcal{F}_{nm}\mathcal{F}_{nm}=\sum_{m}\mathcal{F}_{nm}\mathcal{F}_{mn}=(\mathcal{F}^2)_{nn}$.
Now, we take the average over the variance of the different gradient entries $n$ and use the Pythagorean theorem (i.e. one can sum over the variance of independent variables)
\begin{align*}
\langle\langle(\mathcal{F}\Delta\boldsymbol{\theta})^{(n)})^2\rangle_{\Delta\boldsymbol{\theta}}\rangle_n&=\frac{1}{M}\sum_n\langle((\mathcal{F}\Delta\boldsymbol{\theta})^{(n)})^2\rangle\\
&=\frac{-4\log(A_0)\text{Tr}(\mathcal{F}^2)}{M\text{Tr}(\mathcal{F})}\,,
\end{align*}
where $\langle\langle . \rangle_{\Delta\boldsymbol{\theta}}\rangle_n$ indicates that we average first over $\Delta\boldsymbol{\theta}^{(n)}$ and then $n$.
Finally, we get
\begin{align*}
\langle\langle(\nabla &K_\text{t}(\Delta\boldsymbol{\theta})^{(n)})^2\rangle_{\Delta\boldsymbol{\theta}}\rangle_n\\
&=-\log\left[\frac{ K_\text{t}(\boldsymbol{\theta})}{K_0}\right]\frac{\text{Tr}(\mathcal{F}^2)}{M\text{Tr}(\mathcal{F})}K_0^2e^{-\frac{1}{8}\langle\Delta\boldsymbol{\theta}^\text{T}\mathcal{F}\Delta\boldsymbol{\theta}\rangle}\\
&=\log\left[\frac{K_0}{K_\text{t}(\boldsymbol{\theta})}\right] K_\text{t}(\boldsymbol{\theta})^2\frac{\text{Tr}(\mathcal{F}^2)}{M\text{Tr}(\mathcal{F})}\,.\numberthis\label{eq:var_grad_sup}
\end{align*}
The variance becomes maximal when $ K_\text{t}(\boldsymbol{\theta})=K_0\exp(-\frac{1}{2})\approx0.6K_0$.
The variance of the gradient decays linearly with number of parameters $M$ and is independent of qubit number $N$. 

We now derive the lower bound of the variance.
The variance of the eigenvalues of $\mathcal{F}$ with $M$ eigenvalues $\lambda_n$ is always greater zero
\begin{align*}
\frac{1}{M}\sum_{n=1}^M\lambda_n^2-(\frac{1}{M}\sum_{n=1}^M\lambda_n)^2&\ge0\\
\sum_{n=1}^M\lambda_n^2&\ge\frac{1}{M}(\sum_{n=1}^M\lambda_n)^2\,.
\end{align*}
With the relation $\text{Tr}(\mathcal{F})=\sum_{n=1}^M\lambda_n$ and $\text{Tr}(\mathcal{F}^2)=\sum_{n=1}^M\lambda_n^2$, we find
\begin{equation}
\text{Tr}(\mathcal{F}^2)\ge\frac{1}{M}\text{Tr}(\mathcal{F})^2\,.
\end{equation}
Thus, the variance is lower bounded by
\begin{align*}
\text{var}(\partial_k K_\text{t}(\boldsymbol{\theta}))\ge\frac{\text{Tr}(\mathcal{F}(\boldsymbol{\theta}))}{M^2} K_\text{t}(\boldsymbol{\theta})^2\log\left[\frac{K_0 }{K_\text{t}(\boldsymbol{\theta})}\right]\,.\numberthis\label{eq:var_grad_lower_sup}
\end{align*}

\begin{figure*}[htbp]
	\centering
	\subfigimg[width=0.24\textwidth]{a}{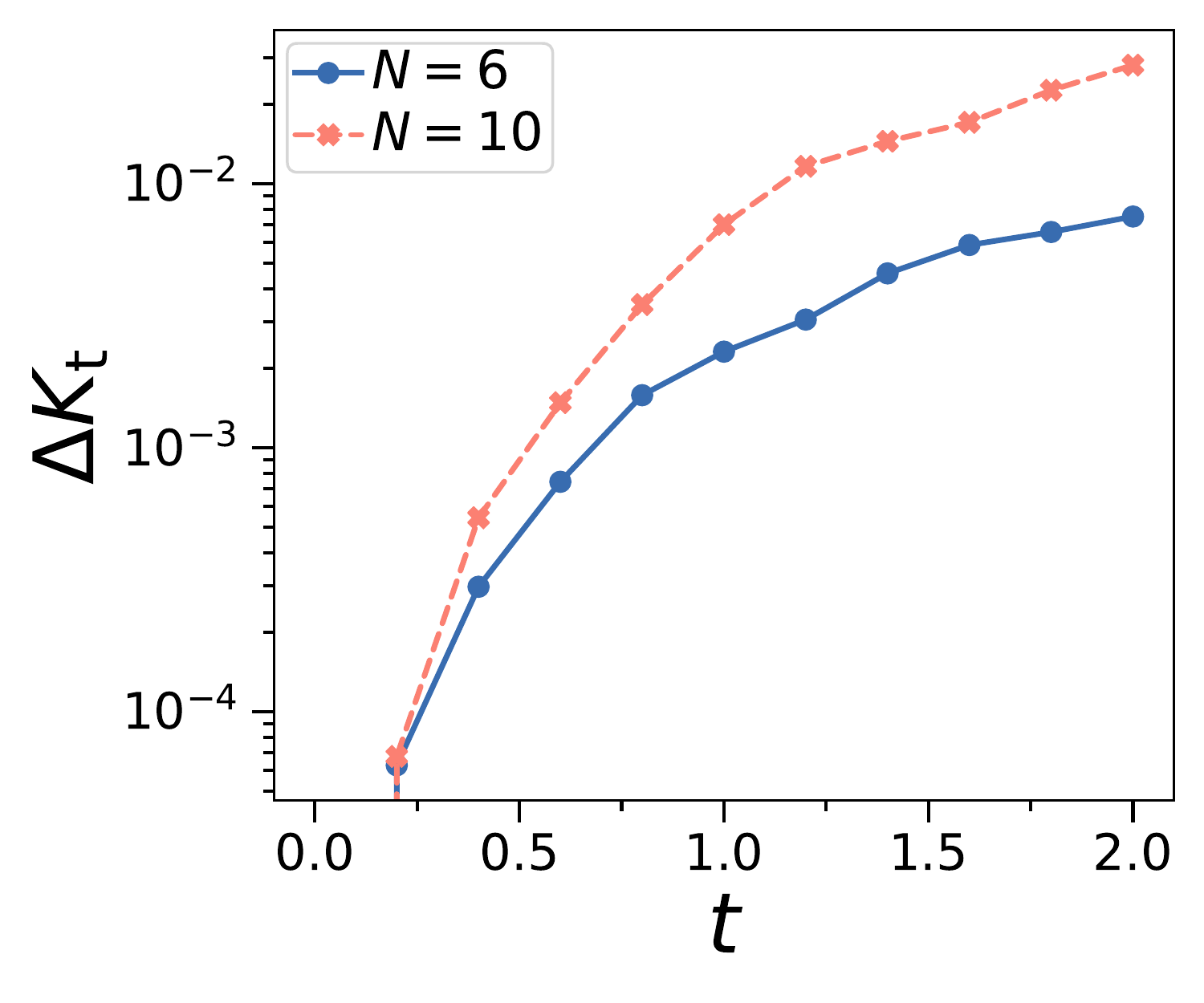}
	\subfigimg[width=0.24\textwidth]{b}{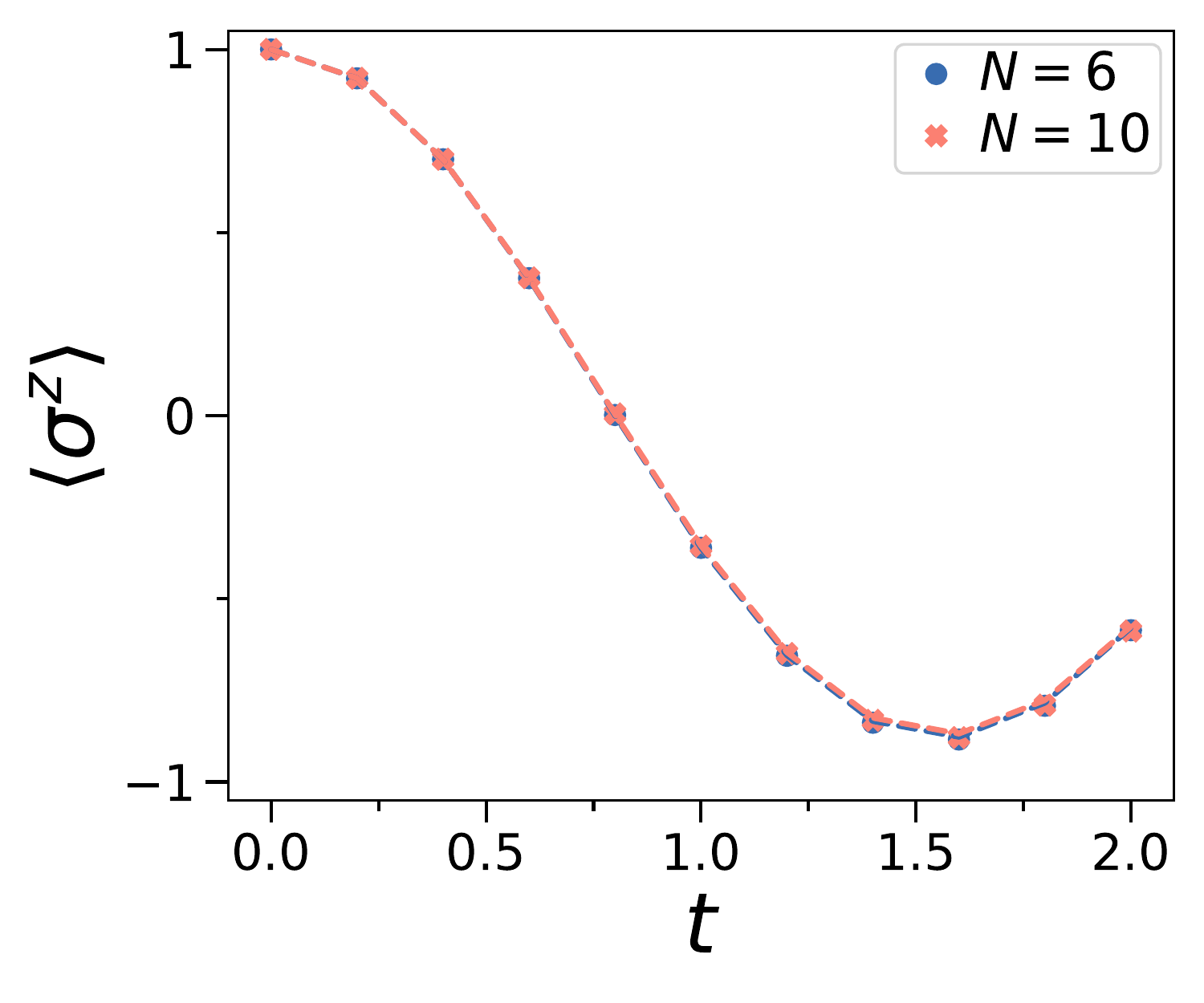}
	\subfigimg[width=0.24\textwidth]{c}{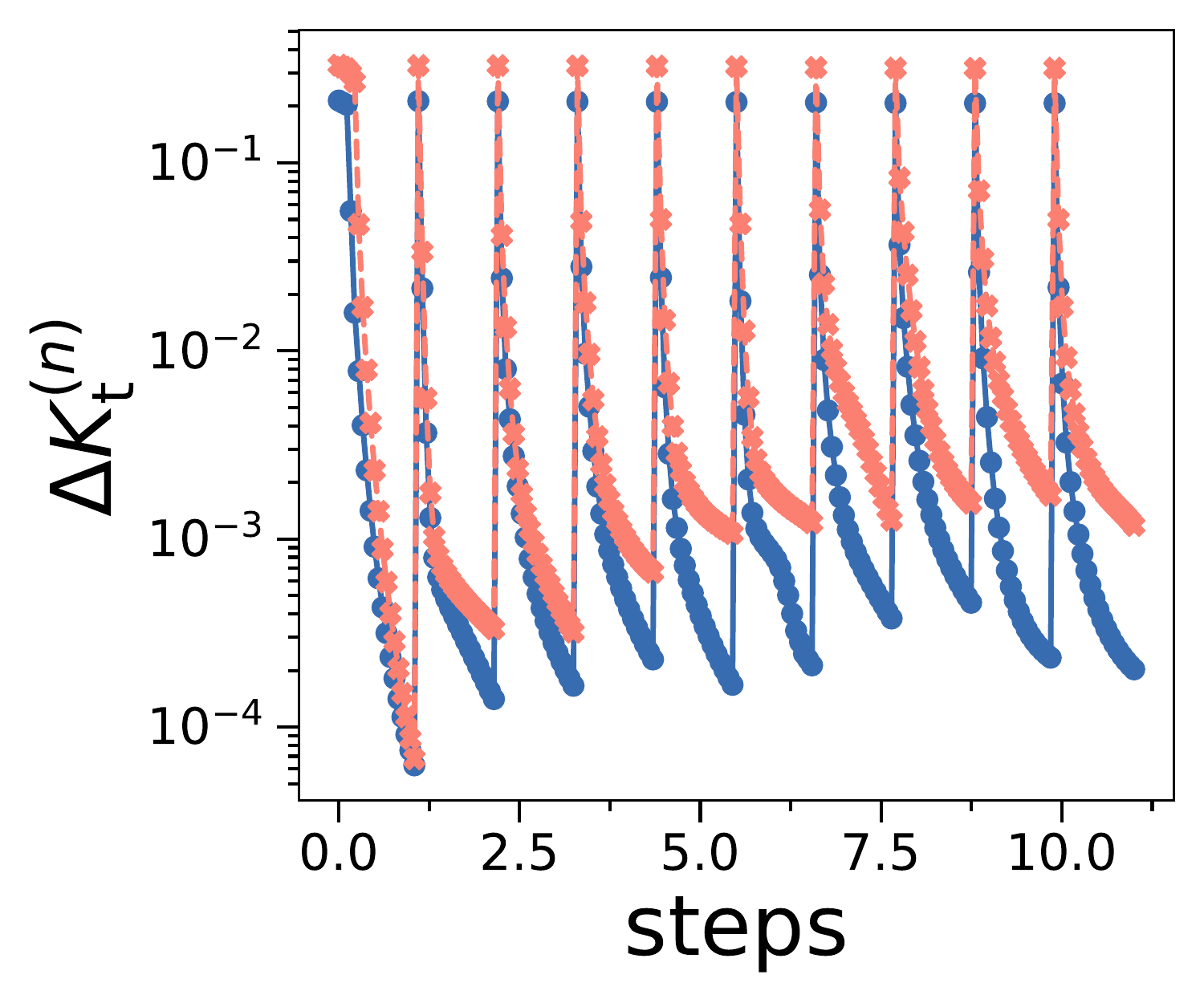}
	\caption{Demonstration of projected variational quantum dynamics using adaptive gradient descent with GQNG ($\beta=\frac{1}{2}$). 
	\idg{a} Fidelity in respect to the exact solution.
	\idg{b} Magnetization found compared to exact solution in dashed lines
	\idg{c} training loss for each Trotter step $N_\text{s}$, each trained with $N_\text{t}$ training iterations.
	We simulate the evolution for total time $T=N_\text{s}\Delta t$ using $N_\text{s}=10$ Trotter steps and $\Delta t=0.2$, with each step trained with $N_\text{t}=20$ training steps.
    The PQC used is YZ-CNOT with $p=N$, $J=0.25$, $h=1$ and initial state is the computational basis state with all zeros.}
	\label{fig:pqvd_sup}
\end{figure*}

\section{Fidelity of time evolution}\label{app:fidelity_time}
Here, we show the lower bound of the fidelity of a time evolved state.
Assume we evolve the state $\ket{\psi}$ with the Hamiltonian $H$ for a time $\Delta t$. The eigenstates of $H$ are given by $\ket{\phi_n}$ and eigenenergies by $\lambda_n$. The initial state is now expressed in terms of the eigenstates $\ket{\psi}=\sum_n \beta_n \ket{\phi_n}$, with overlap $\beta_n$ with the $n$-th eigenstate. We assume that $\ket{\psi}$ has only non-zero overlap $\beta_n\ne0$ with $L$ eigenstates, where $\lambda_1$ is the smallest and $\lambda_L$ the largest eigenvalue with non-zero overlap. We define $\Delta E=\lambda_L-\lambda_1$.
Now, the evolved state is given by $\ket{\psi'}=\exp(-iH\Delta t)\ket{\psi}=\sum_{n=1}^L\beta_n e^{-i\lambda_n\Delta t}\ket{\phi_n}$.
The fidelity between initial and evolved state is given by
\begin{equation}
K=\abs{\braket{\psi'}{\psi}}^2=\abs{\sum_{n=1}^L\abs{\beta_n}^2e^{-i\lambda_n \Delta t}}^2=
\end{equation}
We now derive a lower bound for $K$ for small times $\Delta t$. $K$ is minimal for small $\Delta t$ when $\abs{\beta_1}^2=\frac{1}{2}$ and $\abs{\beta_L}^2=\frac{1}{2}$. This is evident as the relative phase between largest and smallest eigenvalues evolves the fastest. We have in this case
\begin{equation}
K=\frac{1}{2}(1+\cos((\lambda_L-\lambda_1)\Delta t))=\cos^2(\frac{\Delta E\Delta t}{2})\,.
\end{equation}
Now, a lower bound for $K$ can be found via a first order Taylor expansion
\begin{equation}
K>1-\frac{1}{4}(\Delta E\Delta t)^2\,.
\end{equation}

\section{Projected variational quantum dynamics}\label{app:pvqd}
Here, we simulate the dynamics of the transverse Ising model with the projected variational quantum dynamics method~\cite{barison2021efficient}. The Hamiltonian is given by
\begin{equation}
H=J\sum_{n=1}^N \sigma_n^z\sigma_{n+1}^z+h\sum_{i=1}^N\sigma_n^x
\end{equation}
In Fig.\ref{fig:pqvd_sup}a, we show the fidelity in respect to the exact solution.
In Fig.\ref{fig:pqvd_sup}b, we show the dynamics of the magnetization $\langle\sigma^z\rangle=\langle\frac{1}{N}\sum_i\sigma_i^z\rangle$ over time $t$. We observe good match between simulation and exact result.
In Fig.\ref{fig:pqvd_sup}c, we show the training loss for each of the Trotter steps.

\end{document}